\documentclass[sigconf,screen]{acmart}

\setcopyright{none}
\copyrightyear{2024}
\acmYear{2024}
\acmConference[]{ACM}{Survey}{RecSys Poisoning}

\usepackage{graphics}
\DeclareGraphicsExtensions{.eps,.gz,.eps,.pdf,.png,.jpg}
\graphicspath{{./}}

\usepackage{pifont}% http://ctan.org/pkg/pifont
\newcommand{\cmark}{\ding{51}}%

\usepackage{multirow}
\usepackage{amsmath}
\usepackage{adjustbox}
\usepackage{paralist}

\usepackage{parskip}
\usepackage{todonotes}

\newcommand{\editone}[1]{ #1}
\newcommand{\editthree}[1]{ #1}

\newcommand{\sstitle}[1]{\smallskip\noindent\textbf{#1.\/}}

\definecolor{lightgray}{gray}{0.9}

\usepackage{rotating}
\newcommand*{\headformat}[1]{#1}

\newlength{\maxlen}
\newcommand*{\shortheadformat}[1]{#1}

\newlength{\shortmaxlen}
\newcommand*{\head}[1]{%
    \begin{sideways}
      \makebox[\maxlen][l]{\headformat{#1}}
    \end{sideways}}
    
\newcommand*{\shorthead}[1]{%
    \begin{sideways}
      \makebox[\shortmaxlen][l]{\shortheadformat{#1}}
    \end{sideways}}

\newcommand\Mark[1]{\textsuperscript#1}

\def\Snospace~{\S{}}

\usepackage[T1]{fontenc}
\usepackage[utf8]{inputenc}
\usepackage{microtype}

\begin{document}

\setlength{\belowdisplayskip}{3pt}
\setlength{\belowdisplayshortskip}{3pt}
\setlength{\abovedisplayskip}{3pt}
\setlength{\abovedisplayshortskip}{3pt}

%\title{A Survey of Poison Attacks: Creation and Detection}
% \title{A Survey of Poison Attacks and Countermeasures for Recommender Systems}

\title{Manipulating Recommender Systems: A Survey of Poisoning Attacks and Countermeasures}
% \title{\small Supplementary Materials for: \\ Manipulating Recommender Systems: A Survey of Poisoning Attacks and Countermeasures}

\author{
Thanh Toan Nguyen\Mark{1}, 
Quoc Viet Hung Nguyen\Mark{1},
Thanh Tam Nguyen\Mark{1*},%\ua\thanks{\ua Corresponding author},
Thanh Trung Huynh\Mark{2},
Thanh Thi Nguyen\Mark{3},
Matthias Weidlich\Mark{4},
Hongzhi Yin\Mark{5}% <-this % stops a space
%\vspace{1.6mm}\\
%\fontsize{9}{9}\selectfont\rmfamily\itshape
%\small
%\Mark{1}Griffith University (Australia),
%%\Mark{2}Ecole Polytechnique Federale de Lausanne (Switzerland),
%\Mark{3}University of Bremen (Germany),
%\Mark{4}The University of Queensland (Australia)
%%\Mark{3}Hanoi University of Science and Technology,
%%\Mark{4}Huazhong University of Science and Technology
%%\Mark{4}Humboldt-Universit\"at zu Berlin (Germany),
%%\Mark{5}Queen's University Belfast (UK)
}

\affiliation{%
  \institution{
  \Mark{1}Griffith University,
  \Mark{2}\'{E}cole Polytechnique F\'{e}d\'{e}rale de Lausanne,
  \Mark{3}Monash University,
  \Mark{4}Humboldt-Universit\"at zu Berlin,
  \Mark{5}The University of Queensland
  }
%  \institution{\'{E}cole Polytechnique F\'{e}d\'{e}rale de Lausanne}
  \country{}
}

\renewcommand{\shortauthors}{Nguyen, et al.}

\begin{abstract}

Recommender systems have become an integral part of online services to help users locate specific information in a sea of data. However, existing studies show that some recommender systems are vulnerable to poisoning attacks, particularly those that involve learning schemes. A poisoning attack is where an adversary injects carefully crafted data into the process of training a model, with the goal of manipulating the system's final recommendations. Based on recent advancements in artificial intelligence, such attacks have gained importance recently. At present, we do not have a full and clear picture of why adversaries mount such attacks, nor do we have comprehensive knowledge of the full capacity to which such attacks can undermine a model or the impacts that might have. \editone{While numerous countermeasures to poisoning attacks have been developed, they have not yet been systematically linked to the properties of the attacks. Consequently, assessing the respective risks and potential success of mitigation strategies is difficult, if not impossible. This survey aims to fill this gap by primarily focusing on poisoning attacks and their countermeasures. This is in contrast to prior surveys that mainly focus on attacks and their detection methods. Through an exhaustive literature review, we provide a novel taxonomy for poisoning attacks, formalise its dimensions, and accordingly organise 30+ attacks described in the literature. Further, we review 40+ countermeasures to detect and/or prevent poisoning attacks, evaluating their effectiveness against specific types of attacks}. This comprehensive survey should serve as a point of reference for protecting recommender systems against poisoning attacks. The article concludes with a discussion on open issues in the field and impactful directions for future research.
A rich repository of resources associated with poisoning attacks is available at \url{https://github.com/tamlhp/awesome-recsys-poisoning}.

\end{abstract}

\begin{CCSXML}
<ccs2012>
<concept>
<concept_id>10002951.10003317.10003347.10003350</concept_id>
<concept_desc>Information systems~Recommender systems</concept_desc>
<concept_significance>500</concept_significance>
</concept>
<concept>
<concept_id>10002978.10002997</concept_id>
<concept_desc>Security and privacy~Intrusion/anomaly detection and malware mitigation</concept_desc>
<concept_significance>300</concept_significance>
</concept>
<concept>
<concept_id>10010147.10010257</concept_id>
<concept_desc>Computing methodologies~Machine learning</concept_desc>
<concept_significance>300</concept_significance>
</concept>
</ccs2012>
\end{CCSXML}



\maketitle

\section{Introduction}
\label{sec:introduction}

In the era of data deluge, identifying relevant information to support decision making has become a challenging task for the users of online services. By helping users to find useful, personalised information, a recommender system can not only increase a service’s usability, it can also contribute directly to the platform provider’s ultimate success. For this reason, companies like Youtube, Amazon, and eBay are deploying recommender systems as one of the primary means by which their customers locate items of interest – be they videos, individual products, or categories of products ~\cite{aggarwal2016recommender}. 
Recommender systems have actually been around for decades; however, it has only been in the last decade that they have seen great commercial success as evidenced by their enormous growth in recent years. In fact, back in 2018, Industry Arc forecast the recommender system market to increase from US\$1.14 billion to US\$12.03 billion by 2025~\cite{IndustryARC} – a prediction that looks set to prove true. Statistics like this are clear evidence that recommender systems have become an integral part of helping internet users navigate the sea of choices they face every time they go online.
% items, in making their decisions and lessening consumer 
%confusion caused by 
%over-choices.

In many practical scenarios, recommender systems operate in a relatively open environment. That is, these systems rely on data that is generated by other users, whether actively through comments, posts, and ratings, or passively through clicks, views, or purchases. While this openness is key to the success of many systems, given that the quality of a recommendations usually depends on drawing from a large pool of candidate data, it also renders recommender systems prone to manipulation ~\cite{lam2006you, wang2015recommender, polatidis2017recommender, zhang2019deep}. More specifically, recommender systems are known to be vulnerable to poisoning attacks ~\cite{rosenberg2021adversarial}, in which an adversary injects carefully crafted data into the model’s training data so as to change the model’s behaviour. Clearly, such vulnerabilities are unavoidable to some extent, even though the operators of recommender systems are generally acutely aware that such attacks can occur. Yet, poisoning attacks remain a very powerful means of manipulating users. As such, they can seriously undermine the commercial success of any company falling victim to such an attack.
%widely accepted by users 
%and operators of RSs. 
%, encouraging researchers to explore and validate the robustness of 
%contemporary RSs when these systems are posing to different types of poison 
%attacks. 
%On the other hand, many service providers leverage poison attacks to 
A prime example of a poisoning attack is to crowdsource users who will post fraudulent ratings of a company’s products for a small fee per rating. Fraudulent positive ratings will promote a company’s products, while fraudulent negative ones will demote those of the company’s competitors. Marketplaces such as Amazon and eBay suffer from such attacks daily ~\cite{lam2004shilling, sidhu2016attacks, lit2021survey}. The use of fake reviews to increase the number of recommendations of particular movies is another well-documented example of a poisoning attack motivated by economic considerations~\cite{SonyFakeNews}.
% \footnote{\url{http://news.bbc.co.uk/2/hi/entertainment/1368666.stm}}
%Take Sony Pictures as another example; they exploited fake reviews to make 
%some of their movies be recommended to more 
%users~\footnote{http://news.bbc.co.uk/2/hi/entertainment/1368666.stm}. 
Hence, a comprehensive understanding of the different types of poisoning attacks that can be launched against recommender systems – along with their goals, capacities, and impacts – is an important prerequisite for designing robust models.

\subsection{Prior Classifications and Surveys of Recommender Systems}

\begin{table*}[!h]
\centering
	\caption{\editthree{The comparison between our work and existing surveys}.}
	\label{tbl_the_gap}
	\vspace{-1em}
	\footnotesize
	\begin{adjustbox}{max width=0.95\textwidth}
		\begin{tabular}{llcccccccc} 
			\toprule
			\textbf{Category}                  & 
			\textbf{Sub-category}               
			& \editthree{\cite{fan2022survey}} & \editthree{\cite{zhao2023manipulating}} & 
			\cite{wang2015comparative}      & \cite{wang2018comparative}   & 
			\cite{si2020shilling}    & \cite{sundar2020understanding}     &  
			\cite{rezaimehr2021survey}    & \textbf{Ours}    \\  
			\midrule
			\multirow{4}{*}{Attacks}         & Classic heuristic 
			attacks       & \editthree{\cmark}   & \editthree{\cmark}              & \cmark & \cmark & \cmark & \cmark 
			& \cmark & \cmark  \\
			& AI-based attacks  & \editthree{\cmark}  & \editthree{\cmark}                            
			&                       &                       
			&                       &                       
			&                       & \cmark  \\
			& Unified taxonomy  &   &                               
			&                       &                       
			&                       &                       
			&                       & \cmark  \\
			& Formalisation of attack dimensions &  & 
			&                       &                       
			&                       &                       
			&                       & \cmark  \\
			& \editthree{Recommender system domain of applications} &  &                              
			&                       &                       
			&                       &                       
			&                       & \cmark  \\ 
			\midrule
			\multirow{6}{*}{Countermeasures} & Detection of classic 
			heuristic attacks   & \editthree{\cmark} & \editthree{\cmark}  & \cmark & \cmark & \cmark & 
			\cmark & \cmark & \cmark  \\
			& Detection of AI-based attacks & \editthree{\cmark}  & \editthree{\cmark}             
			&                       &                       
			&                       &                       
			&                       & \cmark  \\
                & \editthree{Formalising vulnerabilities in recommender systems} &  &                 
			&                       &                       
			&                       &                       
			&                       & \cmark  \\
			& Prevention against classic heuristic attacks &  &               
			&                       &                       & \cmark & \cmark 
			&                       & \cmark  \\
			& Prevention against AI-based attacks   &    &         & 
			\multicolumn{1}{l}{}  & \multicolumn{1}{l}{}  & 
			\multicolumn{1}{l}{}  & \multicolumn{1}{l}{}  & 
			\multicolumn{1}{l}{}  & \cmark  \\
			& Countermeasures effective against attacks &  & 
			&                       &                       
			&                       &                       
			&                       & \cmark  \\
			& Countermeasures weak against attacks &   &   
			&                       &                       
			&                       &                       
			&                       & \cmark  \\ 
			\midrule
			\multirow{2}{*}{Discussion}        & Research gaps \& future 
			research directions & \editthree{\cmark}     & \editthree{\cmark}              &                       
			&                       & \cmark &                       
			&                       & \cmark  \\
			& Materials collection &  & 
			&                       &                       
			&                       &                       
			&                       & \cmark  \\
			\bottomrule
		\end{tabular}
	\end{adjustbox}
	\vspace{-1em}
\end{table*}

Recommender systems typically fall into one of three categories: i) \emph{content-based filtering}; ii) \emph{collaborative filtering}; or iii) \emph{hybrid systems}. Of these, collaborative filtering (CF) methods~\cite{shi2014collaborative} have been the mainstream of recommender system research, due to the high quality of the resulting recommendations. Therefore, in this survey, we have focused on CF-based recommender systems, breaking them down into the learning scheme used~\cite{aditya2016comparative}, as follows:
%Depends on how a model learns from its underlying data, CF recommender 
%systems can also be further classified into:
\begin{compactitem}
    \item \emph{Memory-based CF recommender systems} rely on nearest neighbour search, which is applied directly to a user’s interaction history. As such, these approaches do not work with an explicit     model. For example, a recommendation algorithm finds the closest users to a given target user, identifies their common preferences, and derives a recommendation for the target user based on those shared preferences.
    
    % Memory based approaches directly works with values of recorded interactions, assuming no model, and are essentially based on nearest neighbours search (for example, find the closest users from a user of interest and suggest the most popular items among these neighbours). 
    
    \item \emph{Model-based CF recommender systems} presume an underlying ``generative'' model that explains the user-item interactions. These systems attempt to estimate a model which generates the recommendations for a given target user.
\end{compactitem}
With the models adopted by CF recommender systems becoming more and more advanced over the past two decades, the attacks on them have changed as well. These attacks can be divided into two broad paradigms: AI attacks and classic heuristic attacks.
%principal eras in developing a poison attack against recommender systems based on the attacks' 
%efficiency.

%\begin{figure}[!h]
%\vspace{-0.5em}
%	\centering
%	\includegraphics[width=1.0\linewidth]{attacks_history.png}
%	\vspace{-0.45cm}
%	\caption{\editthree{The evolution of poisoning attacks over the last seven years w.r.t. the ratio of \textbf{AI-based} attacks and \textbf{classic} attacks. The two paradigms are distinguished based on the methodology followed to train the attack. The number of attacks has increased over time, with a clear shift from classic attacks to AI-based attacks.}} 
%	\label{fig:attacks_history}
%	% \vspacwe{-0.5em}
%\end{figure}

\begin{compactenum}
    \item \emph{Classic heuristic attacks:} The strategy these attacks follow comprises two steps: (1) detecting malicious users, which is formulated as an optimisation problem; and (2) solving the problem, which is done through a heuristic technique. Attacks that fall into this category date back 20 years, but they are still employed these days. While many of the specific methods used by these attacks seem relatively simple, they have proven to be effective. 
    
    \item \emph{AI-based attacks:} Over the last seven years, we have seen the advent of various attacks that train an end-to-end framework to forge user profiles and imitate authentic user behaviours. For instance, several schemes based on generative adversarial networks (GANs) have recently been proposed that automatically mimic the behaviour of genuine users to influence the targeted system ~\cite{tang2020revisiting, lin2020attacking, aggarwal2021generative}. Another technique is to exploit the reward signal in a reinforcement learning scheme as a back-door into the recommender model ~\cite{song2020poisonrec}.  
\end{compactenum}

%As \autoref{fig:attacks_history} shows, the last seven years has seen an increasing number of attacks on recommender systems being proposed. At the same time, there has been clear shift from classic attacks to AI-based attacks. This illustration lists the number of published attacks per year and category, as extracted using bibliometrics analysis from major computer science repositories such as Google Scholar and Scopus~\cite{de2009bibliometrics}.
%Details of of the survey methodology are provided in~\autoref{sec:survey_methodology}. 
%This trend has also been confirmed by security experts: the number of attacks that exploit AI methods can be expected to grow substantially over the coming years~\cite{top5AI}.

 \begin{figure}[!h]
	\centering
% 	\vspace{-1em}
	\includegraphics[width=1.0\linewidth]{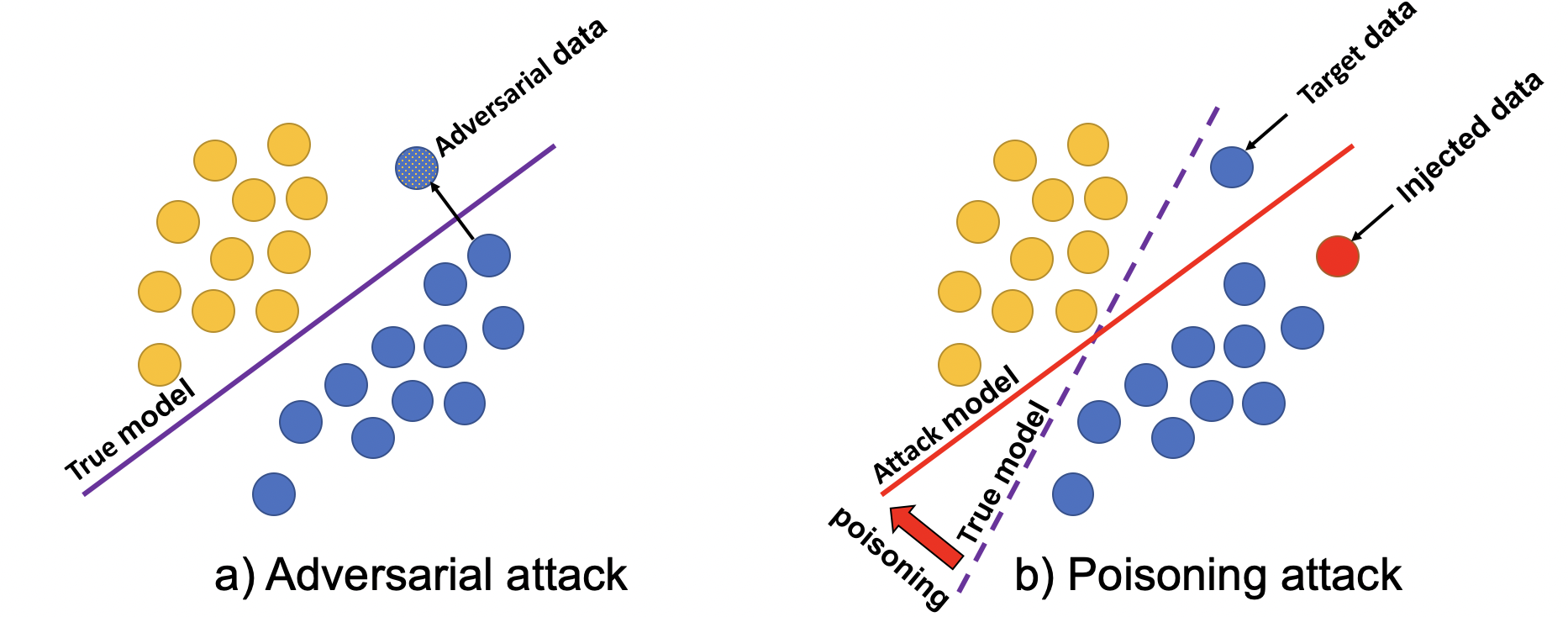}
	\vspace{-1em}
	\caption{Adversarial attacks vs. poisoning attacks}
	\label{fig:poison_adversarial}
	\vspace{-1em}
\end{figure}

While {AI-based poisoning attacks} pose severe threats to recommender  systems, existing surveys of attacks on these  systems primarily focus on classic heuristic attacks and their respective countermeasures ~\cite{wang2015comparative, wang2018comparative, si2020shilling, sundar2020understanding, rezaimehr2021survey}. \editthree{The more recent studies cover a combination of heuristic and AI-based attacks. Additionally, some surveys cover poison attacks in general~\cite{fan2022survey}, while others investigate these attacks in specific domains. So far however, no one has undertaken a survey focussed on recommender systems~\cite{zhao2023manipulating}}.
In this survey, we address this gap by providing a comprehensive overview of the state-of-the-art attacks on recommender systems and the countermeasures one can take to prevent them. As summarised in \autoref{tbl_the_gap}, we also go beyond existing surveys by: providing a generic taxonomy for poisoning attacks; formalising the dimensions of this taxonomy; and linking the attacks to countermeasures. This last exercise not only highlights which measures effectively detect and prevent certain attacks, it also sheds light on the ability of attacks to resist certain countermeasures. Finally, our survey is accompanied by a collection of materials that will enable scientists to kick-start their own research work in the field.

\begin{figure}[!h]
    \centering
    \includegraphics[width=1.0\linewidth]{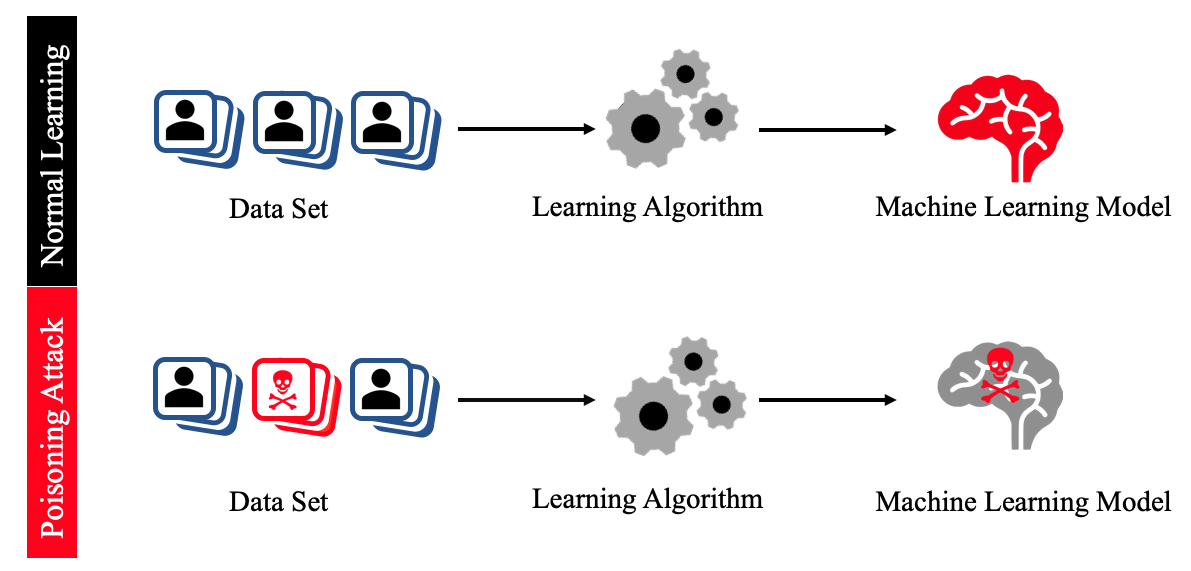}
    \vspace{-2em}
    \caption{The process of a typical poisoning attack}
    \label{fig:poison_attack}
    \vspace{-1em}
\end{figure}

\editone{Several existing surveys consider adversarial attacks on recommender systems~\cite{wang2019security, deldjoo2021survey}. While there are some conceptual similarities between poisoning attacks and adversarial attacks, they do have fundamental differences (as illustrated in \autoref{fig:poison_adversarial}). The aim of an adversarial attacks~\cite{ren2020generating,chang2022example,han2018adversarial,ren2020enhancing} is to find some adversarial samples that corrupt the outcome of a recommender system at the time of inference without altering the underlying model~\cite{wang2019security, deldjoo2021survey}. These attacks manipulate the input data to temporarily deceive the system into producing inaccurate predictions or recommendations during the attack. Poisoning attacks, however, are mounted during the model’s training phase~\cite{baracaldo2017mitigating}. Here, the attacker injects carefully crafted data into the model’s training data with the  goal of fundamentally converting the \emph{true model} into an \emph{attack model} that yields outcomes beneficial to the attacker's goal. By injecting malicious data during training, the attacker aims to bias the system's learning process and manipulate its recommendations in persistently.}

%\begin{figure*}[!h]
%\centering
%  \begin{minipage}{0.47\linewidth}
%	\centering
%% 	\vspace{-1em}
%	\includegraphics[width=1.0\linewidth]{poison_evasion.png}
%	\vspace{-1em}
%	\caption{Adversarial attacks vs. poisoning attacks}
%	\label{fig:poison_adversarial}
%	\vspace{-1em}
%\end{minipage}
%\quad
%\begin{minipage}{0.47\linewidth}
%\vspace{-2em}
%    \centering
%    \includegraphics[width=1.0\linewidth]{example1.png}
%    \vspace{-2em}
%    \caption{The process of a typical poisoning attack}
%    \label{fig:poison_attack}
%    \vspace{-1em}
%\end{minipage}
%\end{figure*}

\editone{To provide a more concrete example, consider an online marketplace that recommends products to users based on their browsing history. In an adversarial attack, an attacker might manipulate the content of a particular product listing or modify the user's browsing history temporarily to promote a specific item. This would influence the recommendations shown to that user during their current session. Conversely, in a poisoning attack, the attacker would inject malicious data into the system's training data, such as fraudulent reviews or manipulated purchase records. In this way, the attacker would bias the model to favour certain products or manipulate the recommendation rankings, thereby impacting the recommendations not just for a single user session but for a broader set of users over an extended period of time. 
Consequently, these two different types of attacks have distinct goals and implications. Adversarial attacks are largely orthogonal to the poisoning attacks covered in this survey.}

\subsection{Main Contributions}

This survey is intended as a point of reference for protecting recommender systems against poisoning attacks. To this end, we have structured the field of poisoning attacks and countermeasures based on a comprehensive literature review. More specifically, this survey contributes the following:

\begin{compactitem}
    \item We discuss the challenges that an attacker faces in the domain of recommender systems, while separating attacks on recommender systems from related approaches in fields such as machine learning and computer vision. For example, recommender systems generally exploit the correlations between user and item data to generate recommendations. These correlations can create a certain robustness in the underlying model that renders attacks difficult.
    
    \item We present the first comprehensive review of poisoning attacks on recommender systems, covering both classic heuristic modes of attack as well as AI-based attacks.  
    To structure and present these 30+ attacks, we devised a taxonomy of five dimensions that provides a holistic view of the full field of this research. 
    
    \item We present an extensive review of 40+ defenses for the identified poisoning attacks. Moreover, we link the attacks with the countermeasures, considering both measures that are effective against certain attacks as well as measures that can be expected to fail against certain attacks.
    
    \item Moving beyond our main objective, we also describe open issues in the field and conclude the survey with directions of promising future research. 

    \item To enable researchers to kickstart their work in the field, we have assembled a public repository~\footnote{\url{https://github.com/tamlhp/awesome-recsys-poisoning}}
    that includes all reviewed papers as well as the program codes and the datasets released in the context of these studies. As such, this repository provides researchers entering the field with a comprehensive collection of material to support their work in securing recommender systems against potential threats. 
    
\end{compactitem}

\subsection{Survey Methodology}
\label{sec:survey_methodology}

We implemented several procedures to ensure a high-quality survey. To identify pertinent articles that establish the state-of-the-art in poisoning attacks, we first extracted papers indexed by the major computer science repositories, including Scopus and DBLP.
%\footnote{\url{https://www.scopus.com}}
%\footnote{\url{https://dblp.uni-trier.de/}} 
Realising that some important work may not be indexed in these databases, we also screened articles using Google Scholar.
%\footnote{\url{https://scholar.google.com/}} 
With DBLP, we queried key terms in the articles’ titles using the repository’s search interface. With Google Scholar and Scopus, we searched the articles’ content. Having collected all relevant articles, we then processed and filtered them.
%We can collect a complete list of all relevant papers. 
Any paper deemed irrelevant was removed. For example, a paper was removed if it focused on poisoning attacks in a field other than recommender systems, such as machine learning or cybersecurity. We also concentrated on publications in top-tier venues (including RecSys, SIGIR, IJCAI, WWW, and KDD) and adopted a lightweight screening procedure for workshop proceedings and publications in entry-level venues.

\subsection{Structure of the Survey}

The rest of this article is organised as follows. \autoref{sec:background} provides an overview of recommender systems and the security landscape in this field. We also highlight the unique challenges in securing recommender systems from poisoning attacks. \autoref{sec:tanonomy} introduces a novel taxonomy for poisoning attacks and formally defines the dimensions of this taxonomy. In \autoref{sec:creation}, we review existing work on poisoning attacks according to this taxonomy, covering both model-agnostic and model-intrinsic techniques. \autoref{sec:countermeasures} describes countermeasures to detect and prevent poisoning attacks. Finally, we highlight research gaps and future research directions in \autoref{sec:future_direction}, before concluding the article in \autoref{sec:conclusion}. 

% \begin{figure}[!h]
%     \centering
%     \includegraphics[width=\linewidth]{survey_structure.png}
%     \caption{The structure of this survey. [\textcolor{red}{need to be updated}]}
%     \label{fig:survey_structure}
% \end{figure}

\section{Background}
\label{sec:background}

\subsection{Overview of Recommender Systems}

Recommender systems based on collaborative filtering (CF)~\cite{shi2014collaborative} are ubiquitous. For example, the recommender systems seen on YouTube, Netflix, Amazon, and Google Play are all CF-based systems. These models analyse the past interactions between users and items to derive a recommendation for a specific target user. Hence, CF-based recommenders can be divided into the following categories based on the underlying strategy(s) used to capture those historical interactions.

\subsubsection{Matrix-factorisation-based}
MF-based recommender systems \cite{koren2009matrix, alhijawi2020recommender} assume that a small number of latent factors are sufficient to represent the users' past behaviour. Based on that assumption, a low-rank matrix is used to estimate the full user-item rating matrix. More specifically, this low-rank matrix serves as the basis for inferring missing values in the full user-item rating matrix. A recommendation for a target user is then derived as a list of the items with the highest prediction scores, even if the user has never interacted with these items before. Some methods go a step further to improve the prediction results by assigning different weights based on the activeness of the users and the popularity of the items ~\cite{xue2017deep}.

\subsubsection{Graph-based}
Graph-based recommender systems model the historic interactions between users and items as a weighted bipartite graph, called the user preference graph ~\cite{huang2002graph, guo2020survey}. A random walk is then performed over the user preference graph to generate the recommendations. The walk starts at a user and returns to that user with a predefined probability. The stationary probability of the consequent random walk is then used to generate the list of recommendations. 

\subsubsection{Association-rule-based}
The idea behind recommender systems based on association rules is to find the co-occurrence patterns in items based on the ratings issued by users ~\cite{ali2016hybrid, anwar2020machine}. For instance, consider an example in which many users have assigned high ratings to two items, item X and item Y. As many users appreciate both items, the system assumes that there must be a hidden relation between the two. Hence, when a user assigns a high rating to item X, item Y will be recommended.

\subsubsection{Neighbourhood-based}
Neighbourhood-based recommender systems~\cite{sachan2013survey, periyasamy2017analysis} make recommendations based on the similarities between users and/or items. With \emph{user-based} similarities, the system finds the nearest users and aggregates their ratings to recommend items. The same process applies to \emph{item-based} similarities. 

\subsubsection{Deep-learning-based}
Deep-learning-based recommender systems involve a variety of deep learning models to model the interactions between users and items ~\cite{mu2018survey, zhang2019deep, ali2020deep} – from multilayer perceptrons ~\cite{pouyanfar2018survey} to autoencoders ~\cite{zhang2020survey} ; from deep reinforcement learning~\cite{pateria2021hierarchical, yu2021reinforcement, padakandla2021survey} to adversarial networks~\cite{cai2021generative}. What all these systems have in common is that they leverage contemporary algorithms in their training schemes, which generally significantly improves the quality of recommendations compared to the other categories of recommender systems.

%\begin{figure}[!h]
%\vspace{-0.5em}
%    \centering
%    \includegraphics[width=0.6\linewidth]{example1.png}
%    \vspace{-0.5em}
%    \caption{A typical poisoning attack process}
%    \label{fig:poison_attack}
%    \vspace{-0.3cm}
%\end{figure}

\subsection{Poisoning Attacks}

In a poisoning attack, an adversary tampers with the training data of a machine learning model to corrupt its integrity. Fig. \autoref{fig:poison_attack} illustrates the typical process of a poisoning attack compared to a normal learning process. In the latter case, a machine learning model is trained on some data that is subsequently used to derive a recommendation~\cite{mehrabi2021survey}. As such, the quality of the machine learning model depends on the quality of the data used for training. In a poisoning attack, data is injected into the training set, and hence the model, to produce unintended or harmful conclusions~\cite{pitropakis2019taxonomy}. In this way, adversaries can launch an attack against even the most advanced training algorithms and the most complex models.

In a poisoning attack on a recommender system, the data injected into the training set will typically relate to fake users and their fake ratings as an attempt to modify the resulting recommendations. Here, the usual goal is to either promote an item (if bolstering one’s own reputation) or demote one (if attacking one’s competitors) ~\cite{zhang2021pipattack}. The general course of action for an adversary is to infiltrate the recommender system by registering a number of fake users. These users will then rate a subset of items to coerce the desired result. Independent of the attack strategy, and no matter whether it is a classic heuristic or an AI-based attack, the fake data, which can be crafted either manually or automatically, will influence any model that learns from the data. As a result, the recommendations derived from the model will be manipulated towards the adversary's goal. 

One way to categorise poisoning attacks on recommender systems is to divide them by the type of recommender system they are designed to target. Are they \emph{model-agnostic}? or \emph{model-intrinsic}? Model-agnostic attacks ignore the underlying model and any algorithms used to build it. For this reason, the effectiveness of these types of attacks is typically limited. Model-intrinsic attacks, however, are optimised for a specific type of training process. As such, these attacks can cause substantial damage to the underlying model.

Some scholars have likened poisoning attacks to \emph{profile pollution} attacks, e.g., ~\cite{meng2014your}. However, there are some notable differences between the two:

\begin{compactitem}
	\item \emph{Profile pollution attacks} are designed to disrupt the rating behaviour of regular users with the intention of compromising the system. These attacks can not only be targeted at recommendation systems but also at other personalised online services, such as information retrieval or web search systems~\cite{thang2015evaluation,nguyen2015tag,zhao2021eires}. For instance, attackers can manipulate a user's browsing history to distort their personalised recommendations. However, executing such attacks usually requires that the adversary exploit vulnerabilities in the cross-site request forgery (CSRF) ~\cite{barth2008robust, maes2009browser}, which limits their scalability.

	\item \emph{Poisoning attacks}, as detailed above, inject fake users together with fake ratings into a system, so that the model learns biased behaviour~\cite{chacon2019deep}. Not only can these attacks be performed at large scales, adversaries can also incorporate multiple different goals into their attacks. 
\end{compactitem}
Between these two types of attacks, we note that poisoning attacks pose a more severe threat to economies and society. From an economic perspective, the motivations to promote/demote products and services on a large scale are very strong. Such attacks can make or break a company. However, they can cause irreversible harm to the fairness and trustworthiness of the targeted recommender system and bring the platform owner into disrepute. From a social perspective, poisoning attacks can manipulate popular belief. For instance, an adversary that compromises a recommender system delivering online news can directly manipulate a community's opinions about anything, including an election or public policy.

\subsection{Characteristics of Poisoning Attacks on Recommender Systems}

It is important to note that poisoning attacks are not simply malicious assaults on an online system. In fact, poisoning attacks are of great significance to the sustainability of machine learning models in that they are the \emph{de facto} standard procedure for evaluating a model's robustness against noise or polluted data. Considering the vital role poisoning attacks play in the field of machine learning, they have been studied for a wide range of machine learning models, including support vector machines~\cite{tavara2019parallel}, decision trees~\cite{fletcher2019decision}, regression models~\cite{branco2016survey}, and neural networks~\cite{goodfellow2016deep}. However, generic poisoning attacks on machine learning models have only limited application to recommender systems for several reasons.
\begin{compactitem}
    % \item 
\item \emph{Data correlation.} While machine learning models learn hidden knowledge from one source of training data, recommender systems learn user preferences from the interactions of two data sources – the users and the items. Hence, a recommender system’s robustness stems from the correlations between these data~\cite{ricci2011introduction}. This makes poisoning attacks on recommender systems different from attacks of traditional machine learning models, most particularly, in that they typically require more effort to execute. For instance, in a poisoning attack on a computer vision system, it may be enough to change a single pixel~\cite{su2019one}, whereas a successful poisoning attack on a recommender system would require the adversary to inject many, many user-item correlations into the model’s dataset.

\item \emph{The lack of prior knowledge.} Another popular approach to poisoning attacks in the field of machine learning is to leverage gradient descent to discover dedicated perturbations, which are subsequently combined with the regular samples. As these combinations are undetectable, they can significantly affect the quality of the learned models. By contrast, recommender systems are typically black-box systems that do not provide access to the underlying model. This means the attack must usually be based only on the training data (i.e., the rating matrix). Additionally, users of recommender systems often have {privacy concerns} and, hence, are hesitant to publish their preferences. This means that an attack will commonly be based on only a small subset of the data on which the recommender model was trained. 

\item \emph{Multiple attack goals.} When attacking a recommender system, an adversary typically has several attack goals in mind. For example, one aim might be to promote a set of items, while another goal is to tarnish the reputation of their competitors’ items. However, fusing multiple goals generally involves a trade-off of some sort since certain actions that are beneficial to one goal may undermine the success of the other goal. Additionally, attacks striving to achieve multiple goals are generally easier to characterise and, hence, to detect. 
%Incorporating multiple attack goals and keeping the undetectable attack at the %same time may introduce extra challenges for poison attacks in recommender system domain. 

\end{compactitem}
\subsection{Challenges when Securing Recommender Systems against Poisoning Attacks}
The general problem of securing machine learning models against poisoning attacks has been studied extensively ~\cite{gardiner2016security, rosenberg2021adversarial, rosenberg2021adversarial, zhang2021systematic}. While the literature includes elaborate countermeasures against poisoning attacks ~\cite{aliwa2021cyberattacks}, challenges remain, especially when trying to secure a recommender system against such an attacks. 
%the online and open nature of %RSs introduce additional unique challenges, as outlined below.   

\begin{compactitem}
   \item \emph{Openness.} Recommender systems are typically public, i.e., they are accessible to large numbers of users, making them very vulnerable to poisoning attacks. The data used to influence the recommendations is also open, in the sense that it cannot be characterised \emph{a priori} in terms of volume or distribution, which offers many degrees of freedom for data manipulation. 

 \item \emph{Concept drift.} To secure recommender systems, many existing methods apply anomaly  detection techniques~\cite{tam2023portable} to identify and filter out fake users. However, recommender systems commonly suffer from a phenomenon called concept drift, meaning that user behaviour constantly evolves due to seasonal or trending preferences. Consequently, real users can easily be misclassified as fake users~\cite{gama2014survey}. 
% Securing RSs on top of these changes without excluding the 
% authentic users is a constant battle to maintain a robust and high-quality RS.

\item \emph{Imbalanced data.} Any classification of regular and fake users is also hampered by the imbalance of the respective classes. Attacks typically have a certain “budget”, meaning there is a limit on how many fake users/ratings an adversary can inject into the system and, moreover, this number of fake users is often only a tiny portion of the overall user base. 

\end{compactitem}

\subsection{\editone{Application Perspectives of Poison Attacks in recommender systems}}
\editone{In this survey, we examine the key domains where recommender systems are widely used: e-commerce~\cite{nguyen2023poisoning}, social media~\cite{zhang2021reverse}, and news recommendations~\cite{yi2022ua}. In e-commerce~\cite{nguyen2023poisoning}, poison attacks can cause biased recommendations, erode user trust, and result in financial losses. Social media platforms~\cite{zhang2021reverse} face the risks of misinformation, polarisation, and the promotion of harmful content. News recommendations~\cite{yi2022ua} can be manipulated to shape opinions and compromise democratic values. Understanding these vulnerabilities helps to develop targeted defences to protect users, maintain trust, and ensure the responsible use of recommender systems across domains.}

\section{Threat model and taxonomy}
\label{sec:tanonomy}

\begin{figure}[!h]
    \centering
    \vspace{-.5em}
    \includegraphics[width=1.0\linewidth]{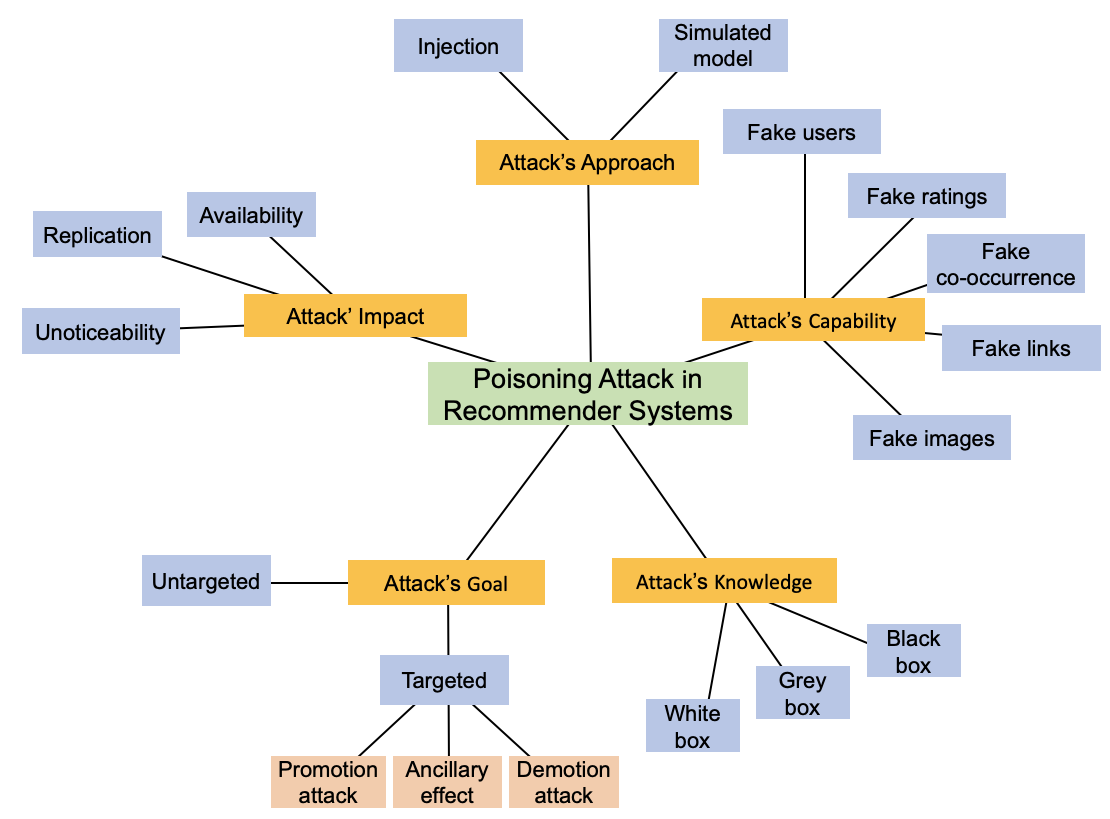}
    \vspace{-1em}
    \caption{\editone{Taxonomy of poisoning attacks on RecSys.}}
    \label{fig:tanonomy}
    \vspace{-1em}
\end{figure}

%Poison attacks can render the recommender system models inaccurate, possibly resulting in poor 
%recommendation outcomes. 
Tabassi et al.~\cite{tabassi2019taxonomy} proposed a general taxonomy for organising poisoning attacks on machine learning models. However, in the light of the above attack characteristics and challenges faced in the context of recommender systems, we present a novel taxonomy that is specifically geared towards poisoning attacks on recommender systems. As illustrated in~\autoref{fig:tanonomy},our taxonomy comprises five distinct dimensions. These five dimensions were included for the following reasons:
%poison attacks in 
%RS domain has unique challenges, presented in the previous section,
 %necessitating a novel taxonomy to describe existing attacks systematically. 
%Here, we propose a novel taxonomy to  categorise poison attacks in RSs into 
%five distinct dimensions, which is presented shown in ~\autoref{fig:tanonomy}. 
%We include these dimensions in our proposed taxonomy for the following reasons:
\begin{compactenum}
    \item The dimensions are specific to the domain of recommender systems. 
%    While the name of the dimensions shares some fleeting similarities among 
%domains, their possible values are different. 
For example, attacks in the general field of machine learning may seek to violate the integrity or confidentiality of the learned models. Yet, in the context of recommender systems, the goals specifically relate to promoting or demoting items. 
    
    \item The dimensions are relevant to the general threat model of data poisoning attacks. These dimensions are important for highlighting the current gaps in research and for outlining directions of future research into poisoning attacks on recommender systems. See ~\autoref{sec:future_direction} for a more detailed discussion on future directions of research.
 \end{compactenum}
% \todo{check value names in taxonomy}

% \todo[inline]{Figure: poison attack -> poisoning attack}
% \todo[inline]{Figure: better fake co-occurrence than co-visitations?}

Notably, this taxonomy applies to both classic heuristic attacks and AI-based attacks. The remaining subsections provide a formal and comprehensive view of the dimensions of the taxonomy and of the threat model associated with poisoning attacks on recommender systems.
(\autoref{sec:attack_goals} - \autoref{sec:attack_approach}). 
The section begins with some general notions and notations related to CF-based recommender systems as a primer for those not familiar with this material.
(\autoref{sec:cf_primer}).    

\subsection{Collaborative Filtering: A Primer}
\label{sec:cf_primer}

Consider a basic CF scenario, where $\mathcal{U}$ is the number of users and $\mathcal{I}$ is the number of items. The tuple $( \langle u,i \rangle, r_{ui})$ indicates that user $u$ has rated the item $i$ with a rating score of $r_{ui}$. The rating matrix $\mathcal{R} \in \mathbb{R}^{|\mathcal{U}| \times |\mathcal{I}|}$ captures all these interactions between users and items. Note that not all entries of the rating matrix $\mathcal{R}$ are actually filled with rating scores. Indeed, the goal of the recommender system is to estimate the missing entries. This estimation task can be formulated as predicting the complete user-item matrix $\Tilde{\mathcal{R}}$ based on prior knowledge of $\mathcal{R}$, where each $\Tilde{r}_{ui} \in \Tilde{\mathcal{R}}$ represents a prediction of the rating score $r_{ui}$. The prediction problem is then formulated as a predictive model to approximate the function: $f:\mathcal{U} \times \mathcal{I} \to \mathbb{R}$ with $f: \langle u,i \rangle 
\mapsto \Tilde{r}_{ui}$, where $\theta$ is the set of model parameters. Before revealing the optimisation problem for this recommendation task, we need to introduce the standard loss functions commonly used in recommender systems.

\sstitle{Common loss functions in recommender systems} In general, loss functions in recommender systems are a means to measuring how close the predicted rating is $\Tilde{r}_{ui}$ to the actual rating $r_{ui}$. The two most common loss functions are:

\begin{compactenum}
    \item \emph{Least squares loss:} which is the maximum likelihood estimation under a Gaussian distribution~\cite{salakhutdinov2008bayesian}. More specifically, the loss function casts the recommendation problem as a regression problem~\cite{hu2008collaborative}, formally defined as
    \begin{equation}
        \ell(r_{ui}, \Tilde{r}_{ui}) = \sum_{ r_{ui} \in \mathcal{R},  \Tilde{r}_{ui} \in \Tilde{\mathcal{R}} } w_{ui} ( r_{ui} -  \Tilde{r}_{ui})^2
    \end{equation}
    where $w_{ui}$ is the contributed weight of the user $u$-item $i$ pair into the overall loss. 
    
    \item \emph{Binary cross-entropy loss:} which is the maximum likelihood estimation under a Bernoulli distribution~\cite{dai2013multivariate}. More precisely, the loss function casts the recommendation problem as a binary logistic regression problem~\cite{mendes2012ensemble}, formally defined as 
    \begin{equation}
        \ell(r_{ui}, \Tilde{r}_{ui}) = \sum_{ r_{ui} \in \mathcal{R},  \Tilde{r}_{ui} \in \Tilde{\mathcal{R}} } r_{ui} \log P_{ui} + (1- r_{ui}) \log (1-P_{}ui)
    \end{equation}
   where $P_{ui}$ is the non-linear activation over the predicted rating $\Tilde{r}_{ui}$. This is then calculated by $P_{ui} = \sigma(\Tilde{r}_{ui}) =  \frac{1}{1+e^{-\Tilde{r}_{ui}}}$. 
   
\end{compactenum}

\sstitle{The optimisation problem} Having chosen a loss function ($\ell$), the prediction task is then formulated as an optimisation problem:
\begin{equation}
\label{eqn:cf_basic}
    \min_\theta \sum_{ r_{ui} \in \mathcal{R} } \ell(f(\langle u,i \rangle;\theta), r_{ui})    
\end{equation}
The predicted matrix $\Tilde{\mathcal{R}}$ is then used to recommend a list of items to a user that the user has not encountered yet. For instance, if the goal is to suggest $K$ items to user $u$, the model will extract the top-$K$ items that: (i) the user has not yet rated; and (ii) have the highest predicted score in the row vector $(\Tilde{r}_{u1}, \Tilde{r}_{u2},\ldots, \Tilde{r}_{uN})$ of $\Tilde{\mathcal{R}}$.

% There two main approaches to CF to build a recommender system: a user-based approach and an item-based approach. In the user-based approach, the user plays the leading role; people with the same taste are grouped, and the recommendations for the user are based on evolution items that are rated by the group with the same preferences as the user (see \autoref{fig:CF_RS}. a). The item-based approach provides recommendations by building neighbourhoods for the items that are preferred by the user (see \autoref{fig:CF_RS}.b).

% \begin{figure}[!h]
%     \centering
%     \includegraphics[width=0.7\linewidth]{recommender_systems.png}
%     \caption{(\textbf{a}) User-based collaborative RS. (\textbf{b}) Item-based collaborative RS.}
%     \label{fig:CF_RS}
% \end{figure}

% \subsection{Threat models}

% To attack the target recommender deployed in the system (a.k.a. \emph{victim model}), the attacker will use their own local model (a.k.a. \emph{surrogate model}), to craft fake users and inject them into the original training data of victim model. Below, we elaborate the threat model from several perspectives.

% \editthree{Although we use collaborative filtering as a premier method here, details of other models will be revealed in a later section}.

\subsection{The Adversary’s Goal}
\label{sec:attack_goals}

Generally, attacks on recommender systems aim to either promote an item, demote an item, or both. However, an attack can either be \emph{non-targeted}, where the goal is a general degrading of performance; or \emph{targeted}, where a specific item or items is the quarry. These attack goals are formally characterised next.

\subsubsection{Untargeted poisoning attacks}
The goal in an untargeted attack is to maximise the error of the recommender system, eventually rendering it useless. Suppose $\mathcal{U'}$ is the set of controlled users injected by an adversary with $\mathcal{R'} \in \mathbb{R}^{|\mathcal{U'}| \times |\mathcal{I}|}$ being their rating scores. An untargeted poisoning attack would be formulated as the following optimisation problem:  
\begin{equation}
%    \begin{split}
%        & 
        \min_{\mathcal{R'}} ||\mathcal{R'}|| 
%        \\
%        & 
\quad
        s.t: f(\langle u,i \rangle;\theta)  \neq r_{ui}, \forall\ r_{ui} \in 
        \mathcal{R} \cup \mathcal{R'}
%    \end{split}
\end{equation}
Alternatively, relying on a maximisation problem instead of the more common CF minimisation (\autoref{eqn:cf_basic}), the adversary will seek to maximise the loss between the predicted and actual rating scores. This leads to the following problem formulation:
\begin{equation}
    \min_{\mathcal{R'}} ||\mathcal{R'}|| \max_{ r_{ui} \in \mathcal{R} \cup \mathcal{R'} } \ell(f(\langle u,i \rangle;\theta), r_{ui})    
\end{equation}
% \todo[inline]{should we still add something before the max, which is choosing 
% the (smallest?) $\mathcal{R'}$ to maximise the loss?}

\subsubsection{Targeted poisoning attacks}
The adversary's goal in a targeted attack is to increase or decrease the popularity of a targeted item. These attacks can be referred to as \emph{promotion} and \emph{demotion} attacks, respectively. Since both attacks are similar and exchangeable, we have only described a promotion attack to keep the discussion concise. For a demotion attack, simply change instances of ‘maximise’ to ‘minimise’ and vice versa. 

Let $t$ be the target item, and let $\mathcal{R'}$ be the rating scores of the controlled users injected by adversary. A promotion attack would then be formulated as follows:
\begin{equation}
%    \begin{split}
%        & 
        \min_{\mathcal{R'}} ||\mathcal{R'}|| 
%        \\
%        & 
\quad
        s.t: f(\langle u,i \rangle;\theta)  = r_{ut}, \forall\ r_{ui} \in 
        \mathcal{R} \cup \mathcal{R'}
%    \end{split}
\end{equation}
Here, the attack mechanism tries to boost the visibility of the target $t$ and, thus, minimise the loss between the predicted and the expected rating score $r_{ut}$ of all users who rated item $t$. This problem is similar to the primary setting of CF (\autoref{eqn:cf_basic}), except that only the target item is involved.
%when it is expressed as an unconstrained optimisation problem, 

\begin{equation}
    \label{eqn:non_resilience}
    \min_{\mathcal{R'}} ||\mathcal{R'}|| \min_{ r_{ut} \in \mathcal{R} \cup \mathcal{R'} } \ell(f(\langle u,t \rangle;\theta), r_{ut})
\end{equation}
While the aim of the promotion attack in \autoref{eqn:non_resilience} is to boost the popularity of the target item, this type of attack is oftentimes referred to as a \emph{non-resilience} targeted attack. The reason is that this kind of attack is easily detected due to the level of skew in the rating distribution, as the model will excessively favour the target item to the exclusion of all else. Conversely, a \emph{resilience} targeted attack strives to promote the target less obviously and without hampering the recommendations for any other items in the system. This attack is formulated as an unconstrained optimisation problem:
\begin{equation}
     \min_{\mathcal{R'}} ||\mathcal{R'}|| \min_{ r \in \mathcal{R} \cup \mathcal{R'} } \ell(f(\langle u,i \rangle;\theta), r_{ui}) + \ell(f(\langle u,t \rangle;\theta), r_{ut})
\end{equation}
Beyond promotion and demotion attacks, a couple of studies have recently shed light on a novel attack objective called the \emph{ancillary effect} ~\cite{rezaimehr2021survey}. These approaches are hybrid attacks that aim to manipulate a group of users or items.

\subsection{\editone{The Adversary’s Knowledge}}

\editone{The characteristics of the attacker's knowledge plays an essential role in the threat model. This is because the impact of an attack will differ significantly depending on the extent of knowledge an attacker has about the system they are trying to undermine. Consider the following types of attacks based on the background knowledge an adversary might have:}   

\begin{compactitem}
    \item \textit{Black-box attack.} \editone{In a black-box attack, the adversary does not know the details of the target system. Specifically, they will not be aware of the architecture of the system, the function $f$ used for prediction or its parameters $\theta$, or the users' past behaviours $\mathcal{R}$.} 
    \item \textit{Grey-box attack.} \editone{In a grey-box attack, the adversary has limited knowledge. As such, she can merely modify the user-item interaction matrix $\mathcal{R}$ by injecting a limited number of controlled users along with their ratings to create a poisoned matrix ${\mathcal{R}'}$.}
\item \textit{White-box attack.} \editone{In this type of attack, the adversary has thorough knowledge of the system, including the prediction function $f$, its parameter set $\theta$, and the entire history of interactions between users and items $\mathcal{R}$.} 
    
\end{compactitem}
In machine learning, both white-box and black-box attacks are widespread, and both have been shown to be efficient attacks during the training phase. In the field of recommender systems, however, \emph{grey-box attacks} are of particular importance, as will be discussed in more detail in \autoref{sec:creation}.

\subsection{The Attack Impact}

Considering the different types of background knowledge possessed by an adversary, there are three general, long-term impacts of a poisoning attack on a recommender system, as listed below.
\begin{compactitem}
	\item \textit{Availability.} Recommender systems make decisions based on the data they have amassed in the past. White-box and grey-box attacks disturb the input data to these algorithms ~\cite{huang2018systematically}. At first, an adversary will inject fake users to manipulate the recommender system and weaken the accuracy of the underlying model. However, this weakened model will ultimately serve as a backdoor, helping the adversary to obtain complete control over the system, thereby compromising its availability.

\item \textit{Replication.} While black-box attacks are less efficient at harming the system's availability~\cite{amir2008byzantine}, they may instead replicate the system. A common tactic is to reverse-engineer the underlying model as a simulation. Once known, the model can then be replicated and exploited. 

	\item \textit{Unnoticeability.} Some poisoning attacks ensure that their approach goes unnoticed by preserving crucial data characteristics when injecting fake data. Conventional techniques to guard the recommender system generally fail to address this particular vulnerability. Even with protection techniques in place, such undetected attacks can often be effective and may cause significant damage. 
	
% 	\todo[inline]{explain that}

\end{compactitem}

\subsection{\editone{The Adversary’s Capabilities}}

\editone{When attacking a recommender system in the real world, an adversary has many tools at her disposal. Commonly, the attacker will possess at least one of the following capabilities.}
\begin{compactitem}
    \item \textit{Fake users.} \editone{Any recommender system can be manipulated if the attacker injects enough fake users. However, in practice, there is a trade-off between the ability of an attacker to execute an attack and a system's robustness. In particular, injecting a large number of fake users is difficult from an operational point of view. Many injections tend to inevitably yield markers that separate controlled users from ordinary ones, so that the injected users can be detected easily. The maximum number of fake users that can be injected as part of a specific attack is called \emph{attack size}, denoted as $\mathcal{U'}$. This number is typically much lower than the overall number of users $\mathcal{U}$ in the system.} 
    
    \item \textit{Fake ratings.} \editone{For each fake user, a number of non-zero fake ratings is injected into the target model. However, there is typically a bound $k$ on the number of ratings that can be injected, called the \emph{profile size} This is because, in real-life, users generally only interact with a few items in a the system~\cite{o2002promoting}. Formally, the fake ratings are denoted as $\mathcal{R}'$. Moreover, the domain for these ratings is given by $\mathcal{B}(\mathcal{R}, k)$, which denotes the space around the actual rating matrix $\mathcal{R}$ of radius $k$, i.e., the maximum number of amended ratings.}    
    
    \item \textit{Fake co-occurrence.} \editone{Instead of inserting fake ratings for fake users directly, an attacker may also manipulate the recommender model by introducing fake co-occurrences between items. That is, given a target item, the poisoning attack is based on injecting visits to other items for users that are already linked to the target item.}   
    
    \item \textit{Fake links.} \editone{Several approaches for recommender systems exploit knowledge graphs as an auxiliary source of information to improve recommendation accuracy~\cite{wu2021poisoning}. As these knowledge graphs frequently depend on third-party data, they represent a vulnerability. That is, an adversary may add fake links to rewrite the knowledge graph’s structure, thereby influencing the recommender system.}
    
    \item \textit{Fake images.} \editone{Many recommender systems use images of items to address the cold start problem. Again, such external data sources provide an angle through which to attack the recommender system~\cite{liu2021adversarial}. An attacker may create dedicated images of items that, once injected into the recommender system, promote particular items and increase their ranking in the given recommendations.}
\end{compactitem}

\subsection{The Attack Approach}
\label{sec:attack_approach}

To achieve their goal, the approach taken by adversaries often depends on their knowledge and capabilities. In general, the field makes a distinction between the following two types of approaches:
\begin{compactitem}
    \item \emph{Injection.} The most common scenario for a poisoning attack is that the adversary has limited knowledge about the targeted system. As such, the adversary will tend to inject a limited number of well-designed users. In addition, these fake users will also assign ratings to several other items so as to disguise the attack and their real target. 

    \item \emph{Simulation.} In a black-box setting, where the adversary does not have enough knowledge to perform an injection, the attack will usually be based on simulating the targeted recommender system. More precisely, a surrogate model is trained using data collected from the recommender system to reproduce the targeted system's behaviour.
\end{compactitem}

% \autoref{fig:threat_model}. 

% \begin{figure}[!h]
%     \centering
%     \includegraphics[width=\linewidth]{threat_model.png}
%     \caption{The simulated model [black-box]. [\textcolor{blue}{redraw}]}
%     \label{fig:threat_model}
% \end{figure}

% \subsection{Backup}

% Recently, we have seen a plethora of real-life examples where adversaries desire to infuence users’ beliefs and decisions for their own malicious purposes: fake social media accounts being created to promote news articles about a political ideology; false online product reviews being posted to bias users’ opinions favorably or against certain products; and so on. Thus, studying the degree to which machine learning models for recommendation can be manip- ulated is important. This is a problem well-aligned with the goals of studying machine learned models’ robustness to adversarial ex- amples, to ultimately build safer artifcial intelligence [23].

\section{Poisoning attacks}
\label{sec:creation}

Poisoning attacks in recommender systems can be divided into two groups: (1) \emph{model-agnostic attacks}, which can be executed to evaluate the trustworthiness of any recommender system; and (2) \emph{model-intrinsic attacks} that target a specific type of recommender system. ~\autoref{tbl:poison_attacks} provides a summary of the surveyed attacks. 

\begin{table*}[t]
	\caption{\editthree{Summary of poisoning attacks}.}
	\label{tbl:poison_attacks}
	\vspace{-0.5em}
	\centering
	\begin{adjustbox}{max width=\textwidth}
		\rowcolors{1}{}{lightgray}
\begin{tabular}{cccccc|ccccccc|ccccccccccc|ccccccc} 
\toprule
\textbf{Name}  &  & \textbf{Authors}         &  & \textbf{Type} & 
\multicolumn{1}{c}{} &  & \multicolumn{5}{c}{\textbf{Attack Goal}} & 
\multicolumn{1}{c}{} &  & 
\multicolumn{9}{c}{\textbf{Knowledge \& Capability}}& 
\multicolumn{1}{c}{} &  & 
\multicolumn{6}{c}{\textbf{Approach \& Impact}}\\
\cline{1-1}\cline{3-3}\cline{5-5}\cline{8-12}\cline{15-23}\cline{26-31}
\rowcolor{white}     &  &                 &  &      & \multicolumn{1}{c}{} &  
& 
Untar.            &  & 
\multicolumn{3}{c}{Tar.}                                          & 
\multicolumn{1}{c}{} &  & 
\multicolumn{3}{c}{Adversary’s}                                            &  
& 
\multicolumn{5}{c}{Adversary’s}

                                      & \multicolumn{1}{c}{} &  & 
\multicolumn{2}{c}{Attack}                    &  & 
\multicolumn{3}{c}{Attack}                                             \\
&  &                 &  &      &                      &  & 
Attack                &  & 
\multicolumn{3}{c}{Attack}                                            &         
             &  & 
\multicolumn{3}{c}{Knowledge}                                         &  & 
\multicolumn{5}{c}{Capability}                                                  
  
                                      &                      &  & 
\multicolumn{2}{c}{Appr.}                  &  & 
\multicolumn{3}{c}{Impact}                                             \\ 
\cline{8-8}\cline{10-12}\cline{15-17}\cline{19-23}\cline{26-27}\cline{29-31}
\rowcolor{white}      &  &                 &  &      &                      &  
&                       &  & \head{Promotion Attack}           & \head{Demotion 
Attack}           & \head{Ancillary Effect}      &                      &  & 
\head{White box}             & \head{Grey box}              & \head{Black 
box}             &  & \head{Fake Users}            & \head{Fake 
Ratings}          & \head{Fake co-occurrence}   & \head{Fake Links}           & 
\head{Fake Images}           &                      &  & 
\head{Injection}             & \head{Simulation}            &  & 
\head{Availability}          & \head{Replication}           & 
\head{Unnoticeability}        \\ 
\midrule
\multicolumn{5}{c}{\textbf{Model-agnostic}}                      &  
&                       &  &                       &                       
&                       &                      &  &                       
&                       &                       &  &                       
&                       &                       &                       
&                       &                      &  &                       
&                       &  &                       & &                      
&                        \\ 
\midrule
MA-01  &  & Lam et al.~\cite{lam2004shilling}      &  & Classic 
&                      &  & \cmark &  & \cmark & \cmark &                       
&                      &  &                       & \cmark 
&                       &  & \cmark & \cmark &                       
&                       &                       &                      &  & 
\cmark &                       &  & \cmark &                       & \cmark  \\
MA-02  &  & Song et al.~\cite{song2020poisonrec}     &  & AI-based 
&                      &  &                       &  & \cmark 
&                       &                       &                      &  
&                       &                       & \cmark &  & \cmark & \cmark 
&                       &                       &                       
&                      &  & \cmark & \cmark &  &                       & \cmark 
&                        \\
MA-03  &  & Tang et al.~\cite{tang2020revisiting}     &  & AI-based 
&                      &  & \cmark &  & \cmark & \cmark &                       
&                      &  &                       & \cmark 
&                       &  & \cmark & \cmark &                       
&                       &                       &                      &  & 
\cmark & \cmark &  &                       & \cmark &                        \\
MA-04  &  & Lin et al.~\cite{lin2020attacking}      &  & AI-based 
&                      &  &                       &  & \cmark 
&                       &                       &                      &  
&                       & \cmark &                       &  & \cmark & \cmark 
&                       &                       &                       
&                      &  & \cmark &                       &  & \cmark 
&                       & \cmark  \\
MA-05  &  & Wu et al.~\cite{wu2021triple}       &  & AI-based 
&                      &  &                       &  & \cmark 
&                       &                       &                      &  & 
\cmark & \cmark &                       &  & \cmark & \cmark 
&                       &                       &                       
&                      &  & \cmark &                       &  & \cmark 
&                       & \cmark  \\
MA-06  &  & Zhang et al.~\cite{zhang2021reverse}  &  & AI-based 
&                      &  & \cmark &  & \cmark &                       
&                       &                      &  &                       
&                       & \cmark &  & \cmark & \cmark &                       
&                       &                       &                      &  & 
\cmark & \cmark &  &                       & \cmark &                        \\
MA-07  &  & Fan et al.~\cite{fan2021attacking}      &  & AI-based 
&                      &  &                       &  & \cmark 
&                       &                       &                      &  
&                       &                       & \cmark &  & \cmark & \cmark 
&                       &                       &                       
&                      &  & \cmark & \cmark &  &                       & \cmark 
& \cmark  \\
MA-08  &  & Barbieri et al.~\cite{barbieri2021simulating} &  & AI-based 
&                      &  &                       &  & \cmark 
&                       &                       &                      &  
&                       & \cmark &                       &  & \cmark & \cmark 
&                       &                       &                       
&                      &  & \cmark &                       &  & \cmark 
&                       &  \\
MA-09  &  & Wu et al.~\cite{wu2021ready}     &  & AI-based 
&                      &  &                       &  & \cmark 
&                       &                       &                      &  
&                       & \cmark &                       &  
&                       &                       & \cmark 
&                       &                       &                      &  & 
\cmark &                       &  & \cmark &                       &   \\ 
MA-10  &  & Chen et al.~\cite{chen2022knowledge}     &  & AI-based 
&                      &  &                       &  & \cmark 
&                       &                       &                      &  
&                       &  &      \cmark                 &  
&         \cmark               &             \cmark           & 
&                       &                       &                      &  & 
\cmark &                       &  & \cmark &                       &   \\ 
MA-11  &  & Zhang et al.~\cite{zhang2022loki}     &  & AI-based 
&                      &  &                       &  & \cmark 
&                       &                       &                      &  
&                       &  &      \cmark                 &  
&         \cmark               &             \cmark           & 
&                       &                       &                      &  & 
 &           \cmark            &  &  &         \cmark              &  \cmark \\ 
MA-12  &  & Lin et al.~\cite{lin2022shilling}     &  & AI-based 
&                      &  &                       &  & \cmark 
&       \cmark                 &                       &                      &  
&                       &  &      \cmark                 &  
&         \cmark               &             \cmark           & 
&                       &                       &                      &  & \cmark
 &                       &  & \cmark &                       &  \cmark \\ 
\midrule
\multicolumn{5}{c}{\textbf{Model-intrinsic}}                    &  
&                       &  &                       &                       
&                       &                      &  &                       
&                       &                       &  &                       
&                       &                       &                       
&                       &                      &  &                       
&                       &  &                       &  &                     
&                        \\ 
\midrule
MI-01  &  & Li et al.~\cite{li2016data}       &  & Classic 
&                      &  & \cmark &  & \cmark & \cmark &                       
&                      &  & \cmark &                       
&                       &  & \cmark & \cmark &                       
&                       &                       &                      &  & 
\cmark &                       &  & \cmark &                       & \cmark  \\
MI-02  &  & Yang et al.~\cite{yang2017fake}     &  & Classic 
&                      &  &                       &  & \cmark & \cmark 
&                       &                      &  & \cmark & \cmark & \cmark &  
&                       &                       & \cmark 
&                       &                       &                      &  & 
\cmark &                       &  & \cmark &                       
&                        \\
MI-03  &  & Fang et al.~\cite{fang2018poisoning}     &  & Classic 
&                      &  &                       &  & \cmark 
&                       &                       &                      &  & 
\cmark & \cmark & \cmark &  & \cmark & \cmark &                       
&                       &                       &                      &  & 
\cmark &                       &  & \cmark &                       & \cmark  \\
MI-04  &  & Chris et al.~\cite{christakopoulou2019adversarial}    &  & 
AI-based &                      &  &                       &  & \cmark 
&                       &                       &                      &  & 
\cmark &                       &                       &  & \cmark & \cmark 
&                       &                       &                       
&                      &  & \cmark &                       &  & \cmark 
&                       & \cmark  \\
MI-05  &  & Hu et al.~\cite{hu2019targeted}       &  & AI-based 
&                      &  &                       &  & \cmark 
&                       &                       &                      &  
&                       &                       &                       &  & 
\cmark & \cmark &                       &                       
&                       &                      &  & \cmark 
&                       &  & \cmark &                       
&                        \\
MI-06  &  & Chen et al.~\cite{chen2019data}     &  & Classic 
&                      &  & \cmark &  & \cmark &                       
&                       &                      &  & \cmark 
&                       &                       &  & \cmark & \cmark 
&                       &                       &                       
&                      &  & \cmark &                       &  & \cmark 
&                       &                        \\
MI-07  &  & Zhang et al.~\cite{zhang2020practical}  &  & AI-based 
&                      &  &                       &  & \cmark 
&                       & \cmark &                      &  
&                       &                       & \cmark &  & \cmark & \cmark 
&                       &                       &                       
&                      &  & \cmark & \cmark &  &                       & \cmark 
&                        \\
MI-08  &  & Fang et al.~\cite{fang2020influence}   &  & Classic 
&                      &  &                       &  & \cmark 
&                       &                       &                      &  & 
\cmark & \cmark &                       &  & \cmark & \cmark 
&                       &                       &                       
&                      &  & \cmark &                       &  & \cmark 
&                       &          \cmark              \\
MI-09  &  & Chen et al.~\cite{chen2021data}     &  & Classic 
&                      &  &                       &  & \cmark 
&                       &                       &                      &  
&                       & \cmark & \cmark &  & \cmark & \cmark 
&                       &                       &                       
&                      &  & \cmark &                       &  & \cmark 
&                       &                        \\
MI-10 &  & Zhang et al.~\cite{zhang2021data}  &  & AI-based 
&                      &  &                       &  & \cmark 
&                       &                       &                      &  
&                       & \cmark &                       &  & \cmark & \cmark 
&                       &                       &                       
&                      &  & \cmark & \cmark &  &                       & \cmark 
&                        \\
MI-11 &  & Yue et al.~\cite{yue2021black}      &  & AI-based 
&                      &  &                       &  & \cmark 
&                       &                       &                      &  
&                       &                       & \cmark &  & \cmark & \cmark 
&                       &                       &                       
&                      &  & \cmark & \cmark &  &                       & \cmark 
&                        \\
MI-12 &  & Wu et al.~\cite{wu2021poisoning}     &  & AI-based 
&                      &  &                       &  & \cmark 
&                       &                       &                      &  & 
\cmark &                       &                       &  
&                       &                       &                       & 
\cmark &                       &                      &  & \cmark 
&                       &  & \cmark &                       
&                        \\
MI-13 &  & Liu et al.~\cite{liu2021adversarial}      &  & AI-based 
&                      &  &                       &  & \cmark 
&                       &                       &                      &  & 
\cmark & \cmark & \cmark &  &                       &                       
&                       &                       & \cmark &                      
&  & \cmark &                       &  & \cmark &                       & 
\cmark  \\
MI-14 &  & Huang et al.~\cite{huang2021data}    &  & AI-based 
&                      &  &                       &  & \cmark 
&                       &                       &                      &  
&                       & \cmark &                       &  & \cmark & \cmark 
&                       &                       &                       
&                      &  & \cmark &                       &  & \cmark 
&                       & \cmark  \\
MI-15 &  & Zhang et al.~\cite{zhang2021pipattack}  &  & AI-based 
&                      &  &                       &  & \cmark 
&                       &                       &                      &  & 
\cmark &                       &                       &  & \cmark & \cmark 
&                       &                       &                       
&                      &  & \cmark &                       &  & \cmark 
&                       & \cmark  \\
MI-16 &  & Rong et al.~\cite{rong2022poisoning}  &  & AI-based 
&                      &  &                       &  & \cmark 
&                       &                       &                      &  & 
 &                       &            \cmark           &  & \cmark & \cmark 
&                       &                       &                       
&                      &  & \cmark &                       &  & \cmark 
&                       &   \\
MI-17 &  & Wu et al.~\cite{wu2022fedattack}  &  & AI-based 
&                      &  &          \cmark             &  &  
&                       &                       &                      &  & 
 &                       &            \cmark           &  & \cmark & \cmark 
&                       &                       &                       
&                      &  & \cmark &                       &  & \cmark 
&                       & \cmark  \\
\editthree{MI-18} &  & \editthree{Yi et al.~\cite{yi2022ua}}  &  & \editthree{AI-based} 
&                     &   &             \editthree{\cmark}          &  & 
&                       &                       &                      &  & 
 &        \editthree{\cmark}                 &                       &  & \editthree{\cmark}   & \editthree{\cmark}
&                       &                       &                       
&                      &  & \editthree{\cmark} &                       &  & \editthree{\cmark} 
&                       & \editthree{\cmark}   \\
\editthree{MI-19} &  & \editthree{Nguyen et al.~\cite{nguyen2023poisoning}}   &  & \editthree{AI-based} 
&                        &  &                     &  &  \editthree{\cmark} 
&                       &                       &                      &  & 
 &             \editthree{\cmark}          &                      &  & \editthree{\cmark} & \editthree{\cmark}
&                       &                       &                       
&                      &  & \editthree{\cmark} &        \editthree{\cmark}               &  & 
&              \editthree{\cmark}          & \editthree{\cmark}  \\
\bottomrule
\end{tabular}
	\end{adjustbox}
	\vspace{-0.5em}
\end{table*}

\subsection{Model-agnostic Poisoning Attacks}
\label{sec:agnostic_attack}

\subsubsection{\editthree{Attack Formulation}}
\editthree{A model-agnostic poisoning attack can be executed against any recommender system, regardless of its underlying algorithm. This attack strategy involves injecting manipulated data into the training set, which causes the framework to learn a biased model. The objective here is to manipulate the recommender system into promotings/demoting specific items. Mathematically, a model-agnostic poisoning attack can be formulated:
}
\editthree{
\begin{equation}
%    \begin{split}
%        & 
        \min ||\mathcal{R'} - \mathcal{R}||_2^2 
%        \\
%        & 
\quad
        s.t.: {r'}_{ui} \in \mathcal{R'}, \forall\ \text{fake user } u \text{ and items } i
%    \end{split}
\end{equation}
}
\editthree{
This formulation considers the original rating matrix, denoted as $\mathcal{R}$, as well as the perturbed rating matrix, denoted as $\mathcal{R}'$. The rating ${r'}_{ui}$ is the score for an item $i$ that the attacker plans to inject on behalf of a fictitious user $u$. The notation $||\cdot||_2^2$ denotes the L2 norm. To ensure the injected rating ${r'}_{ui}$ is a valid rating appearing in the perturbed rating matrix $\mathcal{R}'$, the constraint ${r'}_{ui} \in \mathcal{R}'$ is needed. This constraint reflects the attacker's objective of presenting the injected data as legitimate, thereby deceiving the recommender system into considering it as genuine information.}

\subsubsection{Related Work}
Many model-agnostic poisoning attacks were originally designed to test the general robustness of a recommender system in terms of its trustworthiness. As such, these attacks are designed to be independent of any specific prediction model or class of such models. Some prominent examples of these model-agnostic attacks follow. 

\editthree{
Lam et al.'s shilling attack \cite{lam2004shilling} pioneered the formal definition of a poisoning attack. This attack focuses on promoting specific items in the system without disclosing self-interest (\textbf{MA-01} in \autoref{tbl:poison_attacks}). This classic approach assumes the adversary has knowledge of the rating distributions, and so fake users with manipulated ratings are injected into the training set. Although they are hard to detect, shilling attacks can ultimately corrupt the recommendation model. By contrast, Song et al.'s PoisonRec \cite{song2020poisonrec} (\textbf{MA-02}) is an adaptive poisoning attack with reinforcement learning as the training scheme. By actively injecting interactions by fake users into the system, PoisonRec dynamically improves its own attack strategies through a reward signal. Additionally, a hierarchical tree structure guides the search space sampling process, which optimises the attack's efficiency.} 

\editthree{
Tang et al.'s adversarially-learned injection attacks \cite{tang2020revisiting} (\textbf{MA-03}) shows us an alternative approach. It focuses on automatically learning the behaviour of fake users from a separate model. This method offers an exact solution for generating fake user behaviour and includes scalability schemes for large-scale datasets. While these attacks aim to manipulate the recommendation system, their intent is not to replicate the system entirely, but rather to influence its outputs. Lin et al.'s GAN-based approach \cite{lin2020attacking} (\textbf{MA-04}) combines GANs with the Augmented Shilling Attack framework (AUSH) to allow tailored attacks based on specific budgets and goals. The approach, which has demonstrated some robustness against various defence strategies, and has been tested on a variety of recommender systems. By contrast, Zhang et al.'s black-box attacks \cite{fan2021attacking} (\textbf{MA-07}) and \cite{zhang2021reverse} (\textbf{MA-06}) leverage surrogate models and clone user profiles to promote items in targeted domains. Using deep reinforcement learning, the model’s parameters are adjusted by reward signals.}

\editthree{Wu et al.'s TrialAttack \cite{wu2021triple} (\textbf{MA-05}) employs triple adversarial learning and a flexible end-to-end framework to generate malicious users. By formulating an optimisation problem, TrialAttack creates fake users with different intents, effectively imitating a natural distribution of actual user ratings. Barbieri et al.'s approach \cite{barbieri2021simulating} (\textbf{MA-08}) uses variational autoencoders to approximate real user profiles, which tends to generate more realistic malicious profiles. Wu et al.'s GOAT \cite{wu2021ready} (\textbf{MA-09}) integrates GANs with graph neural networks (GNNs) to strike a balance between feasibility and effectiveness, where high-order relations are modelled between co-rated items. Chen et al.'s KGAttack \cite{chen2022knowledge} (\textbf{MA-10}) incorporates knowledge graphs into a hierarchical policy network to generate fake user profiles. Zhang et al.'s LOKI \cite{zhang2022loki} (\textbf{MA-11}) focuses on complex black-box next-item recommender systems, using a recommender simulator and a result estimator to guide the training of attack agents . Finally, Lin et al.'s Leg-UP \cite{lin2022shilling} (\textbf{MA-11}) uses GANs to discover patterns in user behaviour and generate fraudulent user profiles, addressing visibility concerns in conventional shilling attacks.}

\subsection{Model-intrinsic Poisoning Attacks}
 \label{sec:intrinsic_attack}

Model-intrinsic attacks are designed and optimised for a specific type of recommender system. Here, the different attacks are therefore grouped according to the system type. 

\subsubsection{\editthree{Attack Formulation}}
\editthree{In contrast to model-agnostic poisoning attacks, model-intrinsic poisoning attacks are based on knowledge about the underlying algorithms of the target recommender system. The adversary optimises an objective function to inject manipulated data while maintaining recommendation accuracy but simultaneously minimising the chances of detection. Formally, let $\mathcal{L}$ represent the original loss function used in the collaborative filtering process, which typically measures the squared difference between the predicted ratings and the actual ratings. Conversely, let $\mathcal{A}$ denote the poison attack objective function designed to maximise the impact of any injected poisoned data while minimising the effect of that data on genuine interactions. To encompass both functions, an overall objective function can be defined as a combination of $\mathcal{L}$ and $\mathcal{A}$, as follows:}
\editthree{
\begin{equation}
    \mathcal{E} = \alpha * \mathcal{L} + \beta * \mathcal{A}
\end{equation}
}
\editthree{
This formulation introduces two weighting coefficients, $\alpha$ and $\beta$, to control the balance between accuracy and the impact of the attack. These coefficients determine the trade-off between these two factors. The aim of the optimisation process then becomes one of finding the optimal values for the latent factor matrices of the users and items. Additionally, the injected poison data is also optimised to minimise the objective function $\mathcal{E}$. }
\subsubsection{Matrix-factorisation-based recommender systems} 
\editthree{Li et al. (\textbf{MI-01})~\cite{li2016data} pioneered poisoning attacks and demonstrated the effectiveness of generating undetected fake data. By contrast, Yang et al. ~\cite{yang2017fake} (\textbf{MI-02}) focused on injecting fake item co-occurrence data into the training set. So, while Li et al. seek to manipulate user behaviour, Yang et al. look to exploit patterns of item selection. The former essentially imitates normal user behaviour, while the latter leverages the co-occurrence of items. In another vein, Christakopoulou et al. ~\cite{christakopoulou2019adversarial} (\textbf{MI-04}) introduced a ``learning-to-attack'' model that is based on a repeated general-sum game. This strategy focusses on the dynamic aspects of the attack, whereas Hu et al.’s ~\cite{hu2019targeted} (\textbf{MI-05}) attack was devised specifically for social recommender systems. Their approach is to manipulate users through fake social connections, which highlights the adaptability to social RSs, even without direct connections to the attacker.}

\editthree{Chen et al. ~\cite{chen2019data}) (\textbf{MI-06}) proposed a poisoning attack for cross-domain recommendation. They addressed the challenges of cross-domain recommendations through a dual-level optimisation problem and an approximation scheme. Zhang et al. ~\cite{zhang2020practical} (\textbf{MI-07}), on the other hand, developed LOKI, a black-box attack that relies on deep reinforcement learning. LOKI focuses on the next-top-k recommendations and overcomes access restrictions through training on a recommender simulator. Fang et al. ~\cite{fang2020influence} (\textbf{MI-08}) target influential users in recommender systems based on matrix factorisation to optimise ratings. Their approach is essentially to recommend a target item to regular users. Zhang et al. ~\cite{zhang2021data} (\textbf{MI-10}), instead, focus on handling data incompleteness in recommender systems by using a probabilistic generative model to select less perturbed users and items. Liu et al. ~\cite{liu2021adversarial} (\textbf{MI-13}) address the cold start problem in top-k recommender systems by exploiting images through their Adversarial Item Promotion (AIP) attack. By contrast, Yue et al. ~\cite{yue2021black} (\textbf{MI-11}) developed black-box attacks on sequential recommender systems by extracting model parameters without accessing user-item interaction data.}

\subsubsection{Graph-based recommender systems} 
\editthree{Several poisoning attack methods have been developed for graph-based recommendation systems. For example, Fang et al.’s attack ~\cite{fang2018poisoning} (\textbf{MI-03}) injects fake users and rating scores into the recommender system to manipulate the model. The premise is to optimise the fake rating scores to maximise impact while minimising the chances of being detected. Wu et al.’s attack ~\cite{wu2021poisoning} (\textbf{MI-12}) targets recommender systems that rely on knowledge graphs, using deep reinforcement learning to choose the optimal attack combinations in a step-by-step manner. The knowledge graph is manipulated before the model is trained, which effectively alters the target item's ranking. Lastly, GSPAttack~\cite{nguyen2023poisoning} (\textbf{MI-19}) incorporates generative surrogate-based techniques to fabricate users and user-item interactions, yet recommendation accuracy is maintained by preserving the correlations in the data. Unlike the other methods, GSPAttack emphasises the importance of maintaining high accuracy for non-target items, which is crucial for the success of the poisoning attack. These attacks demonstrate different approaches and considerations in poisoning graph-based recommender systems, showcasing optimisation-based, reinforcement learning-based, and surrogate-based techniques for manipulating recommender systems.}

\subsubsection{Neighbourhood-based recommender systems} 
\editthree{Poisoning attacks targeting neighbourhood-based recommender systems are discussed at some length in Chen et al.~\cite{chen2021data} (\textbf{MI-09}). The presented framework, known as UNAttack, shares similarities with other model-intrinsic poisoning attacks by strategically introducing fabricated users into the target models. This ensures that the recommended items are prioritised for a large number of regular users. Additionally, this research reveals two key findings: (i) recommender systems relying on Euclidean-based similarity measures demonstrate high resilience against such attacks; and (ii) UNAttack can be successfully applied to other recommender systems, including neural CF and Bayesian personalised ranking systems.}
% Poisoning attacks for neighbourhood-based RSs are presented 
% in~\cite{chen2021data} (\textbf{MI-9}). The framework, coined 
% UNAttack, is similar to other model-intrinsic poisoning 
% attacks in that it injects carefully crafted fake users into the target models, 
% so that the target items are recommended to as many regular users as possible. 
% In addition, the work reveals that (i) RSs that are based on 
% Euclidean-based similarity measures have a high robustness; and (ii) UNAttack 
% is transferable to other RSs, including neural CF and Bayesian personalised 
% ranking systems.

\subsubsection{Deep-learning-based recommender systems} 
% \editthree{In this section, we will discuss the workings of a Deep-learning-based Recommendation System and provide an overview of the relevant research in this field.}

% \sstitle{\editthree{Technical Aspects of Deep-learning-based recommender system}}
% \editthree{
% Deep-learning-based recommender systems utilise neural network architectures to learn complex patterns and representations from the user-item interaction data.
% Let's consider a deep learning model with parameters $\theta$. The model takes as input the user and item features (or embeddings) and predicts the rating or the probability of interaction between a user and an item.
% The forward propagation in the deep learning model can be represented mathematically as:
% }

% \editthree{
% \begin{equation}
%     \widehat{R} = f_\theta(user\_embedding, item\_embedding)
% \end{equation}
% }

% \editthree{
% Here, $f_\theta$ denotes the mapping function parameterised by $\theta$ that computes the predicted rating or the probability of interaction between the user and item based on their embeddings.
% }

% \sstitle{\editthree{Related work}}
\editthree{ Huang et al.~\cite{huang2021data} (\textbf{MI-14}) devised a poisoning attack specifically for these deep learning-based recommender systems. The attack involves injecting ratings made by fabricated users into the model. Notably, the definition of these ratings is formulated as an optimisation problem, which can be approximately solved by incorporating numerous heuristic rules. Interestingly, the attack remains effective even when the attackers only have access to a relatively small fraction of user-item interactions.}
% Recent advances in deep 
% learning (DL), led to the development of DL-based RSs. For these systems, Huang 
% et al.~\cite{huang2021data} (\textbf{MI-14}) propose a poisoning attack 
% that injects ratings from fake users 
% into the targeted model. Again, the definition of the respective ratings is 
% phrased as an optimisation problem that can be solved approximately by 
% incorporating a large number of heuristic rules. 
% Intriguingly, the attack remains effective even when the attackers have 
% access only to a relatively insignificant fraction of user-item interactions.

\subsubsection{Federated recommender systems} 
\editthree{PipAttack~\cite{zhang2021pipattack} and FedAttack~\cite{wu2022fedattack} are two notable poisoning attacks designed for federated recommender systems. PipAttack~\cite{zhang2021pipattack}  (\textbf{MI-15}) exploits the popularity bias in data-driven recommender systems, creating fake users that mimic popular items to increase the likelihood that they will be included in recommendation lists. Similarly, Rong et al.~\cite{rong2022poisoning} (\textbf{MI-16}) put forward some poisoning attacks for the federated learning setting that do not require prior knowledge. Rather, they use random approximation and hard user mining to approximate benign user embeddings. Wu et al.’s~\cite{wu2022fedattack} FedAttack (\textbf{MI-17}) is an untargeted attack strategy that employs hard negative sampling. These attacks degrade the performance of the victim model while managing to evade most existing defence and detection mechanisms. Additionally, Yi et al.~\cite{yi2022ua} (\textbf{MI-18}) introduce UA-FedRec, the first study on untargeted attacks in a federated news recommendation system. UA-FedRec undermines the performance of the global model in federated learning for news recommendation with minimal malicious client involvement.}

\begin{table*}[!h]
\caption{\editthree{Domain and Evaluation Comparison.}}
\label{tbl:attack_domains}
\vspace{-0.5em}
\centering
\begin{adjustbox}{max width=\textwidth}
\rowcolors{1}{}{lightgray}
\begin{tabular}{cccccc|cccc|ccccccccccccccllc|ccc|cc}
\textbf{Name}             &                      & \textbf{Authors}                  &                      & \textbf{Year}            & \multicolumn{1}{c}{}  &                      & \multicolumn{2}{c}{\textbf{Type}}                                                     & \multicolumn{1}{c}{}  &                      & \multicolumn{15}{c}{\textbf{Evaluation Metrics}}                                                                                                                                                                                                                                                                                                                                                                                                      & \multicolumn{1}{c}{}  &                      & \textbf{Domains}                                               & \multicolumn{1}{c}{}  &                      & \textbf{Datasets}                                                        \\ 
\cline{1-1}\cline{3-3}\cline{5-5}\cline{8-9}\cline{12-26}\cline{29-29}\cline{32-32}
                          &                      &                                   &                      &                          &                       &                      & \shorthead{explicit}                                  & \shorthead{implicit}                                  &                       &                      & \shorthead{Pre@k}                                     & \shorthead{RMSE}                  & \shorthead{MAE}                   & \shorthead{AR}                    & \shorthead{NI}                    & \shorthead{UI}                    & \shorthead{HR@k}                                      & \shorthead{ASR}                   & \shorthead{Recnum}                & \shorthead{DR}                    & \shorthead{NDCG@k}                                    & \shorthead{Agr@k}                 & \shorthead{ER@k}                                      & \shorthead{\editthree{AUC}}                   & \shorthead{\editthree{MRR}}                   &                       &                      &                                                                &                       &                      &                                                                          \\ 
\hline
\multicolumn{6}{c|}{\textbf{Agnostic}}                                                                                                                         &                      &                                           &                                           &                       &                      &                                           &                       &                       &                       &                       &                       &                                           &                       &                       &                       &                                           &                       &                                           &                       &                       &                       &                      &                                                                &                       &                      &                                                                          \\ 
\hline
MA-01                      &                      & Lam et al.~\cite{lam2004shilling}                         &                      & 2004                     &                       &                      & \cmark                     &                                           &                       &                      &                                           &                       & \cmark &                       &                       &                       &                                           &                       &                       &                       &                                           &                       &                                           &                       &                       &                       &                      & movie                                                          &                       &                      & ML                                                                       \\
MA-02                      &                      & Song et al.~\cite{song2020poisonrec}                       &                      & 2020                     &                       &                      &                                           & \cmark                     &                       &                      &                                           &                       &                       &                       &                       &                       &                                           &                       & \cmark &                       &                                           &                       &                                           &                       &                       &                       &                      & \editthree{game, movie, phone, clothing}                                   &                       &                      & St, AMV, ML                                                              \\
MA-03                      &                      & Tang et al.~\cite{tang2020revisiting}                       &                      & 2020                     &                       &                      & \cmark                     &                                           &                       &                      &                                           &                       &                       &                       &                       &                       & \cmark                     &                       &                       &                       &                                           &                       &                                           &                       &                       &                       &                      & \editthree{POI, location-based}                                            &                       &                      & GOW                                                                      \\
MA-04                      &                      & Lin et al.~\cite{lin2020attacking}                        &                      & 2020                     &                       &                      & \cmark                     &                                           &                       &                      &                                           &                       &                       &                       &                       &                       & \cmark                     &                       &                       &                       &                                           &                       &                                           &                       &                       &                       &                      & \editthree{movie, automotive}                                              &                       &                      & ML, FT, AAT                                                              \\
MA-05                      &                      & Wu et al.~\cite{wu2021triple}                         &                      & 2021                     &                       &                      & \cmark                     &                                           &                       &                      &                                           &                       &                       &                       &                       &                       & \cmark                     &                       &                       &                       & \cmark                     &                       &                                           &                       &                       &                       &                      & movie                                                          &                       &                      & ML, FT                                                                   \\
MA-06                      &                      & Zhang et al.~\cite{zhang2021reverse}                    &                      & 2021                     &                       &                      &                                           & \cmark                     &                       &                      & \cmark                     &                       &                       &                       &                       &                       & \cmark                     &                       &                       &                       &                                           &                       &                                           &                       &                       &                       &                      & \editthree{\begin{tabular}[c]{@{}c@{}}movie, book, digital music,\\
social network, citation network\end{tabular}}    &                       &                      & \begin{tabular}[c]{@{}c@{}}ML, AMB, ADM, NF, \\TW, G+, CIT\end{tabular}  \\
MA-07                      &                      & Fan et al.~\cite{fan2021attacking}                        &                      & 2021                     &                       &                      &                                           & \cmark                     &                       &                      &                                           &                       &                       &                       &                       &                       & \cmark                     &                       &                       &                       & \cmark                     &                       &                                           &                       &                       &                       &                      & movie                                                          &                       &                      & ML, NF                                                                   \\
MA-08                      &                      & Barbieri et al.~\cite{barbieri2021simulating}                   &                      & 2021                     &                       &                      & \cmark                     &                                           &                       &                      &                                           &                       &                       &                       &                       &                       & \cmark                     &                       &                       &                       &                                           &                       &                                           &                       &                       &                       &                      & movie                                                          &                       &                      & ML                                                                       \\
\multicolumn{1}{c}{MA-09}  & \multicolumn{1}{c}{} & \multicolumn{1}{c}{Wu et al.~\cite{wu2021ready} }   & \multicolumn{1}{c}{} & \multicolumn{1}{c}{2021} & \multicolumn{1}{l|}{} & \multicolumn{1}{c}{} & \multicolumn{1}{c}{\cmark} & \multicolumn{1}{c}{}                      & \multicolumn{1}{l|}{} & \multicolumn{1}{c}{} & \multicolumn{1}{c}{\cmark} & \multicolumn{1}{c}{}  & \multicolumn{1}{c}{}  & \multicolumn{1}{c}{}  & \multicolumn{1}{c}{}  & \multicolumn{1}{c}{}  & \multicolumn{1}{c}{\cmark} & \multicolumn{1}{c}{}  & \multicolumn{1}{c}{}  & \multicolumn{1}{c}{}  & \multicolumn{1}{c}{}                      & \multicolumn{1}{c}{}  & \multicolumn{1}{c}{}                      &                       &                       & \multicolumn{1}{l|}{} & \multicolumn{1}{c}{} & \multicolumn{1}{c}{movie, product}                             & \multicolumn{1}{l|}{} & \multicolumn{1}{c}{} & \multicolumn{1}{c}{DB, CI}                                               \\
\multicolumn{1}{c}{MA-10} & \multicolumn{1}{c}{} & \multicolumn{1}{c}{Chen et al.~\cite{chen2022knowledge}}   & \multicolumn{1}{c}{} & \multicolumn{1}{c}{2022} & \multicolumn{1}{l|}{} & \multicolumn{1}{c}{} & \multicolumn{1}{c}{\cmark} & \multicolumn{1}{c}{}                      & \multicolumn{1}{l|}{} & \multicolumn{1}{c}{} & \multicolumn{1}{c}{}                      & \multicolumn{1}{c}{}  & \multicolumn{1}{c}{}  & \multicolumn{1}{c}{}  & \multicolumn{1}{c}{}  & \multicolumn{1}{c}{}  & \multicolumn{1}{c}{\cmark} & \multicolumn{1}{c}{}  & \multicolumn{1}{c}{}  & \multicolumn{1}{c}{}  & \multicolumn{1}{c}{\cmark} & \multicolumn{1}{c}{}  & \multicolumn{1}{c}{}                      &                       &                       & \multicolumn{1}{l|}{} & \multicolumn{1}{c}{} & \multicolumn{1}{c}{movie, book, music}                         & \multicolumn{1}{l|}{} & \multicolumn{1}{c}{} & \multicolumn{1}{c}{ML, BC, LA}                                           \\
\multicolumn{1}{c}{MA-11} & \multicolumn{1}{c}{} & \multicolumn{1}{c}{Zhang et al.~\cite{zhang2022loki}}   & \multicolumn{1}{c}{} & \multicolumn{1}{c}{2022} & \multicolumn{1}{l|}{} & \multicolumn{1}{c}{} & \multicolumn{1}{c}{\cmark} & \multicolumn{1}{c}{}                      & \multicolumn{1}{l|}{} & \multicolumn{1}{c}{} & \multicolumn{1}{c}{}                      & \multicolumn{1}{c}{}  & \multicolumn{1}{c}{}  & \multicolumn{1}{c}{}  & \multicolumn{1}{c}{}  & \multicolumn{1}{c}{}  & \multicolumn{1}{c}{\cmark} & \multicolumn{1}{c}{}  & \multicolumn{1}{c}{}  & \multicolumn{1}{c}{}  & \multicolumn{1}{c}{}                      & \multicolumn{1}{c}{}  & \multicolumn{1}{c}{}                      &                       &                       & \multicolumn{1}{l|}{} & \multicolumn{1}{c}{} & \multicolumn{1}{c}{\editthree{product, game, POI, location-based}}         & \multicolumn{1}{l|}{} & \multicolumn{1}{c}{} & \multicolumn{1}{c}{ABT, St, GOW}                                         \\
MA-12                     &                      & Lin et al.~\cite{lin2022shilling}                        &                      & 2022                     &                       &                      & \cmark                     &                                           &                       &                      &                                           &                       &                       &                       &                       &                       & \cmark                     &                       &                       &                       &                                           &                       &                                           &                       &                       &                       &                      & \editthree{\begin{tabular}[c]{@{}c@{}}movie, POI, location-based,\\
automotive, tool, grocery, app\end{tabular}}      &                       &                      & \editthree{\begin{tabular}[c]{@{}c@{}}ML, FT, YE, AAT,\\
THI, GGF, AA  \end{tabular}}                                           \\ 
\hline
\multicolumn{6}{c|}{\textbf{Instrinsic}}                                                                                                                       &                      &                                           &                                           &                       &                      &                                           &                       &                       &                       &                       &                       &                                           &                       &                       &                       &                                           &                       &                                           &                       &                       &                       &                      &                                                                &                       &                      &                                                                          \\ 
\hline
MI-01                      &                      & Li et al.~\cite{li2016data}                         &                      & 2016                     &                       &                      & \cmark                     &                                           &                       &                      &                                           & \cmark &                       & \cmark &                       &                       &                                           &                       &                       &                       &                                           &                       &                                           &                       &                       &                       &                      & movie                                                          &                       &                      & ML                                                                       \\
MI-02                      &                      & Yang et al.~\cite{yang2017fake}                        &                      & 2017                     &                       &                      &                                           & \cmark                     &                       &                      &                                           &                       &                       &                       & \cmark & \cmark &                                           &                       &                       &                       &                                           &                       &                                           &                       &                       &                       &                      & \editthree{\begin{tabular}[c]{@{}c@{}}video, product, movie, POI, \\ location, social network\end{tabular}}            &                       &                      & \begin{tabular}[c]{@{}c@{}}YT, eB, AMV, \\YE, LI\end{tabular}            \\
MI-03                      &                      & Fang et al.~\cite{fang2018poisoning}                       &                      & 2018                     &                       &                      & \cmark                     &                                           &                       &                      &                                           &                       &                       &                       &                       &                       & \cmark                     &                       &                       &                       &                                           &                       &                                           &                       &                       &                       &                      & movie                                                          &                       &                      & ML, AMV                                                                  \\
MI-04                      &                      & Chris et al.~\cite{christakopoulou2019adversarial}                      &                      & 2019                     &                       &                      & \cmark                     &                                           &                       &                      &                                           &                       &                       &                       &                       &                       & \cmark                     &                       &                       &                       &                                           &                       &                                           &                       &                       &                       &                      & movie                                                          &                       &                      & ML                                                                       \\
MI-05                      &                      & Hu et al.~\cite{hu2019targeted}                         &                      & 2019                     &                       &                      & \cmark                     &                                           &                       &                      &                                           &                       &                       &                       &                       &                       &                                           & \cmark &                       &                       &                                           &                       &                                           &                       &                       &                       &                      & movie                                                          &                       &                      & FT                                                                       \\
MI-06                      &                      & Chen et al.~\cite{chen2019data}                       &                      & 2019                     &                       &                      & \cmark                     &                                           &                       &                      &                                           & \cmark &                       &                       &                       &                       &                                           &                       &                       &                       &                                           &                       &                                           &                       &                       &                       &                      & movie                                                          &                       &                      & NF,ML                                                                    \\
MI-07                      &                      & Zhang et al.~\cite{zhang2020practical}                    &                      & 2020                     &                       &                      & \cmark                     &                                           &                       &                      &                                           &                       &                       &                       &                       &                       &                                           &                       &                       & \cmark &                                           &                       &                                           &                       &                       &                       &                      & product                                                        &                       &                      & ABT                                                                      \\
MI-08                      &                      & Fang et al.~\cite{fang2020influence}                     &                      & 2020                     &                       &                      & \cmark                     &                                           &                       &                      &                                           &                       &                       &                       &                       &                       & \cmark                     &                       &                       &                       &                                           &                       &                                           &                       &                       &                       &                      & tourism, music                                                 &                       &                      & YE, ADM                                                                  \\
MI-09                      &                      & Chen et al.~\cite{chen2021data}                       &                      & 2021                     &                       &                      & \cmark                     &                                           &                       &                      &                                           &                       &                       &                       &                       &                       & \cmark                     &                       &                       &                       &                                           &                       &                                           &                       &                       &                       &                      & movie                                                          &                       &                      & FT, ML, AMV                                                              \\
MI-10                     &                      & Zhang et al.~\cite{zhang2021data}                    &                      & 2021                     &                       &                      &                                           & \cmark                     &                       &                      &                                           &                       &                       &                       &                       &                       & \cmark                     &                       &                       &                       &                                           &                       &                                           &                       &                       &                       &                      & movie                                                          &                       &                      & ML, AIV                                                                  \\
MI-11                     &                      & Yue et al.~\cite{yue2021black}                        &                      & 2021                     &                       &                      & \cmark                     &                                           &                       &                      &                                           &                       &                       &                       &                       &                       & \cmark                     &                       &                       &                       & \cmark                     & \cmark &                                           &                       &                       &                       &                      & movie, product                                                 &                       &                      & ML, ABT                                                                  \\
MI-12                     &                      & Wu et al.~\cite{wu2021poisoning}                       &                      & 2021                     &                       &                      &                                           & \cmark                     &                       &                      &                                           &                       &                       &                       &                       &                       & \cmark                     &                       &                       &                       &                                           &                       &                                           &                       &                       &                       &                      & movie, business                                                &                       &                      & ML, FTr                                                                  \\
MI-13                     &                      & Liu et al.~\cite{liu2021adversarial}                        &                      & 2021                     &                       &                      &                                           & \cmark                     &                       &                      &                                           &                       &                       &                       &                       &                       & \cmark                     &                       &                       &                       &                                           &                       &                                           &                       &                       &                       &                      & \editthree{clothing, fashion}                                              &                       &                      & AMM, TC                                                                  \\
MI-14                     &                      & Huang et al.~\cite{huang2021data}                      &                      & 2021                     &                       &                      & \cmark                     &                                           &                       &                      &                                           &                       &                       &                       &                       &                       & \cmark                     &                       &                       &                       &                                           &                       &                                           &                       &                       &                       &                      & movie, music                                                   &                       &                      & ML, LA                                                                   \\
MI-15                     &                      & Zhang et al.~\cite{zhang2021pipattack}                    &                      & 2022                     &                       &                      & \cmark                     &                                           &                       &                      &                                           &                       &                       &                       &                       &                       & \cmark                     &                       &                       &                       &                                           &                       & \cmark                     &                       &                       &                       &                      & \editthree{movie, phone}                                                   &                       &                      & ML, AMP                                                                  \\
\multicolumn{1}{c}{MI-16} & \multicolumn{1}{c}{} & \multicolumn{1}{c}{Rong et al.~\cite{rong2022poisoning}}   & \multicolumn{1}{c}{} & \multicolumn{1}{c}{2022} & \multicolumn{1}{l|}{} & \multicolumn{1}{c}{} & \multicolumn{1}{c}{\cmark} & \multicolumn{1}{c}{}                      & \multicolumn{1}{l|}{} & \multicolumn{1}{c}{} & \multicolumn{1}{c}{}                      & \multicolumn{1}{c}{}  & \multicolumn{1}{c}{}  & \multicolumn{1}{c}{}  & \multicolumn{1}{c}{}  & \multicolumn{1}{c}{}  & \multicolumn{1}{c}{\cmark} & \multicolumn{1}{c}{}  & \multicolumn{1}{c}{}  & \multicolumn{1}{c}{}  & \multicolumn{1}{c}{}                      & \multicolumn{1}{c}{}  & \multicolumn{1}{c}{\cmark} &                       &                       & \multicolumn{1}{l|}{} & \multicolumn{1}{c}{} & \multicolumn{1}{c}{movie, music}                               & \multicolumn{1}{l|}{} & \multicolumn{1}{c}{} & \multicolumn{1}{c}{ML, ADM}                                              \\
\multicolumn{1}{c}{MI-17} & \multicolumn{1}{c}{} & \multicolumn{1}{c}{Wu et al.~\cite{wu2022fedattack}}     & \multicolumn{1}{c}{} & \multicolumn{1}{c}{2022} & \multicolumn{1}{l|}{} & \multicolumn{1}{c}{} & \multicolumn{1}{c}{\cmark} & \multicolumn{1}{c}{}                      & \multicolumn{1}{l|}{} & \multicolumn{1}{c}{} & \multicolumn{1}{c}{}                      & \multicolumn{1}{c}{}  & \multicolumn{1}{c}{}  & \multicolumn{1}{c}{}  & \multicolumn{1}{c}{}  & \multicolumn{1}{c}{}  & \multicolumn{1}{c}{\cmark} & \multicolumn{1}{c}{}  & \multicolumn{1}{c}{}  & \multicolumn{1}{c}{}  & \multicolumn{1}{c}{\cmark} & \multicolumn{1}{c}{}  & \multicolumn{1}{c}{}                      &                       &                       & \multicolumn{1}{l|}{} & \multicolumn{1}{c}{} & \multicolumn{1}{c}{movie, product}                             & \multicolumn{1}{l|}{} & \multicolumn{1}{c}{} & \multicolumn{1}{c}{ML, ABT}                                              \\
\editthree{MI-18} & \multicolumn{1}{c}{} & \editthree{Yi et al.~\cite{yi2022ua}}     & \multicolumn{1}{c}{} & \editthree{2022} & \multicolumn{1}{l|}{} & \multicolumn{1}{c}{} & \multicolumn{1}{c}{}                      & \editthree{\cmark} & \multicolumn{1}{l|}{} & \multicolumn{1}{c}{} & \multicolumn{1}{c}{}                      & \multicolumn{1}{c}{}  & \multicolumn{1}{c}{}  & \multicolumn{1}{c}{}  & \multicolumn{1}{c}{}  & \multicolumn{1}{c}{}  & \multicolumn{1}{c}{}                      & \multicolumn{1}{c}{}  & \multicolumn{1}{c}{}  & \multicolumn{1}{c}{}  & \editthree{\cmark} & \multicolumn{1}{c}{}  & \multicolumn{1}{c}{}                      & \editthree{\cmark} & \editthree{\cmark} & \multicolumn{1}{l|}{} & \multicolumn{1}{c}{} & \multicolumn{1}{c}{\editthree{news}}                                       & \multicolumn{1}{l|}{} & \multicolumn{1}{c}{} & \multicolumn{1}{c}{\editthree{MIND, Feed}}                         \\
\editthree{MI-19} & \multicolumn{1}{c}{} & \editthree{Nguyen et al.~\cite{nguyen2023poisoning}} & \multicolumn{1}{c}{} & \editthree{2023} & \multicolumn{1}{l|}{} & \multicolumn{1}{c}{} & \editthree{\cmark} & \multicolumn{1}{c}{}                      & \multicolumn{1}{l|}{} & \multicolumn{1}{c}{} & \editthree{\cmark} & \multicolumn{1}{c}{}  & \multicolumn{1}{c}{}  & \multicolumn{1}{c}{}  & \multicolumn{1}{c}{}  & \multicolumn{1}{c}{}  & \editthree{\cmark} & \multicolumn{1}{c}{}  & \multicolumn{1}{c}{}  & \multicolumn{1}{c}{}  & \multicolumn{1}{c}{}                      & \multicolumn{1}{c}{}  & \multicolumn{1}{c}{}                      & \editthree{\cmark} &                       & \multicolumn{1}{l|}{} & \multicolumn{1}{c}{} & \multicolumn{1}{c}{\editthree{app, movie, phone, music}} & \multicolumn{1}{l|}{} & \multicolumn{1}{c}{} & \multicolumn{1}{c}{\editthree{FR, ML, AMP, LA}}                    \\
\hline
\end{tabular}
\end{adjustbox}
\vspace{-0.5em}
\end{table*}

\subsection{Comparison of Poisoning Attacks}
Our goal in this section is to provide a comprehensive comparison of the characteristics of poisoning attacks as found in the literature. Our five characteristics are: the adversary's goal, the impact, the approach, the adversary's capability, and their knowledge. Additionally, we have summarised the studies in terms of the domain(s) considered, the interaction types included in the attack, the evaluation metrics used, and the related datasets.

\subsubsection{The adversary's goal}
\editone{Although the literature includes research on both targeted and untargeted poisoning attacks, targeted attacks dominate the field (see~\autoref{tbl:poison_attacks}). All approaches support targeted attacks that promote or demote specific items, but only a small portion of the published strategies consider untargeted attacks, which seek to reduce the efficiency of a recommender system in general. This imbalance is not surprising, given the much clearer incentives associated with promotion/demotion attacks. As a result, adversaries have paid much more attention to targeted attacks, and accordingly so have scholars. Similarly, only a few works discuss the ancillary aim of manipulating a group of users or items.}
% Existing work facilitates both \emph{untargeted} and \emph{targeted} attacks 
% on RSs. However, we note that targeted attacks dominate the 
% field, of poisoning attacks on RSs, see~\autoref{tbl:poison_attacks}. All 
% approaches support targeted attacks that promote or demote a specific target 
% item, whereas only a minor part of them also consider untargeted attacks, i.e., attacks 
% with the general intent to lower the efficiency of an RS. We attribute this 
% observation to the fact that promotion/demotion attacks come with clearer 
% incentive structures for adversaries and, thus, receive more attention. 
% Moreover, we note that only a few works support an ancillary effect that aims 
% to manipulate a group of users or items.     

\subsubsection{The attack impact}
Turning to the impact of poisoning attacks, we infer from~\autoref{tbl:poison_attacks} that two-thirds of the attacks focus on the availability of the targeted system, and one-third focus on both attacking and replicating the victim model. While the number of studies related to replication attacks is much lower than that of availability attacks, the impact caused by replication attacks is much more severe. The reason being that, in addition to manipulating the target system to perform as desired by the adversary, replication attacks devise a clone of the victim model, which could be deployed in direct competition with the target recommender system~\cite{khan2013detection}.

\subsubsection{The attack approach}
As for the approach followed in a poisoning attack, all approaches rely, in some form, on \emph{injecting} fake users and ratings into the training data. Yet, \autoref{tbl:poison_attacks} highlights that several attacks additionally rely on \emph{simulating} the targeted recommender system, mostly to derive user profiles for injection. 

\subsubsection{The adversary's knowledge}
\editone{Attacks can also be categorised according to the level of knowledge the adversary has of targeted system, i.e., white box attacks, grey box attacks, and black box attacks. Notably, the total rate of attacks in ~\autoref{tbl:poison_attacks} does not sum to 100\% since certain attacks support multiple levels of knowledge. Nonetheless, the table does illustrate that the execution rate is evenly distributed among the three levels.}
% According to our taxonomy, attacks can be classified by the knowledge possessed 
% by an adversary about the targeted system, including \emph{white box} attacks, 
% \emph{grey box} attacks, and \emph{black box} attacks. 
% It is worth noting that the total rate is not 100\% \autoref{tbl:poison_attacks} as some attacks support multiple levels of knowledge (i.e., white box, grey box, black box) . However, the figure shows that the adoption rate is relatively even among the three-level.

\subsubsection{The adversary's capabilities}
\editone{The adversary's capabilities refers to their ability to inject different types of data into the training set. The types include: (i) fake users; (ii) fake ratings; (iii) fake co-occurrences; (iv) fake links; and (v) fake images. Among the discussed attacks, the most prevalent ones involve injecting fake users and ratings into the training data. For the majority of studies, this tactic has been the focus. Fake co-occurrences are considered in a smaller proportion of attacks, while fake links and fake images represent specialised attack vectors, each employed exclusively by a single attack.}
% The adversary's capacity may include injection of (i) fake users, (ii) fake 
% ratings, (iii) fake co-occurrence, (iv) fake links, and (v) fake images. Most of the discussed attacks are the injection of fake users 
% and ratings, which have been 
% incorporated by the majority of the studies. Fake co-occurrences are considered by a 
% small share of the attacks, while fake links and fake 
% images denote specialised attack vectors, each being adopted solely by a single 
% attack.  

\begin{table}[!h]
%\vspace{-0.5em}
\centering
\caption{\editthree{Commonly used evaluation metrics.}}
\label{tbl:metrics}
\vspace{-1em}
\footnotesize
\scalebox{0.9}{
\begin{tabular}{l l} 
\toprule
\textbf{Abbrv} & \multicolumn{1}{c}{\textbf{Term}}                                                                                        \\ 
\midrule
Pre@k                 & \begin{tabular}[c]{@{}l@{}}The fraction of the top-K recommended items \\falling into the recommended ones\end{tabular}  \\ 
%\hline
RMSE                  & Root mean square error                                                                                                   \\ 
%\hline
MAE                   & Mean absolute error                                                                                                      \\ 
%\hline
AR                    & Average rating for specific items                                                                                        \\ 
%\hline
NI                    & The increased ranks of the targeted item                                                                                 \\ 
%\hline
UI                    & User impression/Item popularity                                                                                          \\ 
%\hline
HR@k                  & The hit rate ratio                                                                                                       \\ 
%\hline
ASR                   & The attack success rate                                                                                                  \\ 
%\hline
Recnum                & The number of page view that contains target items                                                                       \\ 
%\hline
DR                    & The average display rate                                                                                                 \\ 
%\hline
nDCG@k                & Normalised discounted cumulative gain                                                                                    \\ 
%\hline
Agr@k                 & Agreement at rank k                                                                                                      \\ 
%\hline
ER@k                  & The exposure rate at rank k                                                                                              \\
\editthree{AUC}                  &      \editthree{Area under the ROC Curve}                                                                                         \\
\editthree{MRR}                 & \editthree{Mean Reciprocal Rank}                                                                                           \\
\bottomrule
\end{tabular}
}
\vspace{-2em}
\end{table}

\begin{table}[!h]
%\vspace{-.5em}
\centering
\caption{\editthree{Commonly used datasets.}}
\label{tbl:datasets}
\vspace{-1em}
\footnotesize
\scalebox{0.9}{
\begin{tabular}{lll@{\hspace{4em}}ll} 
\toprule
\textcolor[rgb]{0.125,0.129,0.141}{\textbf{Abbrv}} & \multicolumn{1}{c}{\textbf{Name}} &  & \textcolor[rgb]{0.125,0.129,0.141}{\textbf{Abbrv}} & \multicolumn{1}{c}{\textbf{Name}}  \\ 
\cmidrule{1-2}\cmidrule{4-5}
YT                                                 & 
YouTube                           &  & 
AMB                                                & Amazon 
Book                        \\ 
%\cmidrule{1-2}\cmidrule{4-5}
eB                                                 & 
Ebay                              &  & 
NF                                                 & 
Netflix                            \\ 
%\cmidrule{1-2}\cmidrule{4-5}
AMV                                                & Amazon 
Movie                      &  & 
TW                                                 & 
Twitter                            \\ 
%\cmidrule{1-2}\cmidrule{4-5}
YE                                                 & 
Yelp                              &  & 
G+~                                                & 
Google+                            \\ 
%\cmidrule{1-2}\cmidrule{4-5}
LI                                                 & 
Linkin                            &  & 
CIT                                                & Citation 
Network                   \\ 
%\cmidrule{1-2}\cmidrule{4-5}
ML                                                 & 
MovieLens                         &  & 
FTr                                                & Fund 
Transaction                   \\ 
%\cmidrule{1-2}\cmidrule{4-5}
FT                                                 & Film 
Trust                        &  & 
DB                                                 & 
Douban                             \\ 
%\cmidrule{1-2}\cmidrule{4-5}
St                                                 & 
Steam                             &  & 
CI                                                 & 
Ciao                               \\ 
%\cmidrule{1-2}\cmidrule{4-5}
GOW                                                & 
Gowalla                           &  & 
AMM                                                & Amazon 
Men                         \\ 
%\cmidrule{1-2}\cmidrule{4-5}
ABT                                                & Amazon 
Beauty                     &  & 
TC                                                 & 
Tradesy                            \\ 
%\cmidrule{1-2}\cmidrule{4-5}
AAT                                                & Amazon 
Automotive                 &  & 
LA                                                 & 
Last.fm                            \\ 
%\cmidrule{1-2}\cmidrule{4-5}
ADM                                                & Amazon Digital 
Music              &  & AMP                                                & 
Amazon Cell-phone                  \\ 
%\cmidrule{1-2}\cmidrule{4-5}
AIV                                                & Amazon Instant 
Video              &  & BC                                                    
&                 Book-Crossing                   \\

\editthree{THI}                                                & \editthree{Tool and Home Improvement}               &  & \editthree{AA}                                                   
&                 \editthree{Apps
for Android }                   \\
 \editthree{GGF}                                                &       \editthree{Grocery and Gourmet Food}         &  &     \editthree{MIND}                                                
&  \editthree{Online News}                                 \\
 \editthree{FR}                                                &       \editthree{App Recommendation}         &  &     \editthree{Feeds}                                                
&  \editthree{Online News}                                  \\
\bottomrule
\end{tabular}
}
\vspace{-.5em}
\end{table}

\subsubsection{Domains}
\editthree{The literature on poison attacks in recommender systems covers a broad range of application domains, including movies~\cite{zhang2021pipattack, rong2022poisoning, wu2022fedattack}, POI and location-based services~\cite{yang2017fake, tang2020revisiting,zhang2022loki,lin2022shilling}, citation networks~\cite{zhang2021reverse}, news~\cite{yi2022ua}, and others. These domains demonstrate the extensive impact of poison attacks, as they can manipulate user preferences, influence market trends, promote malicious applications, manipulate consumer choices in fashion, and disrupt information dissemination across various sectors. The prevalence of poison attacks across so many domains underscores the importance of developing robust countermeasures to safeguard the integrity and trustworthiness of the online data we all rely on these days.}
% Poisoning attacks in RSs have been presented in various domains. However, we 
% note that a few attacks are \emph{generic} in the sense that they have not been 
% developed for particular domains. As illustrated in 
% \autoref{tbl:attack_domains}, there is one model-agnostic approach and one 
% model-intrinsic approach that fall into this class. The other attacks focus on 
% particular, \emph{singular} domains. 
% Here, we notice a 
% considerable focus on RS for the \emph{movie} domain (\textcolor{red}{xx\%}) followed by RS for 
% the \emph{product} domain (\textcolor{red}{xx\%}).

\subsubsection{Interaction types}
Another relevant dimension is the type of interaction the system uses to generate its recommendations. A distinction can be made between the attacks designed to target recommender systems that rely on \emph{explicit} interactions between users and items, and those where these interactions are \emph{implicit}. \autoref{tbl:attack_domains} illustrates that both types of interactions are covered in the literature, although the main focus falls on recommender systems with explicit interactions.
%We may notice that most of the studies have been conducted on an explicit RS 
%model type-- recommender based on explicit interactions between users and 
%items. 

\subsubsection{Evaluation metrics}
\editone{Researchers have used a range of evaluation metrics to assess the efficacy of various poisoning attacks ~\cite{ali2021overview}. An overview of these metrics appears in~\autoref{tbl:metrics}, along with a link to the corresponding attack most commonly being evaluated (see \autoref{tbl:poison_attacks}). Among these metrics, \emph{HR@k} stands out as the most predominant, being adopted by nearly two-thirds of studies. This dominance is not surprising, considering that \emph{HR@k} is commonly used as the primary evaluation metric~\cite{shani2011evaluating}. The second popular metric is \emph{NDCG@k}, with a smaller portion of the studies. Other metrics are less popular and have only been used in one or two studies.} 

\subsubsection{Datasets}
\editone{\autoref{tbl:datasets} provides an overview of contemporary datasets commonly used to evaluate poisoning attacks on recommender systems. The table includes links to the respective studies, which are listed in \autoref{tbl:poison_attacks}. Among the popular datasets, the \emph{MovieLens} dataset has been used in over two-thirds of the studies, while datasets from various product categories of Amazon have been used in half the studies. 
%\footnote{https://movielens.org}
Both the MovieLens and Amazon datasets are widely recognised within this research community ~\cite{das2017survey}, making them valuable starting points. Three studies have used the Netflix datasets to help develop their attacks; two have used Yelp; while datasets derived from other public websites such Twitter, Google+, YouTube, etc. have been used sporadically.}

%\begin{figure}[!h]
%\vspace{-0.3cm}
%    \centering
%    \includegraphics[width=1.0\linewidth]{creation_timeline.png}
%    \vspace{-0.5cm}
%    \caption{\editone{The evolution of studies on poisoning attacks over time, $^{(*)}$ 
%    indicates AI-based methods.}}
%    \label{fig:creation_timeline}
%    \vspace{-0.5cm}
%\end{figure}
%
%\subsection{Summary}
%Finally, \autoref{fig:creation_timeline} illustrates that poisoning attacks have attracted a great deal of attention recently. The timeline indicates the growth in related studies over the past seven years. Here, the early studies, i.e., those published prior to 2018, laid the foundations of research on poisoning attacks designed for recommender systems – most of which adopted classic, heuristic techniques. However, the advance of AI in recent years has led to considerable growth in the overall number of published studies, most of them indeed being AI-based (especially over the period 2020-2023). 

\section{Countermeasures}
\label{sec:countermeasures}

Countermeasures to poisoning attacks can be divided into two subgroups: (1) \emph{detection methods}, i.e., approaches that aim to identify user profiles that have been created in poisoning attacks; and (2) \emph{prevention methods}, i.e., approaches that aim to make a recommender system more robust to poisoning attacks, without explicitly trying to identify manipulated profiles. In this section, we review both types of methods. We then map which countermeasures are effective against which types of poisoning attacks in (\autoref{sec:effective_mapping}), and which countermeasures cannot be expected to work well against certain attacks (\autoref{sec:weak_mapping}). 
%Finally, we summarise our insights in \autoref{sec:countermeasure_summary}.

\subsection{Detection Methods}
\label{sec:detections}

This section begins with a discussion on the kinds of features that are commonly used to detect poisoning attacks. This is followed by a review of the traits most often employed by detection methods as signals for detecting poison attacks. The section concludes with a discussion on the actual methods that have been proposed to differentiate between genuine user profiles and those that have been used as part of an attack.

% \subsubsection{\editthree{Detection Formulation}}
        
\subsubsection{Detection Features}
\label{sec:features}

Existing detection methods can rely on either \emph{model-agnostic} features or \emph{model-intrinsic} features.

% In this work, we discuss the most typical features, and we recommend engaging readers to refer to~\cite{todo} for a complete reference for adopted features in the community.

\sstitle{Model-agnostic features} Model-agnostic features are designed to capture general abnormal behaviour, independent of the type of attack model being used. These include:

\begin{enumerate}
    \item \emph{Rating deviation from mean agreement (RDMA):} This is the average deviation in ratings given by a particular user to a number of specific items compared to other users, weighted by the inverse rating frequency of those items. Formally, let $r_i^u$ be the rating given by a user $u$ to the item $i$, and let $\overline{r_i}$ be the average of the ratings given to item $i$. $n_{r^i}$ represents the number of ratings given to item $i$, while $n_u$ denotes the number of items that user $u$ has rated. Formally: 
    \begin{equation}
        RDMA_u = \frac{ \sum_{i=0}^{n_u} \frac{ | r_u^{i} - \overline{r^i} | }{ 
        n_{r^i} } }{ n_u }.
    \end{equation}
    \item \emph{Weighted deviation from mean agreement (WDMA):} This feature is the weighted variance of RDMA, which captures the cumulative differences in user ratings for sparse items. Using the same notation as above, we have:
    \begin{equation}
        WDMA_u = \frac{ \sum_{i=0}^{n_u} \frac{ | r_u^{i} - \overline{r^i} | }{ 
        n_{r^i}^2 } }{ n_u }.
    \end{equation}    
    \item \emph{Length variance (LengthVar):} This is a measure of the difference in profile length, which is the number of items rated by that profile. It is calculated against the average length of all profiles. With $n_u$ as the length of the profile $u$, and $\overline{n_u}$ as the average length of all profiles. We have:
    \begin{equation}
        LengthVar_u = \frac{ n_u - \overline{n_u} }{  \sum_{u \in U} {(n_u - 
        \overline{n_u})}^2  }.
    \end{equation}
    \item \emph{Degree of similarity with top neighbours (DegSim):} This feature reflects the average similarity of a user to its the top-k neighbours. Here, $k$ denotes the number of neighbours, while $sim_{u,v}$ is the similarity between users $u$ and $v$, computed using, say, Pearson's correlation~\cite{sedgwick2012pearson}. The feature is then formulated as
    \begin{equation}
        DegSim_u = \frac{ \sum_{v=1}^k sum_{u,v} }{k}
    \end{equation}
\end{enumerate}

    \sstitle{Model-intrinsic features} Despite the many benefits of model-agnostic features, it has been proven they are not particularly good at differentiating between genuine and malicious user profiles ~\cite{sundar2020understanding}, especially when a real user exhibits some unusual behaviour. To address this issue, various model-intrinsic features have been proposed. Below, we review some common, representative notions associated with such features:

\begin{enumerate}
    \item \emph{Mean variance (MeanVar)}: This is a measure that partitions malicious profiles into different constituent parts: extreme ratings (for targeted items); other ratings (items to simply fill up the profile, i.e., so-called filler items); and unrated items. This feature is estimated by computing the mean variance between ‘other ratings’ and the average rating of all items. Let $P_{u,F}$ denote the filler items $F$ of a user $u$, and let $r_{u,j}$ be the rating for item $j$ given by user $u$. $\overline{r_u}$ is the mean rating given by user $u$ to items. We have:
    \begin{equation}
        MeanVar = \frac{ \sum_{j \in P_{u,F} {(r_{u,j} \times \overline{r_u})}^2  } }{ |P_{u,F}| }
    \end{equation}
    \item \emph{Filler mean target difference model (FMTS):} This is a measure to assess the degree of difference between the ratings in the targeted partition and those in the filler partition. Adopting the above notation, this feature is formulated as
    \begin{equation}
        FMTD = \Bigg| \frac{ \sum_{i \in P_{u,F} r_{u,i} } }{|P_{u,T}|} -  \frac{ \sum_{k \in P_{u,F} r_{u,k} } }{|P_{u,F}|} \Bigg|
    \end{equation}  
    \item \emph{Filler average correlation (FAC):} This feature reflects the correlation between the ratings in a profile compared to the average rating of all items. It is defined as
    \begin{equation}
       FAC = \frac{ \sum_{i \in I_u} (r_{u,i} - \overline{r_i} ) }{ \sqrt{ \sum_{i \in I_u} {(r_{u,i} - \overline{r_i} )}^2 } }
    \end{equation}
    \item \emph{Filler mean difference (FMD):} This feature measures the average value of the absolute difference of the profile's ratings and the average rating of all items, defined as follows:
    \begin{equation}
        FMD = \frac{1}{U_u} \sum_{i=1}^{|U|} | r_{u,i} - \overline{r_i} |
    \end{equation}
\end{enumerate}
    % \item \emph{\editthree{Latent Features}:} \editthree{Hidden features are meaningful features that deep learning models can automatically extracted from raw data during the learning process.}  

\begin{table*}[!h]
\vspace{-1em}
\caption{\editthree{Overview of methods to detect poisoning attacks on recommender 
systems.}}
\label{tbl:detection_methods}
\vspace{-1em}
\centering
\begin{adjustbox}{max width=\textwidth}
\rowcolors{1}{}{lightgray}
%
\end{adjustbox}
\vspace{-.5em}
\end{table*}

\subsubsection{Detection Traits}
\label{sec:detection_traits}

This subsection extends the foundational framework of detection traits developed by Sundar et al. ~\cite{sundar2020understanding}. Their framework classifies attack detection traits into four groups: \emph{(i) user profiles}, \emph{(ii) target ratings}, \emph{(iii) filler ratings}, and \emph{(iv) side information}. 
%To this, we have {[TODO: explain how you have extended this framework in broad terms]}.
% The traits that are considered by existing methods to detect attacks can be 
% classified into four groups: the \emph{(i) user profile}, the \emph{(ii) target 
% rating}, the \emph{(iii) filler rating}, and some \emph{(iv) side information}. 

\sstitle{User profile} This first group of traits is used by detection methods to differentiate between malicious profiles and genuine profiles. We have added new characteristics for comparison:
\begin{compactenum}
    \item \emph{Similarity:} The more similar a profile is to its neighbours, the higher the probability that the profile has been created as part of an attack. 
    
    \item \emph{Size:} The attack size (i.e., the number of injected profiles). A size of highly similar profiles much smaller than the entire set of user profiles indicates an attack.
%    Moreover, malicious profiles often exhibit high similarity, thus exposing a 
%valuable trait for the detection algorithm. 
    
    \item \emph{Group behaviour:} Commonly, user behaviour has hidden characteristics. For example, a group of malicious users might have a positive correlation in rating variance. Such group behaviour reveals valuable clues to help detect malicious users.
    
    \item \emph{Attributes:} The user attributes of genuine and injected profiles will tend to follow different distributions. For this reason, statistical analysis methods can often reveal abnormal profiles to help detect attacks. 
\end{compactenum}

\sstitle{Target rating} Recall that target ratings are the ratings given to the item that attackers aim to promote or demote. Two traits related to target ratings need to be considered: 

\begin{compactenum}
     \item \emph{Crowdability:} The frequency of ratings of a targeted item is generally abnormally high following a poisoning attack. These peaks in rating frequency can be an indicator of attack. 
     
     \item \emph{Skewed Ratings:} The ultimate goal of the adversary is to manipulate user attitudes toward a targeted item. 
     %Take promoting attack as an example, the targeted item should be given a 
     %high rating to increase the chance to appear in the recommendation list. 
Hence, fake ratings will tend to deviate from average ratings – a finding that can be exploited to detect attacks. 
\end{compactenum}

\sstitle{Filler rating} As mentioned above, filler ratings are ratings assigned to regular items as part of an attack rather than to the target item. Two traits need to be considered here as well.
% These ratings reveal traits that the detection algorithm can leverage to 
%locate attack behaviours. We consider two traits related to filler ratings as 
%follow.

\begin{compactenum}
    \item \emph{Rating:} 
    %Rather than only interacting with target items, a malicious profile gives 
    %ratings to several regular items, the so-called ``filler'' items. 
To disguise an attack by maximising similarity, the ratings assigned to filler items will usually be close to the current average rating. Based on this observation, a good strategy may be to first assess the ratings of filler items and then identify the adversary. 
    
    \item \emph{Length:} The number of items rated by a profile is known as the length of the profile. Given that an adversary will try to create a profile that is as similar to a genuine profile as possible, a malicious profile will usually be much longer than a regular profile. Profile length can therefore give away an attacker.   
\end{compactenum}

\sstitle{Side information} More recently, poisoning attacks have begun to rely on side information to manipulate the target system~\cite{strub2016hybrid}. This opens up another class of traits for consideration.

\begin{compactenum}
    \item \emph{Co-occurrence:} Some recommender systems use co-occurrence graphs as the basis for suggesting items to users~\cite{yang2017fake}. With these systems, an adversary might inject crafted co-occurrence information into the recommender system to manipulate the recommendations. 
%    As such, exploring co-occurrence behaviours constitute a crucial trait for 
%attack detection.  
    
    \item \emph{User-user graph:} The similarity between users is also a characteristic to differentiate malicious from genuine users. Using graph mining algorithms~\cite{zhang2015panther}, the similarity of users can be explored by computing the similarity between nodes of a graph. The resulting graph of user interactions provides a valuable angle from which to detect an attack. 
\end{compactenum}

\subsubsection{Overview of Detection Methods}
\label{sec:detection}

~\autoref{tbl:detection_methods} provides a summary of methods used to detect poisoning attacks. These methods are reviewed in more detail below, starting with the supervised methods, before turning to the semi-supervised and unsupervised ones. 
%on This section aims to comprehensively review methods that build to detect 
%data poisoning attacks in recommender systems. These methods can be classified into three 
%categories: \emph{(i) supervised methods}, \emph{(ii) semi-supervised 
%methods}, 
%and \emph{(iii) unsupervised methods}. We continuously provide information 
%about each category in the following sections.

\sstitle{Supervised methods} In supervised settings, there needs to be a label to be able to distinguish a malicious profile from a genuine profile. To our best knowledge, Chirita et al. ~\cite{chirita2005preventing} (\textbf{CM-1}) were the first to formulate detecting a poisoning attack as a classification problem. Their model verifies success using the RDMA and DegSim features to detect malicious profiles. Additionally, to improve the performance of the classifier, two abstract features increase the generalisability of the mode. However, the main issue with using abstract features is that regular users with unusual behaviours might be misclassified as malicious. Therefore, several research teams i.e., ~\cite{burke2006classification}, \textbf{CM-2} and ~\cite{burke2006detecting}  \textbf{CM-3} have suggested including some more specific features to improve accuracy,. This improved model has subsequently been used to detect a wide range of attacks, including average, segment, random, and bandwagon attacks.

Williams et al.~\cite{williams2007defending, mobasher2007toward} (\textbf{CM-4}) further improved the performance of classifiers by combining a reverse-engineered attack with a mechanism that detects anomalous ratings. These studies go on to confirm that using various features in an aggregated manner, such as RDMA, WDMA, LengthVar, MeanVar, DegSim, FAC, FMD, and FMTD, mean the attacks can be detected more quickly. The main weakness of this approach is that the detection model’s performance strongly depends on the classifier's choices. Hence, the authors recommend using a support vector machine (SVM) to optimise accuracy. Other approaches, such as Zhang et al. ~\cite{zhang2012meta} \textbf{CM-5} and Zhang et al. ~\cite{zhang2014hht} \textbf{CM-6}, incorporate meta-learning over the set of features used in \textbf{CM-4}. This strategy renders the approach more effective compared to SVM classifiers that simply rely on single or majority voting. 

Another problem is class imbalance. To overcome this issue, Zhou et al. ~\cite{zhou2016svm} (\textbf{CM-7}) devised  a dual-phase detection method, called SVM-TIA. The method proceeds in two steps: (1) the Borderline-SMOTE technique~\cite{smiti2020bankruptcy} is used to obtain initial results while alleviating any class imbalances; and (2) these results are then fine-tuned and analysed to discover malicious profiles. The approach mostly relies on model-intrinsic features, including FMTD, FMD, FAC, and MeanVar. 
Yang et al. ~\cite{yang2016re} (\textbf{CM-8}) followed a similar direction to handle class imbalances , also using a two-step process to improve detection accuracy. In the first phase, a statistical analysis of various attack models is applied to extract features from user profiles. Then RAdaBoost, a variant of AdaBoost, efficiently classifies the injected profiles. Hao et al.~\cite{hao2019detecting} \textbf{CM-9} developed an ensemble of detection methods that goes beyond prior approaches by considering the `novelty of items'. Here, the popularity of each item is calculated based on features extracted from the ratings and a user graph, where differences in popularity provide hints as to the targeted items. 

Relying on trust features and time series analysis, Xu et al. ~\cite{xu2019detecting} (\textbf{CM-10}) developed TSA-TF, which is a detection method designed specifically for social recommender systems. Here, suspicious items are first detected by employing a single exponential smoothing technique that results in a set of suspicious profiles. Second, four features are extracted based on rating patterns and the trust relations between users. Finally, an SVM classifies these extracted features to reveal malicious profiles. To overcome the limitations of hand-crafted features, Zhou et al.~\cite{zhou2020recommendation} (\textbf{CM-11}) devised DL-DRA, an approach to detecting poison attacks based on deep learning. The authors propose an end-to-end learning process that learns directly from raw rating data. More precisely, a bicubic interpolation algorithm~\cite{dengwen2010edge} scales down the rating matrix to overcome problems with sparsity. A convolutional neural network (CNN) then extracts any hidden user features based on the resized rating matrix. Finally, these hidden features are parsed through an algorithm designed specifically to detect poisoning attacks.

\sstitle{Semi-supervised Methods} 
In most recommendation systems, the number of labelled users is typically quite small. Further, annotating labels for the remaining users is generally impractical as it costs too much and, moreover, access to the data is usually restricted. In these situations, semi-supervised methods have proven to be quite effective. For example, Wu et al.\cite{wu2012hysad} (\textbf{CM-12}) proposed the first semi-supervised framework for detecting poisoning attacks called HySAD. HySAD uses MC-Relief, a wrapper that selects features by aggregating various popular attack detection metrics. The authors employ a semi-supervised Naïve Bayes (SNB\_$\lambda$)\cite{ximeng2018situation} classifier to categorise both labelled and unlabelled profiles. Similarly, Cao et al.~\cite{cao2013shilling} (\textbf{CM-13}) introduced a semi-supervised algorithm, called Semi-SAD, that can learn from both labelled and unlabelled user profiles. This work combines the Naïve Bayes classifier with a variant of the expectation-maximization algorithm (EM-$\lambda$) in a sequential manner. Specifically, Semi-SAD initially trains the Naïve Bayes classifier using a small set of labelled profiles. The results are then fine-tuned and improved by incorporating unlabelled profiles with the EM-$\lambda$ algorithm.
% In most RSs, there is only a tiny portion of 
% labelled users, whereas obtaining labels for the remaining users is impractical 
% due to high labelling costs and access restrictions. In such scenarios, 
% semi-supervised methods become effective. A first semi-supervised framework for 
% the detection of poisoning attacks, named HySAD, was proposed by Wu et 
% al.~\cite{wu2012hysad} (\textbf{CM-12}). HySAD introduces MC-Relief, a wrapper 
% that selects the features to be used by aggregating a wide range of popular 
% attack detecting metrics. Based thereon, the authors employ a Semi-supervised 
% Na\"ive Bayes (SNB\_$\lambda$)~\cite{ximeng2018situation} classifier to 
% categorise both labelled and unlabelled profiles. Similarly, Cao et 
% al.~\cite{cao2013shilling} propose a semi-supervised algorithm, namely Semi-SAD 
% (\textbf{CM-13}), to learn from both labelled and unlabelled user profiles. The 
% unique twist of this work is the combination of the Na\"ive Bayes classifier 
% and a variant of Expectation-Maximization (i.e., EM-$\lambda$) in a chronic 
% order. In particular, Semi-SAD trains the Na\"ive Bayes classifier on a small 
% set of labelled profiles. The results are then fine-tuned and improved by 
% incorporating unlabelled profiles with the EM-$\lambda$ algorithm. 

\sstitle{Unsupervised methods} In the complete absence of labels, the only alternative is an unsupervised method of detection. The first to propose such an approach was Zhang et al.~\cite{zhang2006attack} (\textbf{CM-14}). Their method builds on the idea that the distribution of ratings over time might reveal various kinds of hidden attacks. On this premise, the authors grouped consecutive ratings into windows of the same size $k$ and found, through a theoretical proof, that some specific window sizes are optimal for detecting an attack, but only if the number of malicious profiles is known. For circumstances when the number is not known, they devised a heuristic algorithm to select the window size dynamically. 

To improve detection accuracy, Mehta et al.~\cite{mehta2007lies, mehta2009unsupervised} applied principle component analysis (PCA) to the problem of profile detection (\textbf{CM-17}). Bryan et al.’s ~\cite{bryan2008unsupervised} algorithm called UnRAP was inspired by the efficacy of a technique developed in the field of gene expression analysis called Hv\_score ~\cite{bryan2006bottom, bryan2008unsupervised}) (\textbf{CM-16}). The novelty of their approach is that they formulate detecting an attack as a problem of detecting anomalous structures. The strategy is so successful that the method can detect some types of attacks with higher confidence than even supervised methods.

Based on the hypothesis that malicious profiles will be similar to each other, many approaches employ clustering to separate malicious profiles from genuine profiles, i.e., ~\cite{bhaumik2011clustering} \textbf{CM-18}, ~\cite{bilge2014novel} \textbf{CM-20}, ~\cite{yang2017spotting} \textbf{CM-26}, ~\cite{zhang2018ud} \textbf{CM-27}. For example, Chung et al.~\cite{chung2013betap} (\textbf{CM-19})  applied the Beta distribution algorithm~\cite{owen2008parameter} to detect poisoned profiles. Wen et al.~\cite{zhou2014detection} (\textbf{CM-21}) put forward De-TIA, which, rather than only considering individual attacks, detects groups of attacks. The central idea of De-TIA is to incorporate both the DegSim metric~\cite{zhou2014attack} and the RDMA metric~\cite{zhou2015shilling} into a single framework to better differentiate between malicious profiles and genuine profiles. The authors also extend their work to the problem of targeted rating pattern analysis. Unlike prior work, which only tends to address a specific type of attack, Zhang et al. ~\cite{zhang2015catch} (\textbf{CM-22}) continuously scan the recommender system to identify attacks as they emerge. Their method looks for the propagation of fraudulent actions; it also estimates the reliability of users in propagation-based systems. 

Adopting a statistical approach, Zhou et al. ~\cite{zhou2015shilling} (\textbf{CM-23}) developed a method of identifying the ratings patterns associated with malicious profiles and their features. Likewise, Xia et al. ~\cite{xia2015novel} focused on detecting anomalous items directly in order to filter out items manipulated by fake profiles (\textbf{[CM-24]}). Separating malicious profiles from genuine profiles can also be done through user-user graphs and correlation analysis ~\cite{yang2016estimating} (\textbf{[CM-25]}).

Zhou et al. ~\cite{zhou2018shilling} (\textbf{CM-28}) detect malicious behaviour by analysing ratings as a time series. In this way, they are able to identify anomalous user groups. Similar ideas are discussed in Cai et al.~\cite{cai2019bs} (\textbf{CM-29}). Their approach, called BS-SC, starts with an in-depth analysis of user behaviour, after which, the hidden features are extracted to discriminate between fake users and genuine users. The second step is to build a similarity matrix behaviour for all the profiles. Cai et al. ~\cite{cai2019detecting} (\textbf{CM-30}) presented a similar approach that involves analysing user behaviour. Unlike most unsupervised methods, which only consider rating distributions to detect malicious profiles, Aktukmak et al. ~\cite{aktukmak2019quick} (\textbf{CM-32}) rely on user features as an additional reference to enhance detection performance. More specifically, these researchers employ a probabilistic factorisation model to project these two data sources into a latent space. Based on the learned latent space, the framework provides a detection mechanism for new users based on anomaly statistics. In addition to protecting recommender systems from poisoning attacks, Cai et al.~\cite{cai2019trustworthy} \textbf{CM-31} proposed an approach called Value-based Neighbour Selection (VNS) that also improves the profitability of e-retailers at the same time. 

Lastly, Yang et al. ~\cite{yang2020identification} (\textbf{CM-33}) published a detection method called IMIA-HCRF that trains models based on various aspects of the users' behaviour. It incorporates the density of user ratings as well as co-visitation behaviour to identify malicious profiles. Exploiting recent advances in deep learning with graphs~\cite{huynh2021network,duong2022deep,nguyen2014reconciling,nguyen2015smart,hung2019handling}, Hao et al. ~\cite{hao2021unsupervised} (\textbf{CM-34}) present a method that involves reconstructing user-user graphs that are subsequently used to identify malicious profiles. Similarly, You et al. ~\cite{you2023anti} (\textbf{CM-35}) propose Anti-FakeU, a framework that integrates confidence scores from a malicious user detector into a neighbourhood aggregation mechanism. Fake users, generated automatically, eliminate the need for labels. The approach involves constructing a user-user graph to capture malicious behaviour patterns and a novel GNN-based detector to identify fake users. 

\subsection{Prevention Methods}
\label{sec:preventions}

\subsubsection{\editthree{Prevention Formulation}}
\editthree{Prevention methods aim to mitigate the impact of poisoning attacks. One effective defence involves a form of robust optimisation~\cite{mehta2008attack}.  Formally, let $\mathcal{R}$ denote the set of observed ratings in the user-item interaction matrix. The objective function is then formulated as}
\editthree{
\begin{equation}
    \mathcal{E} = \mathcal{L} + \lambda * \mathcal{R}
\end{equation}
}
\editthree{
In this formulation, $\mathcal{L}$ denotes the original loss function, which measures the prediction error between the recommended ratings and the actual ratings. The term $\lambda * \mathcal{R}$ introduces a regularisation component into the objective function, promoting the generation of ratings that align with the observed ratings in the training data~\cite{wu2021fight}. The coefficient $\lambda$ determines the balance between accuracy and the system's resilience against poisoning attacks.}

\subsubsection{\editone{Mitigating Challenges to Achieve Effective Robustness}}
\editone{Most prevention methods rely on a combination of techniques. For example, the issue of \emph{openness} is handled through outlier detection, data sanitisation, and preprocessing techniques~\cite{si2020shilling}. Through such methods, manipulated data can be identified and filtered out. As another example, \emph{concept drift} is often handled by using a dynamic learning algorithm that adapts to evolving user behaviour~\cite{sundar2020understanding}. This adaptability allows the system to distinguish between genuine changes and fake users to help preserve accurate recommendations. \emph{Imbalanced data} is generally addressed via ensemble methods~\cite{rezaimehr2021survey} that assign more weight to the minority classes and use an over or undersampling technique to balance the dataset.}

\subsubsection{Related Work}
In addition to detection methods, there is a line of research that aims to make a recommender system more robust to poisoning attacks without explicitly trying to detect them. More precisely, these algorithms strive to address the vulnerabilities in recommender systems that poisoning attacks exploit. This section briefly highlights existing works in this direction. 

\editone{In their initial study, Sandvig et al.~\cite{sandvig2007robustness} (\textbf{CM-36}) found that model-based recommender systems are more stable and robust than memory-based recommender systems. Building on this, they devised a robust recommender system based on association rule mining, which significantly improves upon neighbour- and model-based recommender systems. Their algorithm captures item relationships through co-occurrences in user profiles, leading to more precise recommendations and mitigating the impact of poisoning attacks. Another study by Mehta et al.~\cite{mehta2007robust} (\textbf{CM-37}) explores the use of statistical techniques, specifically robust M-estimators, for achieving robustness in recommender systems. They leverage robust M-estimators to propose a matrix factorisation algorithm that outperforms other latent semantic-based algorithms such as PLSA and SVD. Mehta et al.\cite{mehta2008attack} (\textbf{CM-38}) present an attack-resistant algorithm for SVD-based collaborative filtering that integrates the efficiency of SVD-based detectors into the system. This algorithm improves both accuracy and robustness of the model. Inspired by graph-based models for capturing malicious profiles, GraphRfi ~\cite{zhang2020gcn} (\textbf{CM-39}) introduces a framework based on a graph convolutional network (GCN) for user representation learning. It incorporates dual-task learning with the aim of simultaneously maintaining robust recommendations while also detecting attacks.} 

\editone{Differential privacy~\cite{friedman2010data} has been widely accepted as an excellent method for protecting machine learning systems against poisoning attacks. Hence, there have been several attempts to apply this idea to recommender systems as well, including by Wadhwa et al. ~\cite{wadhwa2020data} (\textbf{CM-40}). These studies demonstrate that differentially-private recommender systems are more robust than other types of systems and that, in most cases, attack utility is reduced. Anelli et al.\cite{anelli2021study} (\textbf{CM-41}) devised the idea of using adversarial training to enhance the robustness of visual recommender systems (VRSs), exploring adversarial attacks and defence strategies specifically tailored to VRSs. Wu et al.\cite{wu2021fight} (\textbf{CM-42}) presented the APT scheme, which also involves adversarial training. This time, fake profiles are simulated and the agents learn to generate fake users with minimal influence on the target recommender system's empirical risk. In turn, using both real and generated data strengthens the recommender system's robustness through dynamic training. The PORE framework, introduced by Jia et al. ~\cite{jia2023pore} (\textbf{CM-43}), builds recommender systems with a proven robustness against untargeted data poisoning attacks. PORE can convert any existing recommender system into a resilient one by limiting fake user ratings, incorporating knowledge from the recommender system algorithm, and establishing guarantees against targeted data poisoning attacks.}

\begin{figure}[!h]
\vspace{-1em}
    \centering
    \includegraphics[width=1.0\linewidth]{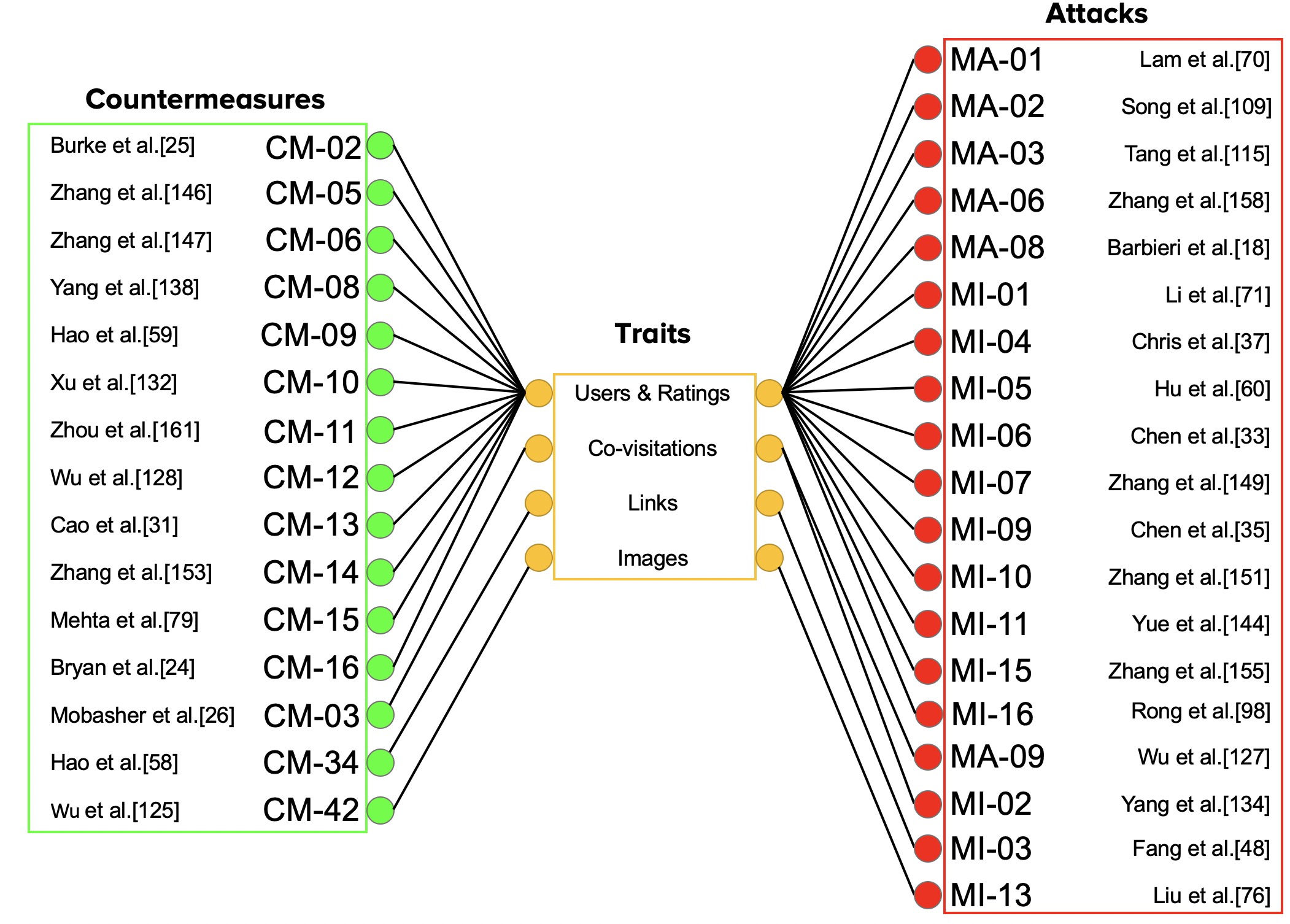}
    \vspace{-2em}
    \caption{Effective countermeasures against poisoning attacks.}
    \label{fig:effective_mapping}
    \vspace{-1em}
\end{figure}

\begin{figure}[!h]
\vspace{-1em}
    \centering
    \includegraphics[width=1.0\linewidth]{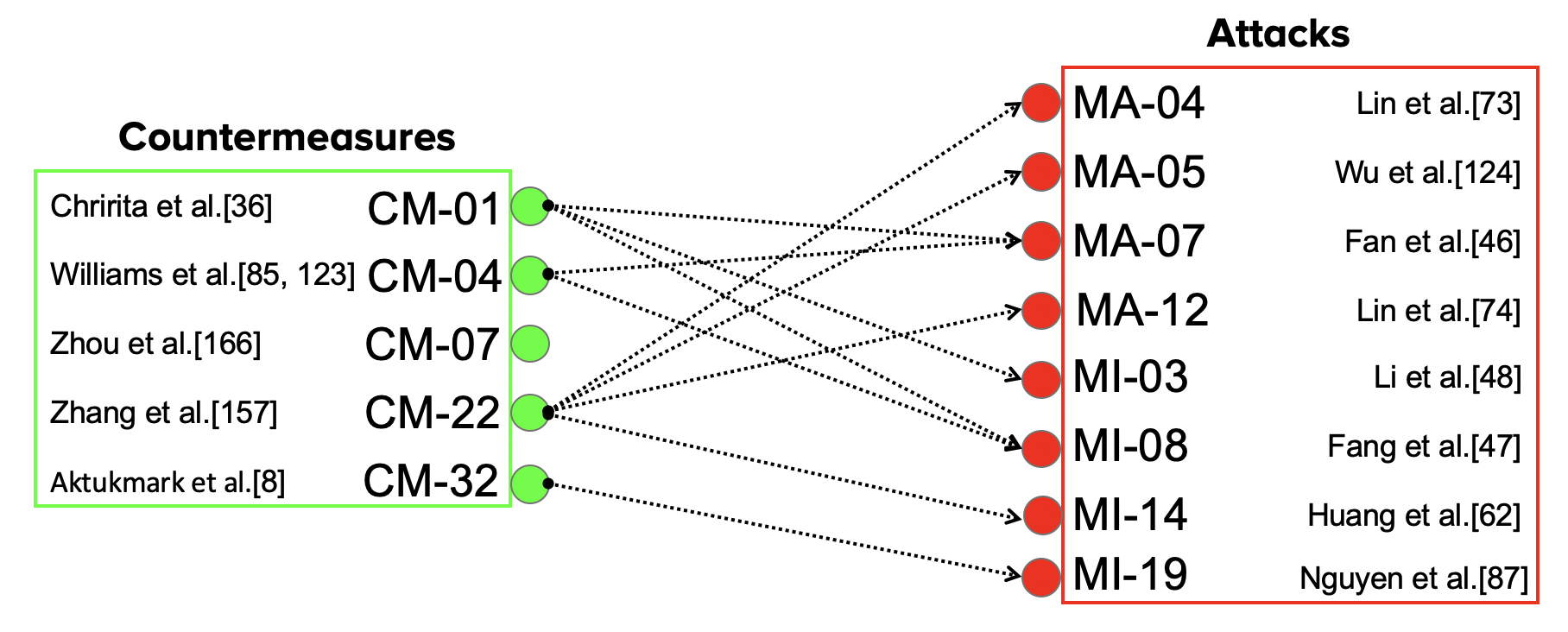}
    \vspace{-1em}
    \caption{Poisoning attacks that are resilient against certain 
    countermeasures.}
    \label{fig:weak_mapping}
%    \vspace{-1em}
  \vspace{-1em}
\end{figure}

In summary, early studies on preventing poisoning attacks primarily focused on traditional analysis techniques, such as statistical and SVD-based methods. Later approaches have witnessed more elaborate foundations, such as GCNs and differential privacy. Moreover, adversarial learning has proven to provide an excellent opportunity to obtain better representations of user profiles, thereby significantly improving the robustness of the systems. 

\subsection{Which Countermeasures are Effective Against Which Attacks? }
\label{sec:effective_mapping}

%Visualising the battlefield has long been an essential factor to identify the 
%most effective responses to the attacks~\cite{evangelopoulou2014attack}. 
Following the idea of attack visualisations~\cite{evangelopoulou2014attack}, we developed a mapping of which countermeasures are effective against which attacks. The maps capture the attack traits and link them to the countermeasures as well as the strategies and capacities of an adversary. The mapping results are shown in~\autoref{fig:effective_mapping}. Here, not surprisingly, we observe a predominant focus on the traits related to users and ratings, which highlights that many of the aforementioned strategies can actually be expected to achieve some level of robustness against a large number of attacks.
%The findings in the mapping is crucial as they provide new researchers with a 
%quick benchmarks' effectiveness against common poisoning attacks in RSs 
%research field. 

%\begin{figure}[!h]
%\vspace{-1em}
%    \centering
%    \includegraphics[width=0.6\linewidth]{weak_mapping.png}
%    \vspace{-1em}
%    \caption{Poisoning attacks that are resilient against certain 
%    countermeasures.}
%    \label{fig:weak_mapping}
%    \vspace{-1em}
%\end{figure}

\subsection{Which Countermeasures are Weak Against Which Attacks?}
\label{sec:weak_mapping}

Most attacks do not only have the goal of promoting or demoting a set of items. Rather, they also strive to disguise themselves from countermeasures. While we have already considered this aspect in our review of poisoning attacks (see the `unnoticeable' column in \autoref{tbl:poison_attacks}), this mapping links these attacks to particular countermeasures, as shown in~\autoref{fig:weak_mapping}. Here, weak countermeasures are listed in the left column, while the right column contains attacks that are resilient against certain countermeasures. In general, there are only a small number of countermeasures that are weak against specific attacks. It is also worth noting that each of the found dependencies constitutes an opportunity for research on how to avoid the corresponding weakness.

%\begin{figure}[!h]
%\vspace{-.5em}
%\centering
%\begin{minipage}{0.47\linewidth}
%\vspace{-1em}
%    \centering
%    \includegraphics[width=1.0\linewidth]{defense_timeline.png}
%\vspace{-1em}
%    \caption{\editthree{The evolution of countermeasures.}}
%    \label{fig:defense_timeline}
%    \vspace{-1em}
%\end{minipage}
%\quad
%\begin{minipage}{0.47\linewidth}
%    \centering
%    \vspace{-1em}
%    \includegraphics[width=0.6\linewidth]{gaps.png}
%    \vspace{-1em}
%    \caption{\editthree{Research trends for poisoning attacks and countermeasures for 
%    recommender systems.}}
%    \label{fig:gap}
%    \vspace{-1em}
%\end{minipage}
%\vspace{-.5em}
%\end{figure}
%
%
%
%\subsection{Summary} 
%\label{sec:countermeasure_summary}
%
%Finally, we provide an overview of the evolution of countermeasures against poisoning attacks in recommender systems. \autoref{fig:defense_timeline} shows a timeline that captures the number of related studies in recent years. From this figure, we can see that countermeasure research received solid attention around 2019, whereas only a few studies have been published very recently. 

\section{Research Gaps, Limitations, and Future Directions}
\label{sec:future_direction}

%In this section, we discuss current research gaps and outline potential directions for future work in the field of poisoning attacks designed for recommender systems.

\subsection{Gaps Related to Poisoning Attacks }
%There are several open problems.

%\begin{compactitem}

\sstitle{From white-box and grey-box to black-box attacks}
    \editone{Existing poisoning attacks often rely on knowledge of the interactions between users and items, as is the case with white-box and grey-box attacks. However, privacy and security concerns typically limit access to targeted recommender systems. Although black-box attacks have gained attention, defining them and finding their countermeasures remains an open question requiring further research.}

\sstitle{From specific to agnostic attacks} 
    \editone{Model-intrinsic attacks currently dominate the field. However, they are not particularly applicable to the real world. Future research could therefore focus on model-agnostic attacks to assess the robustness of emerging recommender systems.} 
    
\sstitle{Side-information fusion} Most existing attacks follow the paradigm of injecting fake users and their interactions along with a small set of filler items into the training data (as discussed in~\autoref{tbl:poison_attacks}). However, more recommender systems are recently beginning to also exploit side information to improve the quality of recommendations, such as domain hypergraphs~\cite{guo2021gcn}, the social properties of users~\cite{yu2021socially}, and spatial information~\cite{wang2020next, sun2020go}. \editthree{Although side information can help, it also introduces additional complexities and potential security risks. By including side information, the attacker can manipulate the recommendations and mislead users. This emerging strategy deserves special attention as it has the potential to undermine the reliability of recommender systems~\cite{ye2021optimal}}. 

%\end{compactitem}

\subsection{Gaps Related to Countermeasures}
%Next, we turn to the research gaps relating to countermeasures.

%\begin{compactitem}
\sstitle{Pre-attack detection} 
     \editone{Most current countermeasures primarily focus on detecting poisoning attacks after they have already caused irreversible damage. Few approaches attempt to detect pre-attack behaviour, which involves identifying the malicious activities aimed at preparing an attack. Future research should not only look to identify the kinds of abnormal behaviour that precede an attack but also design mechanisms to monitor this type of behaviour in real-time. This would see a new category of countermeasures emerge in the form of a prediction method to sit alongside prevention and detection.}
    
\sstitle{Going beyond the main traits} 
     \editone{As shown in~\autoref{tbl:detection_methods}, existing detection methods primarily focus on key indicators such as the ratings of filler and target items. However, supplementary information, such as a user’s attributes, play an equally important role in preventing unforeseen damage to the attacked system. For example, if a new user is created with suspicious attributes, the system should implement a mechanism to verify the user before incorporating their interactions. The same approach can be applied to the attributes of newly added items (filler items) to prevent manipulation based on these items. Therefore, we believe that incorporating additional traits into detection methods to identify and mitigate poisoning attacks represents a promising research direction.}   

\sstitle{Going beyond accuracy} 
    \editone{The pursuit of accuracy has traditionally been the primary goal of attack detection models. However, it has been recognised in recent years that focusing solely on accuracy is insufficient~\cite{han2017survey}. Ensuring fairness and explainability of the models has become equally important~\cite{abdollahi2018transparency, li2021tutorial}. Specifically, explainability plays a crucial role in developing countermeasures by understanding the creation of poisoning attacks and deriving preventive measures against similar attacks. Therefore, there is a need for research on explainability techniques to address vulnerabilities in recommender system models.} 

\sstitle{Overhead optimisation}  
     \editone{Our paper primarily focuses on countering poison attacks in recommender systems. However, we acknowledge that the discussed applications may have broader implications. As a future direction, we propose optimising the overhead associated with feature extraction from historical data. This optimisation is crucial due to the significant performance impact on recommender systems. Real-time detection of fake users is essential for effective countermeasures, necessitating further emphasis on reducing overhead.} 
    
%\end{compactitem}

\subsection{Limitations of the work}
\editthree{Our limitation is the absence of a quantitative performance comparison between the reviewed poison attacks. Due to space constraints, we were unable to provide direct empirical evidence regarding the practical efficacy of these models. However, in future work, we plan to conduct a comprehensive quantitative performance comparison using real-world datasets and established evaluation metrics. By addressing this limitation in a future technical report, in conjunction with the current survey, our aim is to provide a more practical and comprehensive understanding of poison attacks in recommender systems and contribute to advancing this crucial research area.}

\subsection{Linking Poisoning Attacks and Countermeasures}
\editone{Our analysis of published studies on poisoning attacks and countermeasures reveals a concerning trend. While there has been a significant amount of research on novel types of poisoning attacks, there has not been much work in recent years on developing countermeasures. This indicates that state-of-the-art recommendation systems will have various vulnerabilities that can potentially be exploited and where there is no effective countermeasure to stop the attack. Therefore, a significant research gap exists that calls for further investigation on how to assess and enhance the robustness of recommender systems. This research is crucial for maintaining and partially restoring the trustworthiness and fairness of the underlying recommendation models.}

\subsection{Integrating General Vulnerabilities of Recommender Systems}
\editone{In addition to the discussed poisoning attacks, recommender systems are susceptible to vulnerabilities arising from general security threats faced by information systems. However, these vulnerabilities have been largely overlooked and only informally addressed ~\cite{ghaffarian2017software}. Consequently, there is an urgent need to systematically describe, analyse, and establish connections between these general vulnerabilities and attacks on recommender systems. Rigorous methodologies and design principles need to be developed, as such efforts are crucial for quantitatively assessing the impact of attacks and designing effective techniques to mitigate their associated risks.}

\subsection{Tools and Environments for Discovering Vulnerabilities}
\editone{As is evident from~\autoref{tbl:detection_methods}, existing attack detection techniques do not cover all the vulnerabilities in today’s recommender systems. Therefore, building a comprehensive countermeasure means combining different techniques. Tools that enable the seamless integration of various detection techniques would help in this regard. Additionally, practical research on novel attacks and countermeasures heavily relies on simulation environments for contemporary recommender system models, considering the limited access to production systems. Future work should prioritise the development of such simulation environments. These resources will serve as valuable testbeds for both practitioners and researchers, contributing to the advancement of the field as a whole.}

\section{Conclusion}
\label{sec:conclusion}

In this survey, we presented a comprehensive overview of poisoning attacks for recommender systems along with countermeasures to detect and prevent them. We first distinguished poisoning attacks from similar concepts, such as adversarial attacks, before introducing a novel taxonomy for poisoning attacks. We formalised the dimensions of this taxonomy and linked a total of 31 attacks described in the literature to it. We complemented the discussion of attacks with a review of 43 countermeasures that have been proposed to detect or prevent poisoning attacks. Further insights have been provided by linking attacks and countermeasures, highlighting which countermeasures can be expected to be effective against which attacks, and which attacks can be expected to be resilient against which countermeasures. \editthree{We concluded by highlighting research gaps and providing directions for future work}.
By providing a public repository that includes all reviewed papers along with the program codes and datasets released in the context of these studies, we have also provided researchers in the field a comprehensive starting point to address these open challenges.

%\sstitle{Supplementary Materials}
%\editthree{Readers interested in exploring further mathematical and technical aspects of this paper are encouraged to refer to the supplementary materials of the article.}

% \begin{acks}
% \end{acks}


%\bibliographystyle{ACM-Reference-Format}
%\bibliography{../ref,../ref_h,../ren}

\begin{thebibliography}{176}

%%% the Web address.

\ifx \showCODEN    \undefined \def \showCODEN     #1{\unskip}     \fi
\ifx \showDOI      \undefined \def \showDOI       #1{#1}\fi
\ifx \showISBNx    \undefined \def \showISBNx     #1{\unskip}     \fi
\ifx \showISBNxiii \undefined \def \showISBNxiii  #1{\unskip}     \fi
\ifx \showISSN     \undefined \def \showISSN      #1{\unskip}     \fi
\ifx \showLCCN     \undefined \def \showLCCN      #1{\unskip}     \fi
\ifx \shownote     \undefined \def \shownote      #1{#1}          \fi
\ifx \showarticletitle \undefined \def \showarticletitle #1{#1}   \fi
\ifx \showURL      \undefined \def \showURL       {\relax}        \fi
% invisible to TeX
\providecommand\bibfield[2]{#2}
\providecommand\bibinfo[2]{#2}
\providecommand\natexlab[1]{#1}
\providecommand\showeprint[2][]{arXiv:#2}

\bibitem[\protect\citeauthoryear{??}{Ind}{[n.\,d.]}]%
        {IndustryARC}
 \bibinfo{year}{[n.\,d.]}\natexlab{}.
\newblock
\newblock
\urldef\tempurl%
\url{https://www.industryarc.com/Research/Recommendation-Engine-Market-Research-500995}
\showURL{%
\tempurl}


\bibitem[\protect\citeauthoryear{??}{Son}{[n.\,d.]}]%
        {SonyFakeNews}
 \bibinfo{year}{[n.\,d.]}\natexlab{}.
\newblock
\newblock
\urldef\tempurl%
\url{http://news.bbc.co.uk/2/hi/entertainment/1368666.stm}
\showURL{%
\tempurl}


\bibitem[\protect\citeauthoryear{Abdollahi and Nasraoui}{Abdollahi and
  Nasraoui}{2018}]%
        {abdollahi2018transparency}
\bibfield{author}{\bibinfo{person}{Behnoush Abdollahi} {and}
  \bibinfo{person}{Olfa Nasraoui}.} \bibinfo{year}{2018}\natexlab{}.
\newblock \showarticletitle{Transparency in fair machine learning: the case of
  explainable recommender systems}.
\newblock In \bibinfo{booktitle}{\emph{Human and machine learning}}.
  \bibinfo{publisher}{Springer}, \bibinfo{pages}{21--35}.
\newblock


\bibitem[\protect\citeauthoryear{Aditya, Budi, and Munajat}{Aditya
  et~al\mbox{.}}{2016}]%
        {aditya2016comparative}
\bibfield{author}{\bibinfo{person}{PH Aditya}, \bibinfo{person}{Indra Budi},
  {and} \bibinfo{person}{Qorib Munajat}.} \bibinfo{year}{2016}\natexlab{}.
\newblock \showarticletitle{A comparative analysis of memory-based and
  model-based collaborative filtering on the implementation of recommender
  system for E-commerce}. In \bibinfo{booktitle}{\emph{ICACSIS}}.
  \bibinfo{pages}{303--308}.
\newblock


\bibitem[\protect\citeauthoryear{Aggarwal, Mittal, and Battineni}{Aggarwal
  et~al\mbox{.}}{2021}]%
        {aggarwal2021generative}
\bibfield{author}{\bibinfo{person}{Alankrita Aggarwal}, \bibinfo{person}{Mamta
  Mittal}, {and} \bibinfo{person}{Gopi Battineni}.}
  \bibinfo{year}{2021}\natexlab{}.
\newblock \showarticletitle{Generative adversarial network: An overview of
  theory and applications}.
\newblock \bibinfo{journal}{\emph{International Journal of Information
  Management Data Insights}} \bibinfo{volume}{1}, \bibinfo{number}{1}
  (\bibinfo{year}{2021}), \bibinfo{pages}{100004}.
\newblock


\bibitem[\protect\citeauthoryear{Aggarwal et~al\mbox{.}}{Aggarwal
  et~al\mbox{.}}{2016}]%
        {aggarwal2016recommender}
\bibfield{author}{\bibinfo{person}{Charu~C Aggarwal} {et~al\mbox{.}}}
  \bibinfo{year}{2016}\natexlab{}.
\newblock \bibinfo{booktitle}{\emph{Recommender systems}}.
  Vol.~\bibinfo{volume}{1}.
\newblock


\bibitem[\protect\citeauthoryear{Aktukmak, Yilmaz, and Uysal}{Aktukmak
  et~al\mbox{.}}{2019}]%
        {aktukmak2019quick}
\bibfield{author}{\bibinfo{person}{Mehmet Aktukmak}, \bibinfo{person}{Yasin
  Yilmaz}, {and} \bibinfo{person}{Ismail Uysal}.}
  \bibinfo{year}{2019}\natexlab{}.
\newblock \showarticletitle{Quick and accurate attack detection in recommender
  systems through user attributes}. In \bibinfo{booktitle}{\emph{RecSys}}.
  \bibinfo{pages}{348--352}.
\newblock


\bibitem[\protect\citeauthoryear{Alhijawi and Kilani}{Alhijawi and
  Kilani}{2020}]%
        {alhijawi2020recommender}
\bibfield{author}{\bibinfo{person}{Bushra Alhijawi} {and}
  \bibinfo{person}{Yousef Kilani}.} \bibinfo{year}{2020}\natexlab{}.
\newblock \showarticletitle{The recommender system: a survey}.
\newblock \bibinfo{journal}{\emph{IJAIP}} \bibinfo{volume}{15},
  \bibinfo{number}{3} (\bibinfo{year}{2020}), \bibinfo{pages}{229--251}.
\newblock


\bibitem[\protect\citeauthoryear{Ali, Kefalas, Muhammad, Ali, and Imran}{Ali
  et~al\mbox{.}}{2020}]%
        {ali2020deep}
\bibfield{author}{\bibinfo{person}{Zafar Ali}, \bibinfo{person}{Pavlos
  Kefalas}, \bibinfo{person}{Khan Muhammad}, \bibinfo{person}{Bahadar Ali},
  {and} \bibinfo{person}{Muhammad Imran}.} \bibinfo{year}{2020}\natexlab{}.
\newblock \showarticletitle{Deep learning in citation recommendation models
  survey}.
\newblock \bibinfo{journal}{\emph{Expert Systems with Applications}}
  \bibinfo{volume}{162} (\bibinfo{year}{2020}), \bibinfo{pages}{113790}.
\newblock


\bibitem[\protect\citeauthoryear{Ali, Khusro, and Ullah}{Ali
  et~al\mbox{.}}{2016}]%
        {ali2016hybrid}
\bibfield{author}{\bibinfo{person}{Zafar Ali}, \bibinfo{person}{Shah Khusro},
  {and} \bibinfo{person}{Irfan Ullah}.} \bibinfo{year}{2016}\natexlab{}.
\newblock \showarticletitle{A hybrid book recommender system based on table of
  contents (toc) and association rule mining}. In
  \bibinfo{booktitle}{\emph{INFOS}}. \bibinfo{pages}{68--74}.
\newblock


\bibitem[\protect\citeauthoryear{Ali, Ullah, Khan, Ullah~Jan, and Muhammad}{Ali
  et~al\mbox{.}}{2021}]%
        {ali2021overview}
\bibfield{author}{\bibinfo{person}{Zafar Ali}, \bibinfo{person}{Irfan Ullah},
  \bibinfo{person}{Amin Khan}, \bibinfo{person}{Asim Ullah~Jan}, {and}
  \bibinfo{person}{Khan Muhammad}.} \bibinfo{year}{2021}\natexlab{}.
\newblock \showarticletitle{An overview and evaluation of citation
  recommendation models}.
\newblock \bibinfo{journal}{\emph{Scientometrics}}  \bibinfo{volume}{126}
  (\bibinfo{year}{2021}), \bibinfo{pages}{4083--4119}.
\newblock


\bibitem[\protect\citeauthoryear{Aliwa et~al\mbox{.}}{Aliwa
  et~al\mbox{.}}{2021}]%
        {aliwa2021cyberattacks}
\bibfield{author}{\bibinfo{person}{Emad Aliwa} {et~al\mbox{.}}}
  \bibinfo{year}{2021}\natexlab{}.
\newblock \showarticletitle{Cyberattacks and countermeasures for in-vehicle
  networks}.
\newblock \bibinfo{journal}{\emph{CSUR}} \bibinfo{volume}{54},
  \bibinfo{number}{1} (\bibinfo{year}{2021}), \bibinfo{pages}{1--37}.
\newblock


\bibitem[\protect\citeauthoryear{Amir, Coan, Kirsch, and Lane}{Amir
  et~al\mbox{.}}{2008}]%
        {amir2008byzantine}
\bibfield{author}{\bibinfo{person}{Yair Amir}, \bibinfo{person}{Brian Coan},
  \bibinfo{person}{Jonathan Kirsch}, {and} \bibinfo{person}{John Lane}.}
  \bibinfo{year}{2008}\natexlab{}.
\newblock \showarticletitle{Byzantine replication under attack}. In
  \bibinfo{booktitle}{\emph{DSN}}. \bibinfo{pages}{197--206}.
\newblock


\bibitem[\protect\citeauthoryear{Anelli, Deldjoo, Di~Noia, Malitesta, and
  Merra}{Anelli et~al\mbox{.}}{2021}]%
        {anelli2021study}
\bibfield{author}{\bibinfo{person}{Vito~Walter Anelli}, \bibinfo{person}{Yashar
  Deldjoo}, \bibinfo{person}{Tommaso Di~Noia}, \bibinfo{person}{Daniele
  Malitesta}, {and} \bibinfo{person}{Felice~Antonio Merra}.}
  \bibinfo{year}{2021}\natexlab{}.
\newblock \showarticletitle{A study of defensive methods to protect visual
  recommendation against adversarial manipulation of images}. In
  \bibinfo{booktitle}{\emph{SIGIR}}.
\newblock


\bibitem[\protect\citeauthoryear{Anwar, Siddiqui, and Sohail}{Anwar
  et~al\mbox{.}}{2020}]%
        {anwar2020machine}
\bibfield{author}{\bibinfo{person}{Khalid Anwar}, \bibinfo{person}{Jamshed
  Siddiqui}, {and} \bibinfo{person}{Shahab~Saquib Sohail}.}
  \bibinfo{year}{2020}\natexlab{}.
\newblock \showarticletitle{Machine learning-based book recommender system: a
  survey and new perspectives}.
\newblock \bibinfo{journal}{\emph{IJIIDS}} \bibinfo{volume}{13},
  \bibinfo{number}{2-4} (\bibinfo{year}{2020}), \bibinfo{pages}{231--248}.
\newblock


\bibitem[\protect\citeauthoryear{Baracaldo, Chen, Ludwig, and Safavi}{Baracaldo
  et~al\mbox{.}}{2017}]%
        {baracaldo2017mitigating}
\bibfield{author}{\bibinfo{person}{Nathalie Baracaldo}, \bibinfo{person}{Bryant
  Chen}, \bibinfo{person}{Heiko Ludwig}, {and} \bibinfo{person}{Jaehoon~Amir
  Safavi}.} \bibinfo{year}{2017}\natexlab{}.
\newblock \showarticletitle{Mitigating poisoning attacks on machine learning
  models: A data provenance based approach}. In
  \bibinfo{booktitle}{\emph{AISec}}. \bibinfo{pages}{103--110}.
\newblock


\bibitem[\protect\citeauthoryear{Barbieri, Alvim, Braida, and
  Zimbr{\~a}o}{Barbieri et~al\mbox{.}}{2021}]%
        {barbieri2021simulating}
\bibfield{author}{\bibinfo{person}{Julio Barbieri}, \bibinfo{person}{Leandro~GM
  Alvim}, \bibinfo{person}{Filipe Braida}, {and} \bibinfo{person}{Geraldo
  Zimbr{\~a}o}.} \bibinfo{year}{2021}\natexlab{}.
\newblock \showarticletitle{Simulating real profiles for shilling attacks: A
  generative approach}.
\newblock \bibinfo{journal}{\emph{KBS}}  \bibinfo{volume}{230}
  (\bibinfo{year}{2021}), \bibinfo{pages}{107390}.
\newblock


\bibitem[\protect\citeauthoryear{Barth, Jackson, and Mitchell}{Barth
  et~al\mbox{.}}{2008}]%
        {barth2008robust}
\bibfield{author}{\bibinfo{person}{Adam Barth}, \bibinfo{person}{Collin
  Jackson}, {and} \bibinfo{person}{John~C Mitchell}.}
  \bibinfo{year}{2008}\natexlab{}.
\newblock \showarticletitle{Robust defenses for cross-site request forgery}. In
  \bibinfo{booktitle}{\emph{CCS}}. \bibinfo{pages}{75--88}.
\newblock


\bibitem[\protect\citeauthoryear{Bhaumik, Mobasher, and Burke}{Bhaumik
  et~al\mbox{.}}{2011}]%
        {bhaumik2011clustering}
\bibfield{author}{\bibinfo{person}{Runa Bhaumik}, \bibinfo{person}{Bamshad
  Mobasher}, {and} \bibinfo{person}{Robin Burke}.}
  \bibinfo{year}{2011}\natexlab{}.
\newblock \showarticletitle{A clustering approach to unsupervised attack
  detection in collaborative recommender systems}. In
  \bibinfo{booktitle}{\emph{ICDATA}}. \bibinfo{pages}{1}.
\newblock


\bibitem[\protect\citeauthoryear{Bilge et~al\mbox{.}}{Bilge
  et~al\mbox{.}}{2014}]%
        {bilge2014novel}
\bibfield{author}{\bibinfo{person}{Alper Bilge} {et~al\mbox{.}}}
  \bibinfo{year}{2014}\natexlab{}.
\newblock \showarticletitle{A novel shilling attack detection method}.
\newblock \bibinfo{journal}{\emph{Procedia Computer Science}}
  \bibinfo{volume}{31} (\bibinfo{year}{2014}), \bibinfo{pages}{165--174}.
\newblock


\bibitem[\protect\citeauthoryear{Branco et~al\mbox{.}}{Branco
  et~al\mbox{.}}{2016}]%
        {branco2016survey}
\bibfield{author}{\bibinfo{person}{Paula Branco} {et~al\mbox{.}}}
  \bibinfo{year}{2016}\natexlab{}.
\newblock \showarticletitle{A survey of predictive modeling on imbalanced
  domains}.
\newblock \bibinfo{journal}{\emph{CSUR}} \bibinfo{volume}{49},
  \bibinfo{number}{2} (\bibinfo{year}{2016}), \bibinfo{pages}{1--50}.
\newblock


\bibitem[\protect\citeauthoryear{Bryan and Cunningham}{Bryan and
  Cunningham}{2006}]%
        {bryan2006bottom}
\bibfield{author}{\bibinfo{person}{Kenneth Bryan} {and}
  \bibinfo{person}{P{\'a}draig Cunningham}.} \bibinfo{year}{2006}\natexlab{}.
\newblock \showarticletitle{Bottom-up biclustering of expression data}. In
  \bibinfo{booktitle}{\emph{CIBCB}}. \bibinfo{pages}{1--8}.
\newblock


\bibitem[\protect\citeauthoryear{Bryan, O'Mahony, and Cunningham}{Bryan
  et~al\mbox{.}}{2008}]%
        {bryan2008unsupervised}
\bibfield{author}{\bibinfo{person}{Kenneth Bryan}, \bibinfo{person}{Michael
  O'Mahony}, {and} \bibinfo{person}{P{\'a}draig Cunningham}.}
  \bibinfo{year}{2008}\natexlab{}.
\newblock \showarticletitle{Unsupervised retrieval of attack profiles in
  collaborative recommender systems}. In \bibinfo{booktitle}{\emph{RecSys}}.
  \bibinfo{pages}{155--162}.
\newblock


\bibitem[\protect\citeauthoryear{Burke, Mobasher, Williams, and Bhaumik}{Burke
  et~al\mbox{.}}{2006a}]%
        {burke2006classification}
\bibfield{author}{\bibinfo{person}{Robin Burke}, \bibinfo{person}{Bamshad
  Mobasher}, \bibinfo{person}{Chad Williams}, {and} \bibinfo{person}{Runa
  Bhaumik}.} \bibinfo{year}{2006}\natexlab{a}.
\newblock \showarticletitle{Classification features for attack detection in
  collaborative recommender systems}. In \bibinfo{booktitle}{\emph{KDD}}.
  \bibinfo{pages}{542--547}.
\newblock


\bibitem[\protect\citeauthoryear{Burke, Mobasher, Williams, and Bhaumik}{Burke
  et~al\mbox{.}}{2006b}]%
        {burke2006detecting}
\bibfield{author}{\bibinfo{person}{Robin Burke}, \bibinfo{person}{Bamshad
  Mobasher}, \bibinfo{person}{Chad Williams}, {and} \bibinfo{person}{Runa
  Bhaumik}.} \bibinfo{year}{2006}\natexlab{b}.
\newblock \showarticletitle{Detecting profile injection attacks in
  collaborative recommender systems}. In \bibinfo{booktitle}{\emph{CEC/EEE)}}.
  \bibinfo{pages}{23--23}.
\newblock


\bibitem[\protect\citeauthoryear{Cai and Zhang}{Cai and Zhang}{2019a}]%
        {cai2019bs}
\bibfield{author}{\bibinfo{person}{Hongyun Cai} {and} \bibinfo{person}{Fuzhi
  Zhang}.} \bibinfo{year}{2019}\natexlab{a}.
\newblock \showarticletitle{BS-SC: An Unsupervised Approach for Detecting
  Shilling Profiles in Collaborative Recommender Systems}.
\newblock \bibinfo{journal}{\emph{IEEE Transactions on Knowledge and Data
  Engineering}} (\bibinfo{year}{2019}).
\newblock


\bibitem[\protect\citeauthoryear{Cai and Zhang}{Cai and Zhang}{2019b}]%
        {cai2019detecting}
\bibfield{author}{\bibinfo{person}{Hongyun Cai} {and} \bibinfo{person}{Fuzhi
  Zhang}.} \bibinfo{year}{2019}\natexlab{b}.
\newblock \showarticletitle{Detecting shilling attacks in recommender systems
  based on analysis of user rating behavior}.
\newblock \bibinfo{journal}{\emph{KBS}}  \bibinfo{volume}{177}
  (\bibinfo{year}{2019}), \bibinfo{pages}{22--43}.
\newblock


\bibitem[\protect\citeauthoryear{Cai and Zhu}{Cai and Zhu}{2019}]%
        {cai2019trustworthy}
\bibfield{author}{\bibinfo{person}{Yuanfeng Cai} {and} \bibinfo{person}{Dan
  Zhu}.} \bibinfo{year}{2019}\natexlab{}.
\newblock \showarticletitle{Trustworthy and profit: A new value-based neighbor
  selection method in recommender systems under shilling attacks}.
\newblock \bibinfo{journal}{\emph{Decision Support Systems}}
  \bibinfo{volume}{124} (\bibinfo{year}{2019}), \bibinfo{pages}{113112}.
\newblock


\bibitem[\protect\citeauthoryear{Cai, Xiong, Xu, Wang, Li, and Pan}{Cai
  et~al\mbox{.}}{2021}]%
        {cai2021generative}
\bibfield{author}{\bibinfo{person}{Zhipeng Cai}, \bibinfo{person}{Zuobin
  Xiong}, \bibinfo{person}{Honghui Xu}, \bibinfo{person}{Peng Wang},
  \bibinfo{person}{Wei Li}, {and} \bibinfo{person}{Yi Pan}.}
  \bibinfo{year}{2021}\natexlab{}.
\newblock \showarticletitle{Generative adversarial networks: A survey toward
  private and secure applications}.
\newblock \bibinfo{journal}{\emph{CSUR}} \bibinfo{volume}{54},
  \bibinfo{number}{6} (\bibinfo{year}{2021}), \bibinfo{pages}{1--38}.
\newblock


\bibitem[\protect\citeauthoryear{Cao, Wu, Mao, and Zhang}{Cao
  et~al\mbox{.}}{2013}]%
        {cao2013shilling}
\bibfield{author}{\bibinfo{person}{Jie Cao}, \bibinfo{person}{Zhiang Wu},
  \bibinfo{person}{Bo Mao}, {and} \bibinfo{person}{Yanchun Zhang}.}
  \bibinfo{year}{2013}\natexlab{}.
\newblock \showarticletitle{Shilling attack detection utilizing semi-supervised
  learning method for collaborative recommender system}.
\newblock \bibinfo{journal}{\emph{World Wide Web}} \bibinfo{volume}{16},
  \bibinfo{number}{5-6} (\bibinfo{year}{2013}), \bibinfo{pages}{729--748}.
\newblock


\bibitem[\protect\citeauthoryear{Chacon, Silva, and Rad}{Chacon
  et~al\mbox{.}}{2019}]%
        {chacon2019deep}
\bibfield{author}{\bibinfo{person}{Henry Chacon}, \bibinfo{person}{Samuel
  Silva}, {and} \bibinfo{person}{Paul Rad}.} \bibinfo{year}{2019}\natexlab{}.
\newblock \showarticletitle{Deep learning poison data attack detection}. In
  \bibinfo{booktitle}{\emph{ICTAI}}. \bibinfo{pages}{971--978}.
\newblock


\bibitem[\protect\citeauthoryear{Chang, Ren, Nguyen, Nejdl, and Schuller}{Chang
  et~al\mbox{.}}{2022}]%
        {chang2022example}
\bibfield{author}{\bibinfo{person}{Yi Chang}, \bibinfo{person}{Zhao Ren},
  \bibinfo{person}{Thanh~Tam Nguyen}, \bibinfo{person}{Wolfgang Nejdl}, {and}
  \bibinfo{person}{Bj{\"o}rn~W Schuller}.} \bibinfo{year}{2022}\natexlab{}.
\newblock \showarticletitle{Example-based Explanations with Adversarial Attacks
  for Respiratory Sound Analysis}. In \bibinfo{booktitle}{\emph{Interspeech}}.
  \bibinfo{pages}{1--5}.
\newblock


\bibitem[\protect\citeauthoryear{Chen and Li}{Chen and Li}{2019}]%
        {chen2019data}
\bibfield{author}{\bibinfo{person}{Huiyuan Chen} {and} \bibinfo{person}{Jing
  Li}.} \bibinfo{year}{2019}\natexlab{}.
\newblock \showarticletitle{Data poisoning attacks on cross-domain
  recommendation}. In \bibinfo{booktitle}{\emph{CIKM}}.
  \bibinfo{pages}{2177--2180}.
\newblock


\bibitem[\protect\citeauthoryear{Chen, Fan, Zhu, Zhao, Yuan, Li, and
  Huang}{Chen et~al\mbox{.}}{2022}]%
        {chen2022knowledge}
\bibfield{author}{\bibinfo{person}{Jingfan Chen}, \bibinfo{person}{Wenqi Fan},
  \bibinfo{person}{Guanghui Zhu}, \bibinfo{person}{Xiangyu Zhao},
  \bibinfo{person}{Chunfeng Yuan}, \bibinfo{person}{Qing Li}, {and}
  \bibinfo{person}{Yihua Huang}.} \bibinfo{year}{2022}\natexlab{}.
\newblock \showarticletitle{Knowledge-enhanced Black-box Attacks for
  Recommendations}. In \bibinfo{booktitle}{\emph{KDD}}.
  \bibinfo{pages}{108--117}.
\newblock


\bibitem[\protect\citeauthoryear{Chen, Xu, Xie, Huang, and Zheng}{Chen
  et~al\mbox{.}}{2021}]%
        {chen2021data}
\bibfield{author}{\bibinfo{person}{Liang Chen}, \bibinfo{person}{Yangjun Xu},
  \bibinfo{person}{Fenfang Xie}, \bibinfo{person}{Min Huang}, {and}
  \bibinfo{person}{Zibin Zheng}.} \bibinfo{year}{2021}\natexlab{}.
\newblock \showarticletitle{Data poisoning attacks on neighborhood-based
  recommender systems}.
\newblock \bibinfo{journal}{\emph{Transactions on Emerging Telecommunications
  Technologies}} \bibinfo{volume}{32}, \bibinfo{number}{6}
  (\bibinfo{year}{2021}), \bibinfo{pages}{e3872}.
\newblock


\bibitem[\protect\citeauthoryear{Chirita, Nejdl, and Zamfir}{Chirita
  et~al\mbox{.}}{2005}]%
        {chirita2005preventing}
\bibfield{author}{\bibinfo{person}{Paul-Alexandru Chirita},
  \bibinfo{person}{Wolfgang Nejdl}, {and} \bibinfo{person}{Cristian Zamfir}.}
  \bibinfo{year}{2005}\natexlab{}.
\newblock \showarticletitle{Preventing shilling attacks in online recommender
  systems}. In \bibinfo{booktitle}{\emph{WIDM}}. \bibinfo{pages}{67--74}.
\newblock


\bibitem[\protect\citeauthoryear{Christakopoulou and Banerjee}{Christakopoulou
  and Banerjee}{2019}]%
        {christakopoulou2019adversarial}
\bibfield{author}{\bibinfo{person}{Konstantina Christakopoulou} {and}
  \bibinfo{person}{Arindam Banerjee}.} \bibinfo{year}{2019}\natexlab{}.
\newblock \showarticletitle{Adversarial attacks on an oblivious recommender}.
  In \bibinfo{booktitle}{\emph{RecSys}}. \bibinfo{pages}{322--330}.
\newblock


\bibitem[\protect\citeauthoryear{Chung, Hsu, and Huang}{Chung
  et~al\mbox{.}}{2013}]%
        {chung2013betap}
\bibfield{author}{\bibinfo{person}{Chen-Yao Chung}, \bibinfo{person}{Ping-Yu
  Hsu}, {and} \bibinfo{person}{Shih-Hsiang Huang}.}
  \bibinfo{year}{2013}\natexlab{}.
\newblock \showarticletitle{$\beta$P: A novel approach to filter out malicious
  rating profiles from recommender systems}.
\newblock \bibinfo{journal}{\emph{Decision Support Systems}}
  \bibinfo{volume}{55}, \bibinfo{number}{1} (\bibinfo{year}{2013}),
  \bibinfo{pages}{314--325}.
\newblock


\bibitem[\protect\citeauthoryear{Dai, Ding, and Wahba}{Dai
  et~al\mbox{.}}{2013}]%
        {dai2013multivariate}
\bibfield{author}{\bibinfo{person}{Bin Dai}, \bibinfo{person}{Shilin Ding},
  {and} \bibinfo{person}{Grace Wahba}.} \bibinfo{year}{2013}\natexlab{}.
\newblock \showarticletitle{Multivariate bernoulli distribution}.
\newblock \bibinfo{journal}{\emph{Bernoulli}} \bibinfo{volume}{19},
  \bibinfo{number}{4} (\bibinfo{year}{2013}), \bibinfo{pages}{1465--1483}.
\newblock


\bibitem[\protect\citeauthoryear{Das, Sahoo, and Datta}{Das
  et~al\mbox{.}}{2017}]%
        {das2017survey}
\bibfield{author}{\bibinfo{person}{Debashis Das}, \bibinfo{person}{Laxman
  Sahoo}, {and} \bibinfo{person}{Sujoy Datta}.}
  \bibinfo{year}{2017}\natexlab{}.
\newblock \showarticletitle{A survey on recommendation system}.
\newblock \bibinfo{journal}{\emph{IJCA}} \bibinfo{volume}{160},
  \bibinfo{number}{7} (\bibinfo{year}{2017}).
\newblock


\bibitem[\protect\citeauthoryear{Deldjoo, Noia, and Merra}{Deldjoo
  et~al\mbox{.}}{2021}]%
        {deldjoo2021survey}
\bibfield{author}{\bibinfo{person}{Yashar Deldjoo}, \bibinfo{person}{Tommaso~Di
  Noia}, {and} \bibinfo{person}{Felice~Antonio Merra}.}
  \bibinfo{year}{2021}\natexlab{}.
\newblock \showarticletitle{A survey on adversarial recommender systems: from
  attack/defense strategies to generative adversarial networks}.
\newblock \bibinfo{journal}{\emph{CSUR}} \bibinfo{volume}{54},
  \bibinfo{number}{2} (\bibinfo{year}{2021}), \bibinfo{pages}{1--38}.
\newblock


\bibitem[\protect\citeauthoryear{Dengwen}{Dengwen}{2010}]%
        {dengwen2010edge}
\bibfield{author}{\bibinfo{person}{Zhou Dengwen}.}
  \bibinfo{year}{2010}\natexlab{}.
\newblock \showarticletitle{An edge-directed bicubic interpolation algorithm}.
  In \bibinfo{booktitle}{\emph{CISP}}, Vol.~\bibinfo{volume}{3}.
  \bibinfo{pages}{1186--1189}.
\newblock


\bibitem[\protect\citeauthoryear{Duong, Nguyen, Hoang, Yin, Weidlich, and
  Nguyen}{Duong et~al\mbox{.}}{2022}]%
        {duong2022deep}
\bibfield{author}{\bibinfo{person}{Chi~Thang Duong}, \bibinfo{person}{Thanh~Tam
  Nguyen}, \bibinfo{person}{Trung-Dung Hoang}, \bibinfo{person}{Hongzhi Yin},
  \bibinfo{person}{Matthias Weidlich}, {and} \bibinfo{person}{Quoc Viet~Hung
  Nguyen}.} \bibinfo{year}{2022}\natexlab{}.
\newblock \showarticletitle{Deep MinCut: Learning Node Embeddings from
  Detecting Communities}.
\newblock \bibinfo{journal}{\emph{Pattern Recognition}} (\bibinfo{year}{2022}),
  \bibinfo{pages}{109126}.
\newblock


\bibitem[\protect\citeauthoryear{Evangelopoulou and Johnson}{Evangelopoulou and
  Johnson}{2014}]%
        {evangelopoulou2014attack}
\bibfield{author}{\bibinfo{person}{M Evangelopoulou} {and} \bibinfo{person}{CW
  Johnson}.} \bibinfo{year}{2014}\natexlab{}.
\newblock \showarticletitle{Attack visualisation for cyber-security situation
  awareness}.
\newblock  (\bibinfo{year}{2014}).
\newblock


\bibitem[\protect\citeauthoryear{Fan, Yan, Li, Qu, and Xiao}{Fan
  et~al\mbox{.}}{2022}]%
        {fan2022survey}
\bibfield{author}{\bibinfo{person}{Jiaxin Fan}, \bibinfo{person}{Qi Yan},
  \bibinfo{person}{Mohan Li}, \bibinfo{person}{Guanqun Qu}, {and}
  \bibinfo{person}{Yang Xiao}.} \bibinfo{year}{2022}\natexlab{}.
\newblock \showarticletitle{A Survey on Data Poisoning Attacks and Defenses}.
  In \bibinfo{booktitle}{\emph{2022 7th IEEE International Conference on Data
  Science in Cyberspace (DSC)}}. IEEE, \bibinfo{pages}{48--55}.
\newblock


\bibitem[\protect\citeauthoryear{Fan, Derr, Zhao, Ma, Liu, Wang, Tang, and
  Li}{Fan et~al\mbox{.}}{2021}]%
        {fan2021attacking}
\bibfield{author}{\bibinfo{person}{Wenqi Fan}, \bibinfo{person}{Tyler Derr},
  \bibinfo{person}{Xiangyu Zhao}, \bibinfo{person}{Yao Ma},
  \bibinfo{person}{Hui Liu}, \bibinfo{person}{Jianping Wang},
  \bibinfo{person}{Jiliang Tang}, {and} \bibinfo{person}{Qing Li}.}
  \bibinfo{year}{2021}\natexlab{}.
\newblock \showarticletitle{Attacking Black-box Recommendations via Copying
  Cross-domain User Profiles}. In \bibinfo{booktitle}{\emph{ICDE}}.
  \bibinfo{pages}{1583--1594}.
\newblock


\bibitem[\protect\citeauthoryear{Fang, Gong, and Liu}{Fang
  et~al\mbox{.}}{2020}]%
        {fang2020influence}
\bibfield{author}{\bibinfo{person}{Minghong Fang},
  \bibinfo{person}{Neil~Zhenqiang Gong}, {and} \bibinfo{person}{Jia Liu}.}
  \bibinfo{year}{2020}\natexlab{}.
\newblock \showarticletitle{Influence function based data poisoning attacks to
  top-n recommender systems}. In \bibinfo{booktitle}{\emph{WWW}}.
  \bibinfo{pages}{3019--3025}.
\newblock


\bibitem[\protect\citeauthoryear{Fang, Yang, Gong, and Liu}{Fang
  et~al\mbox{.}}{2018}]%
        {fang2018poisoning}
\bibfield{author}{\bibinfo{person}{Minghong Fang}, \bibinfo{person}{Guolei
  Yang}, \bibinfo{person}{Neil~Zhenqiang Gong}, {and} \bibinfo{person}{Jia
  Liu}.} \bibinfo{year}{2018}\natexlab{}.
\newblock \showarticletitle{Poisoning attacks to graph-based recommender
  systems}. In \bibinfo{booktitle}{\emph{ACSAC}}. \bibinfo{pages}{381--392}.
\newblock


\bibitem[\protect\citeauthoryear{Fletcher et~al\mbox{.}}{Fletcher
  et~al\mbox{.}}{2019}]%
        {fletcher2019decision}
\bibfield{author}{\bibinfo{person}{Sam Fletcher} {et~al\mbox{.}}}
  \bibinfo{year}{2019}\natexlab{}.
\newblock \showarticletitle{Decision tree classification with differential
  privacy: A survey}.
\newblock \bibinfo{journal}{\emph{CSUR}} \bibinfo{volume}{52},
  \bibinfo{number}{4} (\bibinfo{year}{2019}), \bibinfo{pages}{1--33}.
\newblock


\bibitem[\protect\citeauthoryear{Friedman and Schuster}{Friedman and
  Schuster}{2010}]%
        {friedman2010data}
\bibfield{author}{\bibinfo{person}{Arik Friedman} {and} \bibinfo{person}{Assaf
  Schuster}.} \bibinfo{year}{2010}\natexlab{}.
\newblock \showarticletitle{Data mining with differential privacy}. In
  \bibinfo{booktitle}{\emph{KDD}}. \bibinfo{pages}{493--502}.
\newblock


\bibitem[\protect\citeauthoryear{Gama, {\v{Z}}liobait{\.e}, Bifet, Pechenizkiy,
  and Bouchachia}{Gama et~al\mbox{.}}{2014}]%
        {gama2014survey}
\bibfield{author}{\bibinfo{person}{Jo{\~a}o Gama}, \bibinfo{person}{Indr{\.e}
  {\v{Z}}liobait{\.e}}, \bibinfo{person}{Albert Bifet}, \bibinfo{person}{Mykola
  Pechenizkiy}, {and} \bibinfo{person}{Abdelhamid Bouchachia}.}
  \bibinfo{year}{2014}\natexlab{}.
\newblock \showarticletitle{A survey on concept drift adaptation}.
\newblock \bibinfo{journal}{\emph{CSUR}} \bibinfo{volume}{46},
  \bibinfo{number}{4} (\bibinfo{year}{2014}), \bibinfo{pages}{1--37}.
\newblock


\bibitem[\protect\citeauthoryear{Gardiner and Nagaraja}{Gardiner and
  Nagaraja}{2016}]%
        {gardiner2016security}
\bibfield{author}{\bibinfo{person}{Joseph Gardiner} {and}
  \bibinfo{person}{Shishir Nagaraja}.} \bibinfo{year}{2016}\natexlab{}.
\newblock \showarticletitle{On the security of machine learning in malware c\&c
  detection: A survey}.
\newblock \bibinfo{journal}{\emph{CSUR}} \bibinfo{volume}{49},
  \bibinfo{number}{3} (\bibinfo{year}{2016}), \bibinfo{pages}{1--39}.
\newblock


\bibitem[\protect\citeauthoryear{Ghaffarian and Shahriari}{Ghaffarian and
  Shahriari}{2017}]%
        {ghaffarian2017software}
\bibfield{author}{\bibinfo{person}{Seyed~Mohammad Ghaffarian} {and}
  \bibinfo{person}{Hamid~Reza Shahriari}.} \bibinfo{year}{2017}\natexlab{}.
\newblock \showarticletitle{Software vulnerability analysis and discovery using
  machine-learning and data-mining techniques: A survey}.
\newblock \bibinfo{journal}{\emph{ACM Computing Surveys (CSUR)}}
  \bibinfo{volume}{50}, \bibinfo{number}{4} (\bibinfo{year}{2017}),
  \bibinfo{pages}{1--36}.
\newblock


\bibitem[\protect\citeauthoryear{Goodfellow, Bengio, and Courville}{Goodfellow
  et~al\mbox{.}}{2016}]%
        {goodfellow2016deep}
\bibfield{author}{\bibinfo{person}{Ian Goodfellow}, \bibinfo{person}{Yoshua
  Bengio}, {and} \bibinfo{person}{Aaron Courville}.}
  \bibinfo{year}{2016}\natexlab{}.
\newblock \bibinfo{booktitle}{\emph{Deep learning}}.
\newblock \bibinfo{publisher}{MIT press}.
\newblock


\bibitem[\protect\citeauthoryear{Guo, Tang, Chen, Zhu, Nguyen, and Yin}{Guo
  et~al\mbox{.}}{2021}]%
        {guo2021gcn}
\bibfield{author}{\bibinfo{person}{Lei Guo}, \bibinfo{person}{Li Tang},
  \bibinfo{person}{Tong Chen}, \bibinfo{person}{Lei Zhu}, \bibinfo{person}{Quoc
  Viet~Hung Nguyen}, {and} \bibinfo{person}{Hongzhi Yin}.}
  \bibinfo{year}{2021}\natexlab{}.
\newblock \showarticletitle{DA-GCN: A Domain-aware Attentive Graph Convolution
  Network for Shared-account Cross-domain Sequential Recommendation}.
\newblock \bibinfo{journal}{\emph{arXiv preprint arXiv:2105.03300}}
  (\bibinfo{year}{2021}).
\newblock


\bibitem[\protect\citeauthoryear{Guo, Zhuang, Qin, Zhu, Xie, Xiong, and He}{Guo
  et~al\mbox{.}}{2020}]%
        {guo2020survey}
\bibfield{author}{\bibinfo{person}{Qingyu Guo}, \bibinfo{person}{Fuzhen
  Zhuang}, \bibinfo{person}{Chuan Qin}, \bibinfo{person}{Hengshu Zhu},
  \bibinfo{person}{Xing Xie}, \bibinfo{person}{Hui Xiong}, {and}
  \bibinfo{person}{Qing He}.} \bibinfo{year}{2020}\natexlab{}.
\newblock \showarticletitle{A survey on knowledge graph-based recommender
  systems}.
\newblock \bibinfo{journal}{\emph{IEEE Transactions on Knowledge and Data
  Engineering}} (\bibinfo{year}{2020}).
\newblock


\bibitem[\protect\citeauthoryear{Han and Yamana}{Han and Yamana}{2017}]%
        {han2017survey}
\bibfield{author}{\bibinfo{person}{Jungkyu Han} {and} \bibinfo{person}{Hayato
  Yamana}.} \bibinfo{year}{2017}\natexlab{}.
\newblock \showarticletitle{A survey on recommendation methods beyond
  accuracy}.
\newblock \bibinfo{journal}{\emph{IEICE TRANSACTIONS on Information and
  Systems}} \bibinfo{volume}{100}, \bibinfo{number}{12} (\bibinfo{year}{2017}),
  \bibinfo{pages}{2931--2944}.
\newblock


\bibitem[\protect\citeauthoryear{Han, Zhang, Ren, Ringeval, and Schuller}{Han
  et~al\mbox{.}}{2018}]%
        {han2018adversarial}
\bibfield{author}{\bibinfo{person}{Jing Han}, \bibinfo{person}{Zixing Zhang},
  \bibinfo{person}{Zhao Ren}, \bibinfo{person}{Fabien Ringeval}, {and}
  \bibinfo{person}{Bj{\"o}rn Schuller}.} \bibinfo{year}{2018}\natexlab{}.
\newblock \showarticletitle{Towards conditional adversarial training for
  predicting emotions from speech}. In \bibinfo{booktitle}{\emph{Proc.\
  ICASSP}}. \bibinfo{address}{Calgary, Canada}, \bibinfo{pages}{6822--6826}.
\newblock


\bibitem[\protect\citeauthoryear{Hao and Zhang}{Hao and Zhang}{2021}]%
        {hao2021unsupervised}
\bibfield{author}{\bibinfo{person}{Yaojun Hao} {and} \bibinfo{person}{Fuzhi
  Zhang}.} \bibinfo{year}{2021}\natexlab{}.
\newblock \showarticletitle{An unsupervised detection method for shilling
  attacks based on deep learning and community detection}.
\newblock \bibinfo{journal}{\emph{Soft Computing}} \bibinfo{volume}{25},
  \bibinfo{number}{1} (\bibinfo{year}{2021}), \bibinfo{pages}{477--494}.
\newblock


\bibitem[\protect\citeauthoryear{Hao, Zhang, Wang, Zhao, and Cao}{Hao
  et~al\mbox{.}}{2019}]%
        {hao2019detecting}
\bibfield{author}{\bibinfo{person}{Yaojun Hao}, \bibinfo{person}{Fuzhi Zhang},
  \bibinfo{person}{Jian Wang}, \bibinfo{person}{Qingshan Zhao}, {and}
  \bibinfo{person}{Jianfang Cao}.} \bibinfo{year}{2019}\natexlab{}.
\newblock \showarticletitle{Detecting shilling attacks with automatic features
  from multiple views}.
\newblock \bibinfo{journal}{\emph{Security and Communication Networks}}
  \bibinfo{volume}{2019} (\bibinfo{year}{2019}).
\newblock


\bibitem[\protect\citeauthoryear{Hu, Guo, Pan, and Gong}{Hu
  et~al\mbox{.}}{2019}]%
        {hu2019targeted}
\bibfield{author}{\bibinfo{person}{Rui Hu}, \bibinfo{person}{Yuanxiong Guo},
  \bibinfo{person}{Miao Pan}, {and} \bibinfo{person}{Yanmin Gong}.}
  \bibinfo{year}{2019}\natexlab{}.
\newblock \showarticletitle{Targeted poisoning attacks on social recommender
  systems}. In \bibinfo{booktitle}{\emph{GLOBECOM}}. \bibinfo{pages}{1--6}.
\newblock


\bibitem[\protect\citeauthoryear{Hu, Koren, et~al\mbox{.}}{Hu
  et~al\mbox{.}}{2008}]%
        {hu2008collaborative}
\bibfield{author}{\bibinfo{person}{Yifan Hu}, \bibinfo{person}{Yehuda Koren},
  {et~al\mbox{.}}} \bibinfo{year}{2008}\natexlab{}.
\newblock \showarticletitle{Collaborative filtering for implicit feedback
  datasets}. In \bibinfo{booktitle}{\emph{ICDM}}. \bibinfo{pages}{263--272}.
\newblock


\bibitem[\protect\citeauthoryear{Huang, Mu, Gong, Li, Liu, and Xu}{Huang
  et~al\mbox{.}}{2021}]%
        {huang2021data}
\bibfield{author}{\bibinfo{person}{Hai Huang}, \bibinfo{person}{Jiaming Mu},
  \bibinfo{person}{Neil~Zhenqiang Gong}, \bibinfo{person}{Qi Li},
  \bibinfo{person}{Bin Liu}, {and} \bibinfo{person}{Mingwei Xu}.}
  \bibinfo{year}{2021}\natexlab{}.
\newblock \showarticletitle{Data poisoning attacks to deep learning based
  recommender systems}.
\newblock \bibinfo{journal}{\emph{arXiv preprint arXiv:2101.02644}}
  (\bibinfo{year}{2021}).
\newblock


\bibitem[\protect\citeauthoryear{Huang, Siegel, and Madnick}{Huang
  et~al\mbox{.}}{2018}]%
        {huang2018systematically}
\bibfield{author}{\bibinfo{person}{Keman Huang}, \bibinfo{person}{Michael
  Siegel}, {and} \bibinfo{person}{Stuart Madnick}.}
  \bibinfo{year}{2018}\natexlab{}.
\newblock \showarticletitle{Systematically understanding the cyber attack
  business: A survey}.
\newblock \bibinfo{journal}{\emph{CSUR}} \bibinfo{volume}{51},
  \bibinfo{number}{4} (\bibinfo{year}{2018}), \bibinfo{pages}{1--36}.
\newblock


\bibitem[\protect\citeauthoryear{Huang, Chung, Ong, and Chen}{Huang
  et~al\mbox{.}}{2002}]%
        {huang2002graph}
\bibfield{author}{\bibinfo{person}{Zan Huang}, \bibinfo{person}{Wingyan Chung},
  \bibinfo{person}{Thian-Huat Ong}, {and} \bibinfo{person}{Hsinchun Chen}.}
  \bibinfo{year}{2002}\natexlab{}.
\newblock \showarticletitle{A graph-based recommender system for digital
  library}. In \bibinfo{booktitle}{\emph{Proceedings of the 2nd ACM/IEEE-CS
  joint conference on Digital libraries}}. \bibinfo{pages}{65--73}.
\newblock


\bibitem[\protect\citeauthoryear{Hung, Weidlich, Tam, Mikl{\'o}s, Aberer, Gal,
  and Stantic}{Hung et~al\mbox{.}}{2019}]%
        {hung2019handling}
\bibfield{author}{\bibinfo{person}{Nguyen Quoc~Viet Hung},
  \bibinfo{person}{Matthias Weidlich}, \bibinfo{person}{Nguyen~Thanh Tam},
  \bibinfo{person}{Zolt{\'a}n Mikl{\'o}s}, \bibinfo{person}{Karl Aberer},
  \bibinfo{person}{Avigdor Gal}, {and} \bibinfo{person}{Bela Stantic}.}
  \bibinfo{year}{2019}\natexlab{}.
\newblock \showarticletitle{Handling probabilistic integrity constraints in
  pay-as-you-go reconciliation of data models}.
\newblock \bibinfo{journal}{\emph{Information Systems}}  \bibinfo{volume}{83}
  (\bibinfo{year}{2019}), \bibinfo{pages}{166--180}.
\newblock


\bibitem[\protect\citeauthoryear{Huynh, Duong, Nguyen, Van, Sattar, Yin, and
  Nguyen}{Huynh et~al\mbox{.}}{2021}]%
        {huynh2021network}
\bibfield{author}{\bibinfo{person}{Thanh~Trung Huynh},
  \bibinfo{person}{Chi~Thang Duong}, \bibinfo{person}{Thanh~Tam Nguyen},
  \bibinfo{person}{Vinh~Tong Van}, \bibinfo{person}{Abdul Sattar},
  \bibinfo{person}{Hongzhi Yin}, {and} \bibinfo{person}{Quoc Viet~Hung
  Nguyen}.} \bibinfo{year}{2021}\natexlab{}.
\newblock \showarticletitle{Network alignment with holistic embeddings}.
\newblock \bibinfo{journal}{\emph{TKDE}} \bibinfo{volume}{35},
  \bibinfo{number}{2} (\bibinfo{year}{2021}), \bibinfo{pages}{1881--1894}.
\newblock


\bibitem[\protect\citeauthoryear{Jia, Liu, Hu, and Gong}{Jia
  et~al\mbox{.}}{2023}]%
        {jia2023pore}
\bibfield{author}{\bibinfo{person}{Jinyuan Jia}, \bibinfo{person}{Yupei Liu},
  \bibinfo{person}{Yuepeng Hu}, {and} \bibinfo{person}{Neil~Zhenqiang Gong}.}
  \bibinfo{year}{2023}\natexlab{}.
\newblock \showarticletitle{PORE: Provably Robust Recommender Systems against
  Data Poisoning Attacks}.
\newblock \bibinfo{journal}{\emph{arXiv preprint arXiv:2303.14601}}
  (\bibinfo{year}{2023}).
\newblock


\bibitem[\protect\citeauthoryear{Khan, Aalsalem, Saad, and Xiang}{Khan
  et~al\mbox{.}}{2013}]%
        {khan2013detection}
\bibfield{author}{\bibinfo{person}{Wazir~Zada Khan},
  \bibinfo{person}{Mohammed~Y Aalsalem}, \bibinfo{person}{Mohammed Naufal
  Bin~Mohammed Saad}, {and} \bibinfo{person}{Yang Xiang}.}
  \bibinfo{year}{2013}\natexlab{}.
\newblock \showarticletitle{Detection and mitigation of node replication
  attacks in wireless sensor networks: a survey}.
\newblock \bibinfo{journal}{\emph{IJDSN}} \bibinfo{volume}{9},
  \bibinfo{number}{5} (\bibinfo{year}{2013}), \bibinfo{pages}{149023}.
\newblock


\bibitem[\protect\citeauthoryear{Koren et~al\mbox{.}}{Koren
  et~al\mbox{.}}{2009}]%
        {koren2009matrix}
\bibfield{author}{\bibinfo{person}{Yehuda Koren} {et~al\mbox{.}}}
  \bibinfo{year}{2009}\natexlab{}.
\newblock \showarticletitle{Matrix factorization techniques for recommender
  systems}.
\newblock \bibinfo{journal}{\emph{Computer}} \bibinfo{volume}{42},
  \bibinfo{number}{8} (\bibinfo{year}{2009}), \bibinfo{pages}{30--37}.
\newblock


\bibitem[\protect\citeauthoryear{Lam, Frankowski, Riedl, et~al\mbox{.}}{Lam
  et~al\mbox{.}}{2006}]%
        {lam2006you}
\bibfield{author}{\bibinfo{person}{Shyong~K Lam}, \bibinfo{person}{Dan
  Frankowski}, \bibinfo{person}{John Riedl}, {et~al\mbox{.}}}
  \bibinfo{year}{2006}\natexlab{}.
\newblock \showarticletitle{Do you trust your recommendations? An exploration
  of security and privacy issues in recommender systems}. In
  \bibinfo{booktitle}{\emph{ETRICS}}. \bibinfo{pages}{14--29}.
\newblock


\bibitem[\protect\citeauthoryear{Lam and Riedl}{Lam and Riedl}{2004}]%
        {lam2004shilling}
\bibfield{author}{\bibinfo{person}{Shyong~K Lam} {and} \bibinfo{person}{John
  Riedl}.} \bibinfo{year}{2004}\natexlab{}.
\newblock \showarticletitle{Shilling recommender systems for fun and profit}.
  In \bibinfo{booktitle}{\emph{WWW}}. \bibinfo{pages}{393--402}.
\newblock


\bibitem[\protect\citeauthoryear{Li, Wang, Singh, and Vorobeychik}{Li
  et~al\mbox{.}}{2016}]%
        {li2016data}
\bibfield{author}{\bibinfo{person}{Bo Li}, \bibinfo{person}{Yining Wang},
  \bibinfo{person}{Aarti Singh}, {and} \bibinfo{person}{Yevgeniy Vorobeychik}.}
  \bibinfo{year}{2016}\natexlab{}.
\newblock \showarticletitle{Data poisoning attacks on factorization-based
  collaborative filtering}. In \bibinfo{booktitle}{\emph{NIPS}},
  Vol.~\bibinfo{volume}{29}. \bibinfo{pages}{1885--1893}.
\newblock


\bibitem[\protect\citeauthoryear{Li, Ge, and Zhang}{Li et~al\mbox{.}}{2021}]%
        {li2021tutorial}
\bibfield{author}{\bibinfo{person}{Yunqi Li}, \bibinfo{person}{Yingqiang Ge},
  {and} \bibinfo{person}{Yongfeng Zhang}.} \bibinfo{year}{2021}\natexlab{}.
\newblock \showarticletitle{Tutorial on Fairness of Machine Learning in
  Recommender Systems}. In \bibinfo{booktitle}{\emph{SIGIR}}.
  \bibinfo{pages}{2654--2657}.
\newblock


\bibitem[\protect\citeauthoryear{Lin, Chen, Li, Xiao, Li, and Yang}{Lin
  et~al\mbox{.}}{2020}]%
        {lin2020attacking}
\bibfield{author}{\bibinfo{person}{Chen Lin}, \bibinfo{person}{Si Chen},
  \bibinfo{person}{Hui Li}, \bibinfo{person}{Yanghua Xiao},
  \bibinfo{person}{Lianyun Li}, {and} \bibinfo{person}{Qian Yang}.}
  \bibinfo{year}{2020}\natexlab{}.
\newblock \showarticletitle{Attacking recommender systems with augmented user
  profiles}. In \bibinfo{booktitle}{\emph{CIKM}}. \bibinfo{pages}{855--864}.
\newblock


\bibitem[\protect\citeauthoryear{Lin, Chen, Zeng, Zhang, Gao, and Li}{Lin
  et~al\mbox{.}}{2022}]%
        {lin2022shilling}
\bibfield{author}{\bibinfo{person}{Chen Lin}, \bibinfo{person}{Si Chen},
  \bibinfo{person}{Meifang Zeng}, \bibinfo{person}{Sheng Zhang},
  \bibinfo{person}{Min Gao}, {and} \bibinfo{person}{Hui Li}.}
  \bibinfo{year}{2022}\natexlab{}.
\newblock \showarticletitle{Shilling Black-Box Recommender Systems by Learning
  to Generate Fake User Profiles}.
\newblock \bibinfo{journal}{\emph{TNNLS}} (\bibinfo{year}{2022}).
\newblock


\bibitem[\protect\citeauthoryear{Lit, Kim, and Sy}{Lit et~al\mbox{.}}{2021}]%
        {lit2021survey}
\bibfield{author}{\bibinfo{person}{Yanyan Lit}, \bibinfo{person}{Sara Kim},
  {and} \bibinfo{person}{Eric Sy}.} \bibinfo{year}{2021}\natexlab{}.
\newblock \showarticletitle{A Survey on Amazon Alexa Attack Surfaces}. In
  \bibinfo{booktitle}{\emph{CCNC}}. \bibinfo{pages}{1--7}.
\newblock


\bibitem[\protect\citeauthoryear{Liu and Larson}{Liu and Larson}{2021}]%
        {liu2021adversarial}
\bibfield{author}{\bibinfo{person}{Zhuoran Liu} {and} \bibinfo{person}{Martha
  Larson}.} \bibinfo{year}{2021}\natexlab{}.
\newblock \showarticletitle{Adversarial Item Promotion: Vulnerabilities at the
  Core of Top-N Recommenders that Use Images to Address Cold Start}. In
  \bibinfo{booktitle}{\emph{WWW}}. \bibinfo{pages}{3590--3602}.
\newblock


\bibitem[\protect\citeauthoryear{Maes, Heyman, Desmet, and Joosen}{Maes
  et~al\mbox{.}}{2009}]%
        {maes2009browser}
\bibfield{author}{\bibinfo{person}{Wim Maes}, \bibinfo{person}{Thomas Heyman},
  \bibinfo{person}{Lieven Desmet}, {and} \bibinfo{person}{Wouter Joosen}.}
  \bibinfo{year}{2009}\natexlab{}.
\newblock \showarticletitle{Browser protection against cross-site request
  forgery}. In \bibinfo{booktitle}{\emph{SecuCode}}. \bibinfo{pages}{3--10}.
\newblock


\bibitem[\protect\citeauthoryear{Mehrabi, Morstatter, Saxena, Lerman, and
  Galstyan}{Mehrabi et~al\mbox{.}}{2021}]%
        {mehrabi2021survey}
\bibfield{author}{\bibinfo{person}{Ninareh Mehrabi}, \bibinfo{person}{Fred
  Morstatter}, \bibinfo{person}{Nripsuta Saxena}, \bibinfo{person}{Kristina
  Lerman}, {and} \bibinfo{person}{Aram Galstyan}.}
  \bibinfo{year}{2021}\natexlab{}.
\newblock \showarticletitle{A survey on bias and fairness in machine learning}.
\newblock \bibinfo{journal}{\emph{CSUR}} \bibinfo{volume}{54},
  \bibinfo{number}{6} (\bibinfo{year}{2021}), \bibinfo{pages}{1--35}.
\newblock


\bibitem[\protect\citeauthoryear{Mehta, Hofmann, and Fankhauser}{Mehta
  et~al\mbox{.}}{2007a}]%
        {mehta2007lies}
\bibfield{author}{\bibinfo{person}{Bhaskar Mehta}, \bibinfo{person}{Thomas
  Hofmann}, {and} \bibinfo{person}{Peter Fankhauser}.}
  \bibinfo{year}{2007}\natexlab{a}.
\newblock \showarticletitle{Lies and propaganda: detecting spam users in
  collaborative filtering}. In \bibinfo{booktitle}{\emph{IUI}}.
  \bibinfo{pages}{14--21}.
\newblock


\bibitem[\protect\citeauthoryear{Mehta, Hofmann, and Nejdl}{Mehta
  et~al\mbox{.}}{2007b}]%
        {mehta2007robust}
\bibfield{author}{\bibinfo{person}{Bhaskar Mehta}, \bibinfo{person}{Thomas
  Hofmann}, {and} \bibinfo{person}{Wolfgang Nejdl}.}
  \bibinfo{year}{2007}\natexlab{b}.
\newblock \showarticletitle{Robust collaborative filtering}. In
  \bibinfo{booktitle}{\emph{RecSys}}. \bibinfo{pages}{49--56}.
\newblock


\bibitem[\protect\citeauthoryear{Mehta and Nejdl}{Mehta and Nejdl}{2008}]%
        {mehta2008attack}
\bibfield{author}{\bibinfo{person}{Bhaskar Mehta} {and}
  \bibinfo{person}{Wolfgang Nejdl}.} \bibinfo{year}{2008}\natexlab{}.
\newblock \showarticletitle{Attack resistant collaborative filtering}. In
  \bibinfo{booktitle}{\emph{SIGIR}}. \bibinfo{pages}{75--82}.
\newblock


\bibitem[\protect\citeauthoryear{Mehta and Nejdl}{Mehta and Nejdl}{2009}]%
        {mehta2009unsupervised}
\bibfield{author}{\bibinfo{person}{Bhaskar Mehta} {and}
  \bibinfo{person}{Wolfgang Nejdl}.} \bibinfo{year}{2009}\natexlab{}.
\newblock \showarticletitle{Unsupervised strategies for shilling detection and
  robust collaborative filtering}.
\newblock \bibinfo{journal}{\emph{User Modeling and User-Adapted Interaction}}
  \bibinfo{volume}{19}, \bibinfo{number}{1} (\bibinfo{year}{2009}),
  \bibinfo{pages}{65--97}.
\newblock


\bibitem[\protect\citeauthoryear{Mendes-Moreira, Soares, Jorge, and
  Sousa}{Mendes-Moreira et~al\mbox{.}}{2012}]%
        {mendes2012ensemble}
\bibfield{author}{\bibinfo{person}{Joao Mendes-Moreira},
  \bibinfo{person}{Carlos Soares}, \bibinfo{person}{Al{\'\i}pio~M{\'a}rio
  Jorge}, {and} \bibinfo{person}{Jorge Freire~De Sousa}.}
  \bibinfo{year}{2012}\natexlab{}.
\newblock \showarticletitle{Ensemble approaches for regression: A survey}.
\newblock \bibinfo{journal}{\emph{CSUR}} \bibinfo{volume}{45},
  \bibinfo{number}{1} (\bibinfo{year}{2012}), \bibinfo{pages}{1--40}.
\newblock


\bibitem[\protect\citeauthoryear{Meng, Xing, Sheth, Weinsberg, and Lee}{Meng
  et~al\mbox{.}}{2014}]%
        {meng2014your}
\bibfield{author}{\bibinfo{person}{Wei Meng}, \bibinfo{person}{Xinyu Xing},
  \bibinfo{person}{Anmol Sheth}, \bibinfo{person}{Udi Weinsberg}, {and}
  \bibinfo{person}{Wenke Lee}.} \bibinfo{year}{2014}\natexlab{}.
\newblock \showarticletitle{Your online interests: Pwned! a pollution attack
  against targeted advertising}. In \bibinfo{booktitle}{\emph{SIGSAC}}.
  \bibinfo{pages}{129--140}.
\newblock


\bibitem[\protect\citeauthoryear{Mobasher, Burke, Bhaumik, and
  Williams}{Mobasher et~al\mbox{.}}{2007}]%
        {mobasher2007toward}
\bibfield{author}{\bibinfo{person}{Bamshad Mobasher}, \bibinfo{person}{Robin
  Burke}, \bibinfo{person}{Runa Bhaumik}, {and} \bibinfo{person}{Chad
  Williams}.} \bibinfo{year}{2007}\natexlab{}.
\newblock \showarticletitle{Toward trustworthy recommender systems: An analysis
  of attack models and algorithm robustness}.
\newblock \bibinfo{journal}{\emph{TOIT}} \bibinfo{volume}{7},
  \bibinfo{number}{4} (\bibinfo{year}{2007}), \bibinfo{pages}{23--es}.
\newblock


\bibitem[\protect\citeauthoryear{Mu}{Mu}{2018}]%
        {mu2018survey}
\bibfield{author}{\bibinfo{person}{Ruihui Mu}.}
  \bibinfo{year}{2018}\natexlab{}.
\newblock \showarticletitle{A survey of recommender systems based on deep
  learning}.
\newblock \bibinfo{journal}{\emph{Ieee Access}}  \bibinfo{volume}{6}
  (\bibinfo{year}{2018}), \bibinfo{pages}{69009--69022}.
\newblock


\bibitem[\protect\citeauthoryear{Nguyen, Do, Nguyen, and Aberer}{Nguyen
  et~al\mbox{.}}{2015a}]%
        {nguyen2015tag}
\bibfield{author}{\bibinfo{person}{Quoc Viet~Hung Nguyen},
  \bibinfo{person}{Son~Thanh Do}, \bibinfo{person}{Thanh~Tam Nguyen}, {and}
  \bibinfo{person}{Karl Aberer}.} \bibinfo{year}{2015}\natexlab{a}.
\newblock \showarticletitle{Tag-based paper retrieval: minimizing user effort
  with diversity awareness}. In \bibinfo{booktitle}{\emph{International
  Conference on Database Systems for Advanced Applications}}.
  \bibinfo{pages}{510--528}.
\newblock


\bibitem[\protect\citeauthoryear{Nguyen, Nguyen, Chau, Wijaya, Mikl{\'o}s,
  Aberer, Gal, and Weidlich}{Nguyen et~al\mbox{.}}{2015b}]%
        {nguyen2015smart}
\bibfield{author}{\bibinfo{person}{Quoc Viet~Hung Nguyen},
  \bibinfo{person}{Thanh~Tam Nguyen}, \bibinfo{person}{Vinh~Tuan Chau},
  \bibinfo{person}{Tri~Kurniawan Wijaya}, \bibinfo{person}{Zolt{\'a}n
  Mikl{\'o}s}, \bibinfo{person}{Karl Aberer}, \bibinfo{person}{Avigdor Gal},
  {and} \bibinfo{person}{Matthias Weidlich}.} \bibinfo{year}{2015}\natexlab{b}.
\newblock \showarticletitle{SMART: A tool for analyzing and reconciling schema
  matching networks}. In \bibinfo{booktitle}{\emph{ICDE}}.
  \bibinfo{pages}{1488--1491}.
\newblock


\bibitem[\protect\citeauthoryear{Nguyen, Nguyen~Thanh, Mikl{\'o}s, and
  Aberer}{Nguyen et~al\mbox{.}}{2014}]%
        {nguyen2014reconciling}
\bibfield{author}{\bibinfo{person}{Quoc Viet~Hung Nguyen}, \bibinfo{person}{Tam
  Nguyen~Thanh}, \bibinfo{person}{Zolt{\'a}n Mikl{\'o}s}, {and}
  \bibinfo{person}{Karl Aberer}.} \bibinfo{year}{2014}\natexlab{}.
\newblock \showarticletitle{Reconciling schema matching networks through
  crowdsourcing}.
\newblock \bibinfo{journal}{\emph{EAI Endorsed Transactions on Collaborative
  Computing}} \bibinfo{volume}{1}, \bibinfo{number}{2} (\bibinfo{year}{2014}),
  \bibinfo{pages}{e2}.
\newblock


\bibitem[\protect\citeauthoryear{Nguyen~Thanh, Quach, Nguyen, Huynh, Vu,
  Nguyen, Jo, and Nguyen}{Nguyen~Thanh et~al\mbox{.}}{2023}]%
        {nguyen2023poisoning}
\bibfield{author}{\bibinfo{person}{Toan Nguyen~Thanh}, \bibinfo{person}{Nguyen
  Duc~Khang Quach}, \bibinfo{person}{Thanh~Tam Nguyen},
  \bibinfo{person}{Thanh~Trung Huynh}, \bibinfo{person}{Viet~Hung Vu},
  \bibinfo{person}{Phi~Le Nguyen}, \bibinfo{person}{Jun Jo}, {and}
  \bibinfo{person}{Quoc Viet~Hung Nguyen}.} \bibinfo{year}{2023}\natexlab{}.
\newblock \showarticletitle{Poisoning GNN-based recommender systems with
  generative surrogate-based attacks}.
\newblock \bibinfo{journal}{\emph{ACM Transactions on Information Systems}}
  \bibinfo{volume}{41}, \bibinfo{number}{3} (\bibinfo{year}{2023}),
  \bibinfo{pages}{1--24}.
\newblock


\bibitem[\protect\citeauthoryear{Owen}{Owen}{2008}]%
        {owen2008parameter}
\bibfield{author}{\bibinfo{person}{Claire~B Owen}.}
  \bibinfo{year}{2008}\natexlab{}.
\newblock \bibinfo{booktitle}{\emph{Parameter estimation for the beta
  distribution}}.
\newblock \bibinfo{publisher}{Brigham Young University}.
\newblock


\bibitem[\protect\citeauthoryear{O’Mahony, Hurley, and Silvestre}{O’Mahony
  et~al\mbox{.}}{2002}]%
        {o2002promoting}
\bibfield{author}{\bibinfo{person}{Michael~P O’Mahony},
  \bibinfo{person}{Neil~J Hurley}, {and} \bibinfo{person}{Guenole Silvestre}.}
  \bibinfo{year}{2002}\natexlab{}.
\newblock \showarticletitle{Promoting recommendations: An attack on
  collaborative filtering}. In \bibinfo{booktitle}{\emph{DEXA}}.
  \bibinfo{pages}{494--503}.
\newblock


\bibitem[\protect\citeauthoryear{Padakandla}{Padakandla}{2021}]%
        {padakandla2021survey}
\bibfield{author}{\bibinfo{person}{Sindhu Padakandla}.}
  \bibinfo{year}{2021}\natexlab{}.
\newblock \showarticletitle{A survey of reinforcement learning algorithms for
  dynamically varying environments}.
\newblock \bibinfo{journal}{\emph{CSUR}} \bibinfo{volume}{54},
  \bibinfo{number}{6} (\bibinfo{year}{2021}), \bibinfo{pages}{1--25}.
\newblock


\bibitem[\protect\citeauthoryear{Pateria, Subagdja, Tan, and Quek}{Pateria
  et~al\mbox{.}}{2021}]%
        {pateria2021hierarchical}
\bibfield{author}{\bibinfo{person}{Shubham Pateria}, \bibinfo{person}{Budhitama
  Subagdja}, \bibinfo{person}{Ah-hwee Tan}, {and} \bibinfo{person}{Chai Quek}.}
  \bibinfo{year}{2021}\natexlab{}.
\newblock \showarticletitle{Hierarchical reinforcement learning: A
  comprehensive survey}.
\newblock \bibinfo{journal}{\emph{CSUR}} \bibinfo{volume}{54},
  \bibinfo{number}{5} (\bibinfo{year}{2021}), \bibinfo{pages}{1--35}.
\newblock


\bibitem[\protect\citeauthoryear{Periyasamy, Jaiganesh, Ponnambalam, Rajasekar,
  and Arputharaj}{Periyasamy et~al\mbox{.}}{2017}]%
        {periyasamy2017analysis}
\bibfield{author}{\bibinfo{person}{Kola Periyasamy},
  \bibinfo{person}{Jayadharini Jaiganesh}, \bibinfo{person}{Kanchan
  Ponnambalam}, \bibinfo{person}{Jeevitha Rajasekar}, {and}
  \bibinfo{person}{Kannan Arputharaj}.} \bibinfo{year}{2017}\natexlab{}.
\newblock \showarticletitle{Analysis and Performance Evaluation of Cosine
  Neighbourhood Recommender System.}
\newblock \bibinfo{journal}{\emph{IAJIT}} \bibinfo{volume}{14},
  \bibinfo{number}{5} (\bibinfo{year}{2017}).
\newblock


\bibitem[\protect\citeauthoryear{Pitropakis, Panaousis, Giannetsos,
  Anastasiadis, and Loukas}{Pitropakis et~al\mbox{.}}{2019}]%
        {pitropakis2019taxonomy}
\bibfield{author}{\bibinfo{person}{Nikolaos Pitropakis},
  \bibinfo{person}{Emmanouil Panaousis}, \bibinfo{person}{Thanassis
  Giannetsos}, \bibinfo{person}{Eleftherios Anastasiadis}, {and}
  \bibinfo{person}{George Loukas}.} \bibinfo{year}{2019}\natexlab{}.
\newblock \showarticletitle{A taxonomy and survey of attacks against machine
  learning}.
\newblock \bibinfo{journal}{\emph{Computer Science Review}}
  \bibinfo{volume}{34} (\bibinfo{year}{2019}), \bibinfo{pages}{100199}.
\newblock


\bibitem[\protect\citeauthoryear{Polatidis, Pimenidis, Pavlidis, and
  Mouratidis}{Polatidis et~al\mbox{.}}{2017}]%
        {polatidis2017recommender}
\bibfield{author}{\bibinfo{person}{Nikolaos Polatidis}, \bibinfo{person}{Elias
  Pimenidis}, \bibinfo{person}{Michalis Pavlidis}, {and}
  \bibinfo{person}{Haralambos Mouratidis}.} \bibinfo{year}{2017}\natexlab{}.
\newblock \showarticletitle{Recommender systems meeting security: From product
  recommendation to cyber-attack prediction}. In
  \bibinfo{booktitle}{\emph{EANN}}. \bibinfo{pages}{508--519}.
\newblock


\bibitem[\protect\citeauthoryear{Pouyanfar, Sadiq, Yan, Tian, Tao, Reyes,
  et~al\mbox{.}}{Pouyanfar et~al\mbox{.}}{2018}]%
        {pouyanfar2018survey}
\bibfield{author}{\bibinfo{person}{Samira Pouyanfar}, \bibinfo{person}{Saad
  Sadiq}, \bibinfo{person}{Yilin Yan}, \bibinfo{person}{Haiman Tian},
  \bibinfo{person}{Yudong Tao}, \bibinfo{person}{Maria~Presa Reyes},
  {et~al\mbox{.}}} \bibinfo{year}{2018}\natexlab{}.
\newblock \showarticletitle{A survey on deep learning: Algorithms, techniques,
  and applications}.
\newblock \bibinfo{journal}{\emph{CSUR}} \bibinfo{volume}{51},
  \bibinfo{number}{5} (\bibinfo{year}{2018}), \bibinfo{pages}{1--36}.
\newblock


\bibitem[\protect\citeauthoryear{Ren, Baird, Han, Zhang, and Schuller}{Ren
  et~al\mbox{.}}{2020a}]%
        {ren2020generating}
\bibfield{author}{\bibinfo{person}{Zhao Ren}, \bibinfo{person}{Alice Baird},
  \bibinfo{person}{Jing Han}, \bibinfo{person}{Zixing Zhang}, {and}
  \bibinfo{person}{Bj{\"o}rn Schuller}.} \bibinfo{year}{2020}\natexlab{a}.
\newblock \showarticletitle{{Generating and protecting against adversarial
  attacks for deep speech-based emotion recognition models}}. In
  \bibinfo{booktitle}{\emph{Proc.\ ICASSP}}. \bibinfo{address}{Barcelona,
  Spain}, \bibinfo{pages}{7184--7188}.
\newblock


\bibitem[\protect\citeauthoryear{Ren, Han, Cummins, and Schuller}{Ren
  et~al\mbox{.}}{2020b}]%
        {ren2020enhancing}
\bibfield{author}{\bibinfo{person}{Zhao Ren}, \bibinfo{person}{Jing Han},
  \bibinfo{person}{Nicholas Cummins}, {and} \bibinfo{person}{Bj{\"o}rn
  Schuller}.} \bibinfo{year}{2020}\natexlab{b}.
\newblock \showarticletitle{{Enhancing transferability of black-box adversarial
  attacks via lifelong learning for speech emotion recognition models}}. In
  \bibinfo{booktitle}{\emph{Proc.\ INTERSPEECH}}. \bibinfo{address}{Shanghai,
  China}, \bibinfo{pages}{496--500}.
\newblock


\bibitem[\protect\citeauthoryear{Rezaimehr and Dadkhah}{Rezaimehr and
  Dadkhah}{2021}]%
        {rezaimehr2021survey}
\bibfield{author}{\bibinfo{person}{Fatemeh Rezaimehr} {and}
  \bibinfo{person}{Chitra Dadkhah}.} \bibinfo{year}{2021}\natexlab{}.
\newblock \showarticletitle{A survey of attack detection approaches in
  collaborative filtering recommender systems}.
\newblock \bibinfo{journal}{\emph{Artificial Intelligence Review}}
  \bibinfo{volume}{54}, \bibinfo{number}{3} (\bibinfo{year}{2021}),
  \bibinfo{pages}{2011--2066}.
\newblock


\bibitem[\protect\citeauthoryear{Ricci, Rokach, and Shapira}{Ricci
  et~al\mbox{.}}{2011}]%
        {ricci2011introduction}
\bibfield{author}{\bibinfo{person}{Francesco Ricci}, \bibinfo{person}{Lior
  Rokach}, {and} \bibinfo{person}{Bracha Shapira}.}
  \bibinfo{year}{2011}\natexlab{}.
\newblock \showarticletitle{Introduction to recommender systems handbook}.
\newblock In \bibinfo{booktitle}{\emph{Recommender systems handbook}}.
  \bibinfo{pages}{1--35}.
\newblock


\bibitem[\protect\citeauthoryear{Rong, He, and Chen}{Rong
  et~al\mbox{.}}{2022}]%
        {rong2022poisoning}
\bibfield{author}{\bibinfo{person}{Dazhong Rong}, \bibinfo{person}{Qinming He},
  {and} \bibinfo{person}{Jianhai Chen}.} \bibinfo{year}{2022}\natexlab{}.
\newblock \showarticletitle{Poisoning Deep Learning based Recommender Model in
  Federated Learning Scenarios}.
\newblock \bibinfo{journal}{\emph{arXiv preprint arXiv:2204.13594}}
  (\bibinfo{year}{2022}).
\newblock


\bibitem[\protect\citeauthoryear{Rosenberg, Shabtai, Elovici, and
  Rokach}{Rosenberg et~al\mbox{.}}{2021}]%
        {rosenberg2021adversarial}
\bibfield{author}{\bibinfo{person}{Ishai Rosenberg}, \bibinfo{person}{Asaf
  Shabtai}, \bibinfo{person}{Yuval Elovici}, {and} \bibinfo{person}{Lior
  Rokach}.} \bibinfo{year}{2021}\natexlab{}.
\newblock \showarticletitle{Adversarial machine learning attacks and defense
  methods in the cyber security domain}.
\newblock \bibinfo{journal}{\emph{CSUR}} \bibinfo{volume}{54},
  \bibinfo{number}{5} (\bibinfo{year}{2021}), \bibinfo{pages}{1--36}.
\newblock


\bibitem[\protect\citeauthoryear{Sachan and Richariya}{Sachan and
  Richariya}{2013}]%
        {sachan2013survey}
\bibfield{author}{\bibinfo{person}{Atisha Sachan} {and} \bibinfo{person}{Vineet
  Richariya}.} \bibinfo{year}{2013}\natexlab{}.
\newblock \showarticletitle{A survey on recommender systems based on
  collaborative filtering technique}.
\newblock \bibinfo{journal}{\emph{IJIET}} \bibinfo{volume}{2},
  \bibinfo{number}{2} (\bibinfo{year}{2013}), \bibinfo{pages}{8--14}.
\newblock


\bibitem[\protect\citeauthoryear{Salakhutdinov and Mnih}{Salakhutdinov and
  Mnih}{2008}]%
        {salakhutdinov2008bayesian}
\bibfield{author}{\bibinfo{person}{Ruslan Salakhutdinov} {and}
  \bibinfo{person}{Andriy Mnih}.} \bibinfo{year}{2008}\natexlab{}.
\newblock \showarticletitle{Bayesian probabilistic matrix factorization using
  Markov chain Monte Carlo}. In \bibinfo{booktitle}{\emph{ICML}}.
  \bibinfo{pages}{880--887}.
\newblock


\bibitem[\protect\citeauthoryear{Sandvig, Mobasher, and Burke}{Sandvig
  et~al\mbox{.}}{2007}]%
        {sandvig2007robustness}
\bibfield{author}{\bibinfo{person}{Jeff~J Sandvig}, \bibinfo{person}{Bamshad
  Mobasher}, {and} \bibinfo{person}{Robin Burke}.}
  \bibinfo{year}{2007}\natexlab{}.
\newblock \showarticletitle{Robustness of collaborative recommendation based on
  association rule mining}. In \bibinfo{booktitle}{\emph{RecSys}}.
  \bibinfo{pages}{105--112}.
\newblock


\bibitem[\protect\citeauthoryear{Sedgwick}{Sedgwick}{2012}]%
        {sedgwick2012pearson}
\bibfield{author}{\bibinfo{person}{Philip Sedgwick}.}
  \bibinfo{year}{2012}\natexlab{}.
\newblock \showarticletitle{Pearson’s correlation coefficient}.
\newblock \bibinfo{journal}{\emph{Bmj}}  \bibinfo{volume}{345}
  (\bibinfo{year}{2012}).
\newblock


\bibitem[\protect\citeauthoryear{Shani and Gunawardana}{Shani and
  Gunawardana}{2011}]%
        {shani2011evaluating}
\bibfield{author}{\bibinfo{person}{Guy Shani} {and} \bibinfo{person}{Asela
  Gunawardana}.} \bibinfo{year}{2011}\natexlab{}.
\newblock \showarticletitle{Evaluating recommendation systems}.
\newblock In \bibinfo{booktitle}{\emph{Recommender systems handbook}}.
  \bibinfo{pages}{257--297}.
\newblock


\bibitem[\protect\citeauthoryear{Shi, Larson, and Hanjalic}{Shi
  et~al\mbox{.}}{2014}]%
        {shi2014collaborative}
\bibfield{author}{\bibinfo{person}{Yue Shi}, \bibinfo{person}{Martha Larson},
  {and} \bibinfo{person}{Alan Hanjalic}.} \bibinfo{year}{2014}\natexlab{}.
\newblock \showarticletitle{Collaborative filtering beyond the user-item
  matrix: A survey of the state of the art and future challenges}.
\newblock \bibinfo{journal}{\emph{CSUR}} \bibinfo{volume}{47},
  \bibinfo{number}{1} (\bibinfo{year}{2014}), \bibinfo{pages}{1--45}.
\newblock


\bibitem[\protect\citeauthoryear{Si and Li}{Si and Li}{2020}]%
        {si2020shilling}
\bibfield{author}{\bibinfo{person}{Mingdan Si} {and} \bibinfo{person}{Qingshan
  Li}.} \bibinfo{year}{2020}\natexlab{}.
\newblock \showarticletitle{Shilling attacks against collaborative recommender
  systems: a review}.
\newblock \bibinfo{journal}{\emph{Artificial Intelligence Review}}
  \bibinfo{volume}{53}, \bibinfo{number}{1} (\bibinfo{year}{2020}),
  \bibinfo{pages}{291--319}.
\newblock


\bibitem[\protect\citeauthoryear{Sidhu, Sakhuja, and Zhou}{Sidhu
  et~al\mbox{.}}{2016}]%
        {sidhu2016attacks}
\bibfield{author}{\bibinfo{person}{Jaspuneet Sidhu}, \bibinfo{person}{Rohit
  Sakhuja}, {and} \bibinfo{person}{David Zhou}.}
  \bibinfo{year}{2016}\natexlab{}.
\newblock \bibinfo{title}{Attacks on eBay}.
\newblock
\newblock


\bibitem[\protect\citeauthoryear{Smiti and Soui}{Smiti and Soui}{2020}]%
        {smiti2020bankruptcy}
\bibfield{author}{\bibinfo{person}{Salima Smiti} {and} \bibinfo{person}{Makram
  Soui}.} \bibinfo{year}{2020}\natexlab{}.
\newblock \showarticletitle{Bankruptcy prediction using deep learning approach
  based on borderline SMOTE}.
\newblock \bibinfo{journal}{\emph{Information Systems Frontiers}}
  \bibinfo{volume}{22}, \bibinfo{number}{5} (\bibinfo{year}{2020}),
  \bibinfo{pages}{1067--1083}.
\newblock


\bibitem[\protect\citeauthoryear{Song, Li, Hu, Wu, Li, Li, and Gao}{Song
  et~al\mbox{.}}{2020}]%
        {song2020poisonrec}
\bibfield{author}{\bibinfo{person}{Junshuai Song}, \bibinfo{person}{Zhao Li},
  \bibinfo{person}{Zehong Hu}, \bibinfo{person}{Yucheng Wu},
  \bibinfo{person}{Zhenpeng Li}, \bibinfo{person}{Jian Li}, {and}
  \bibinfo{person}{Jun Gao}.} \bibinfo{year}{2020}\natexlab{}.
\newblock \showarticletitle{Poisonrec: an adaptive data poisoning framework for
  attacking black-box recommender systems}. In
  \bibinfo{booktitle}{\emph{ICDE}}. \bibinfo{pages}{157--168}.
\newblock


\bibitem[\protect\citeauthoryear{Strub et~al\mbox{.}}{Strub
  et~al\mbox{.}}{2016}]%
        {strub2016hybrid}
\bibfield{author}{\bibinfo{person}{Florian Strub} {et~al\mbox{.}}}
  \bibinfo{year}{2016}\natexlab{}.
\newblock \showarticletitle{Hybrid recommender system based on autoencoders}.
  In \bibinfo{booktitle}{\emph{DLRS}}. \bibinfo{pages}{11--16}.
\newblock


\bibitem[\protect\citeauthoryear{Su, Vargas, and Sakurai}{Su
  et~al\mbox{.}}{2019}]%
        {su2019one}
\bibfield{author}{\bibinfo{person}{Jiawei Su},
  \bibinfo{person}{Danilo~Vasconcellos Vargas}, {and} \bibinfo{person}{Kouichi
  Sakurai}.} \bibinfo{year}{2019}\natexlab{}.
\newblock \showarticletitle{One pixel attack for fooling deep neural networks}.
\newblock \bibinfo{journal}{\emph{IEEE Transactions on Evolutionary
  Computation}} \bibinfo{volume}{23}, \bibinfo{number}{5}
  (\bibinfo{year}{2019}), \bibinfo{pages}{828--841}.
\newblock


\bibitem[\protect\citeauthoryear{Sun, Qian, Chen, Liang, Nguyen, and Yin}{Sun
  et~al\mbox{.}}{2020}]%
        {sun2020go}
\bibfield{author}{\bibinfo{person}{Ke Sun}, \bibinfo{person}{Tieyun Qian},
  \bibinfo{person}{Tong Chen}, \bibinfo{person}{Yile Liang},
  \bibinfo{person}{Quoc Viet~Hung Nguyen}, {and} \bibinfo{person}{Hongzhi
  Yin}.} \bibinfo{year}{2020}\natexlab{}.
\newblock \showarticletitle{Where to go next: Modeling long-and short-term user
  preferences for point-of-interest recommendation}. In
  \bibinfo{booktitle}{\emph{AAAI}}, Vol.~\bibinfo{volume}{34}.
  \bibinfo{pages}{214--221}.
\newblock


\bibitem[\protect\citeauthoryear{Sundar, Li, Zou, Gao, and Russomanno}{Sundar
  et~al\mbox{.}}{2020}]%
        {sundar2020understanding}
\bibfield{author}{\bibinfo{person}{Agnideven~Palanisamy Sundar},
  \bibinfo{person}{Feng Li}, \bibinfo{person}{Xukai Zou},
  \bibinfo{person}{Tianchong Gao}, {and} \bibinfo{person}{Evan~D Russomanno}.}
  \bibinfo{year}{2020}\natexlab{}.
\newblock \showarticletitle{Understanding shilling attacks and their detection
  traits: a comprehensive survey}.
\newblock \bibinfo{journal}{\emph{IEEE Access}}  \bibinfo{volume}{8}
  (\bibinfo{year}{2020}), \bibinfo{pages}{171703--171715}.
\newblock


\bibitem[\protect\citeauthoryear{Tabassi, Burns, Hadjimichael, Molina-Markham,
  and Sexton}{Tabassi et~al\mbox{.}}{2019}]%
        {tabassi2019taxonomy}
\bibfield{author}{\bibinfo{person}{Elham Tabassi}, \bibinfo{person}{Kevin
  Burns}, \bibinfo{person}{Michael Hadjimichael}, \bibinfo{person}{Andres
  Molina-Markham}, {and} \bibinfo{person}{Julian Sexton}.}
  \bibinfo{year}{2019}\natexlab{}.
\newblock \showarticletitle{A taxonomy and terminology of adversarial machine
  learning}.
\newblock \bibinfo{journal}{\emph{NIST IR}} (\bibinfo{year}{2019}).
\newblock


\bibitem[\protect\citeauthoryear{Tam, Ren, Nguyen, Jo, Nguyen, and Yin}{Tam
  et~al\mbox{.}}{2023}]%
        {tam2023portable}
\bibfield{author}{\bibinfo{person}{Thanh~Nguyen Tam}, \bibinfo{person}{Zhao
  Ren}, \bibinfo{person}{Thanh~Toan Nguyen}, \bibinfo{person}{Jun Jo},
  \bibinfo{person}{Quoc Viet~Hung Nguyen}, {and} \bibinfo{person}{Hongzhi
  Yin}.} \bibinfo{year}{2023}\natexlab{}.
\newblock \showarticletitle{{Portable graph-based rumour detection against
  multi-modal heterophily}}.
\newblock \bibinfo{journal}{\emph{Knowledge-Based Systems}}
  (\bibinfo{date}{Dec.} \bibinfo{year}{2023}).
\newblock


\bibitem[\protect\citeauthoryear{Tang, Wen, and Wang}{Tang
  et~al\mbox{.}}{2020}]%
        {tang2020revisiting}
\bibfield{author}{\bibinfo{person}{Jiaxi Tang}, \bibinfo{person}{Hongyi Wen},
  {and} \bibinfo{person}{Ke Wang}.} \bibinfo{year}{2020}\natexlab{}.
\newblock \showarticletitle{Revisiting adversarially learned injection attacks
  against recommender systems}. In \bibinfo{booktitle}{\emph{RecSys}}.
  \bibinfo{pages}{318--327}.
\newblock


\bibitem[\protect\citeauthoryear{Tavara}{Tavara}{2019}]%
        {tavara2019parallel}
\bibfield{author}{\bibinfo{person}{Shirin Tavara}.}
  \bibinfo{year}{2019}\natexlab{}.
\newblock \showarticletitle{Parallel computing of support vector machines: a
  survey}.
\newblock \bibinfo{journal}{\emph{CSUR}} \bibinfo{volume}{51},
  \bibinfo{number}{6} (\bibinfo{year}{2019}), \bibinfo{pages}{1--38}.
\newblock


\bibitem[\protect\citeauthoryear{Thang, Tam, Hung, and Aberer}{Thang
  et~al\mbox{.}}{2015}]%
        {thang2015evaluation}
\bibfield{author}{\bibinfo{person}{Duong~Chi Thang},
  \bibinfo{person}{Nguyen~Thanh Tam}, \bibinfo{person}{Nguyen Quoc~Viet Hung},
  {and} \bibinfo{person}{Karl Aberer}.} \bibinfo{year}{2015}\natexlab{}.
\newblock \showarticletitle{An evaluation of diversification techniques}. In
  \bibinfo{booktitle}{\emph{International Conference on Database and Expert
  Systems Applications}}. \bibinfo{pages}{215--231}.
\newblock


\bibitem[\protect\citeauthoryear{Wadhwa, Agrawal, Chaudhari, Sharma, and
  Achan}{Wadhwa et~al\mbox{.}}{2020}]%
        {wadhwa2020data}
\bibfield{author}{\bibinfo{person}{Soumya Wadhwa}, \bibinfo{person}{Saurabh
  Agrawal}, \bibinfo{person}{Harsh Chaudhari}, \bibinfo{person}{Deepthi
  Sharma}, {and} \bibinfo{person}{Kannan Achan}.}
  \bibinfo{year}{2020}\natexlab{}.
\newblock \showarticletitle{Data poisoning attacks against differentially
  private recommender systems}. In \bibinfo{booktitle}{\emph{SIGIR}}.
  \bibinfo{pages}{1617--1620}.
\newblock


\bibitem[\protect\citeauthoryear{Wang and Tang}{Wang and Tang}{2015}]%
        {wang2015recommender}
\bibfield{author}{\bibinfo{person}{Jun Wang} {and} \bibinfo{person}{Qiang
  Tang}.} \bibinfo{year}{2015}\natexlab{}.
\newblock \showarticletitle{Recommender systems and their security concerns}.
\newblock  (\bibinfo{year}{2015}).
\newblock


\bibitem[\protect\citeauthoryear{Wang, Yin, Chen, Huang, Wang, Zhao, and
  Viet~Hung}{Wang et~al\mbox{.}}{2020}]%
        {wang2020next}
\bibfield{author}{\bibinfo{person}{Qinyong Wang}, \bibinfo{person}{Hongzhi
  Yin}, \bibinfo{person}{Tong Chen}, \bibinfo{person}{Zi Huang},
  \bibinfo{person}{Hao Wang}, \bibinfo{person}{Yanchang Zhao}, {and}
  \bibinfo{person}{Nguyen~Quoc Viet~Hung}.} \bibinfo{year}{2020}\natexlab{}.
\newblock \showarticletitle{Next point-of-interest recommendation on
  resource-constrained mobile devices}. In \bibinfo{booktitle}{\emph{WWW}}.
  \bibinfo{pages}{906--916}.
\newblock


\bibitem[\protect\citeauthoryear{Wang, Li, Kuang, Tan, and Li}{Wang
  et~al\mbox{.}}{2019}]%
        {wang2019security}
\bibfield{author}{\bibinfo{person}{Xianmin Wang}, \bibinfo{person}{Jing Li},
  \bibinfo{person}{Xiaohui Kuang}, \bibinfo{person}{Yu-an Tan}, {and}
  \bibinfo{person}{Jin Li}.} \bibinfo{year}{2019}\natexlab{}.
\newblock \showarticletitle{The security of machine learning in an adversarial
  setting: A survey}.
\newblock \bibinfo{journal}{\emph{J. Parallel and Distrib. Comput.}}
  \bibinfo{volume}{130} (\bibinfo{year}{2019}), \bibinfo{pages}{12--23}.
\newblock


\bibitem[\protect\citeauthoryear{Wang, Qian, Li, and Zhang}{Wang
  et~al\mbox{.}}{2018}]%
        {wang2018comparative}
\bibfield{author}{\bibinfo{person}{Youquan Wang}, \bibinfo{person}{Liqiang
  Qian}, \bibinfo{person}{Fanzhang Li}, {and} \bibinfo{person}{Lu Zhang}.}
  \bibinfo{year}{2018}\natexlab{}.
\newblock \showarticletitle{A comparative study on shilling detection methods
  for trustworthy recommendations}.
\newblock \bibinfo{journal}{\emph{Journal of Systems Science and Systems
  Engineering}} \bibinfo{volume}{27}, \bibinfo{number}{4}
  (\bibinfo{year}{2018}), \bibinfo{pages}{458--478}.
\newblock


\bibitem[\protect\citeauthoryear{Wang, Zhang, Tao, Wu, and Cao}{Wang
  et~al\mbox{.}}{2015}]%
        {wang2015comparative}
\bibfield{author}{\bibinfo{person}{Youquan Wang}, \bibinfo{person}{Lu Zhang},
  \bibinfo{person}{Haicheng Tao}, \bibinfo{person}{Zhiang Wu}, {and}
  \bibinfo{person}{Jie Cao}.} \bibinfo{year}{2015}\natexlab{}.
\newblock \showarticletitle{A comparative study of shilling attack detectors
  for recommender systems}. In \bibinfo{booktitle}{\emph{ICSSSM}}.
  \bibinfo{pages}{1--6}.
\newblock


\bibitem[\protect\citeauthoryear{Williams, Mobasher, and Burke}{Williams
  et~al\mbox{.}}{2007}]%
        {williams2007defending}
\bibfield{author}{\bibinfo{person}{Chad~A Williams}, \bibinfo{person}{Bamshad
  Mobasher}, {and} \bibinfo{person}{Robin Burke}.}
  \bibinfo{year}{2007}\natexlab{}.
\newblock \showarticletitle{Defending recommender systems: detection of profile
  injection attacks}.
\newblock \bibinfo{journal}{\emph{Service Oriented Computing and Applications}}
  \bibinfo{volume}{1}, \bibinfo{number}{3} (\bibinfo{year}{2007}),
  \bibinfo{pages}{157--170}.
\newblock


\bibitem[\protect\citeauthoryear{Wu, Lian, Ge, Zhu, and Chen}{Wu
  et~al\mbox{.}}{2021c}]%
        {wu2021triple}
\bibfield{author}{\bibinfo{person}{Chenwang Wu}, \bibinfo{person}{Defu Lian},
  \bibinfo{person}{Yong Ge}, \bibinfo{person}{Zhihao Zhu}, {and}
  \bibinfo{person}{Enhong Chen}.} \bibinfo{year}{2021}\natexlab{c}.
\newblock \showarticletitle{Triple Adversarial Learning for Influence based
  Poisoning Attack in Recommender Systems}. In \bibinfo{booktitle}{\emph{KDD}}.
  \bibinfo{pages}{1830--1840}.
\newblock


\bibitem[\protect\citeauthoryear{Wu, Lian, Ge, Zhu, Chen, and Yuan}{Wu
  et~al\mbox{.}}{2021d}]%
        {wu2021fight}
\bibfield{author}{\bibinfo{person}{Chenwang Wu}, \bibinfo{person}{Defu Lian},
  \bibinfo{person}{Yong Ge}, \bibinfo{person}{Zhihao Zhu},
  \bibinfo{person}{Enhong Chen}, {and} \bibinfo{person}{Senchao Yuan}.}
  \bibinfo{year}{2021}\natexlab{d}.
\newblock \showarticletitle{Fight Fire with Fire: Towards Robust Recommender
  Systems via Adversarial Poisoning Training}. In
  \bibinfo{booktitle}{\emph{SIGIR}}. \bibinfo{pages}{1074--1083}.
\newblock


\bibitem[\protect\citeauthoryear{Wu, Wu, Qi, Huang, and Xie}{Wu
  et~al\mbox{.}}{2022}]%
        {wu2022fedattack}
\bibfield{author}{\bibinfo{person}{Chuhan Wu}, \bibinfo{person}{Fangzhao Wu},
  \bibinfo{person}{Tao Qi}, \bibinfo{person}{Yongfeng Huang}, {and}
  \bibinfo{person}{Xing Xie}.} \bibinfo{year}{2022}\natexlab{}.
\newblock \showarticletitle{FedAttack: Effective and Covert Poisoning Attack on
  Federated Recommendation via Hard Sampling}.
\newblock \bibinfo{journal}{\emph{arXiv preprint arXiv:2202.04975}}
  (\bibinfo{year}{2022}).
\newblock


\bibitem[\protect\citeauthoryear{Wu, Gao, Yu, Wang, et~al\mbox{.}}{Wu
  et~al\mbox{.}}{2021b}]%
        {wu2021ready}
\bibfield{author}{\bibinfo{person}{Fan Wu}, \bibinfo{person}{Min Gao},
  \bibinfo{person}{Junliang Yu}, \bibinfo{person}{Zongwei Wang},
  {et~al\mbox{.}}} \bibinfo{year}{2021}\natexlab{b}.
\newblock \showarticletitle{Ready for emerging threats to recommender systems?
  A graph convolution-based generative shilling attack}.
\newblock \bibinfo{journal}{\emph{Information Sciences}}  \bibinfo{volume}{578}
  (\bibinfo{year}{2021}), \bibinfo{pages}{683--701}.
\newblock


\bibitem[\protect\citeauthoryear{Wu, Wu, Cao, and Tao}{Wu
  et~al\mbox{.}}{2012}]%
        {wu2012hysad}
\bibfield{author}{\bibinfo{person}{Zhiang Wu}, \bibinfo{person}{Junjie Wu},
  \bibinfo{person}{Jie Cao}, {and} \bibinfo{person}{Dacheng Tao}.}
  \bibinfo{year}{2012}\natexlab{}.
\newblock \showarticletitle{HySAD: A semi-supervised hybrid shilling attack
  detector for trustworthy product recommendation}. In
  \bibinfo{booktitle}{\emph{KDD}}. \bibinfo{pages}{985--993}.
\newblock


\bibitem[\protect\citeauthoryear{Wu, Chen, and Huang}{Wu
  et~al\mbox{.}}{2021a}]%
        {wu2021poisoning}
\bibfield{author}{\bibinfo{person}{Zih-Wun Wu}, \bibinfo{person}{Chiao-Ting
  Chen}, {and} \bibinfo{person}{Szu-Hao Huang}.}
  \bibinfo{year}{2021}\natexlab{a}.
\newblock \showarticletitle{Poisoning attacks against knowledge graph-based
  recommendation systems using deep reinforcement learning}.
\newblock \bibinfo{journal}{\emph{Neural Computing and Applications}}
  (\bibinfo{year}{2021}), \bibinfo{pages}{1--19}.
\newblock


\bibitem[\protect\citeauthoryear{Xia et~al\mbox{.}}{Xia et~al\mbox{.}}{2015}]%
        {xia2015novel}
\bibfield{author}{\bibinfo{person}{Hui Xia} {et~al\mbox{.}}}
  \bibinfo{year}{2015}\natexlab{}.
\newblock \showarticletitle{A novel item anomaly detection approach against
  shilling attacks in collaborative recommendation systems using the dynamic
  time interval segmentation technique}.
\newblock \bibinfo{journal}{\emph{Information Sciences}}  \bibinfo{volume}{306}
  (\bibinfo{year}{2015}), \bibinfo{pages}{150--165}.
\newblock


\bibitem[\protect\citeauthoryear{Ximeng, Rennong, and Ying}{Ximeng
  et~al\mbox{.}}{2018}]%
        {ximeng2018situation}
\bibfield{author}{\bibinfo{person}{XU Ximeng}, \bibinfo{person}{YANG Rennong},
  {and} \bibinfo{person}{FU Ying}.} \bibinfo{year}{2018}\natexlab{}.
\newblock \showarticletitle{Situation assessment for air combat based on novel
  semi-supervised naive Bayes}.
\newblock \bibinfo{journal}{\emph{Journal of Systems Engineering and
  Electronics}} \bibinfo{volume}{29}, \bibinfo{number}{4}
  (\bibinfo{year}{2018}), \bibinfo{pages}{768--779}.
\newblock


\bibitem[\protect\citeauthoryear{Xu and Zhang}{Xu and Zhang}{2019}]%
        {xu2019detecting}
\bibfield{author}{\bibinfo{person}{Yishu Xu} {and} \bibinfo{person}{Fuzhi
  Zhang}.} \bibinfo{year}{2019}\natexlab{}.
\newblock \showarticletitle{Detecting shilling attacks in social recommender
  systems based on time series analysis and trust features}.
\newblock \bibinfo{journal}{\emph{KBS}}  \bibinfo{volume}{178}
  (\bibinfo{year}{2019}), \bibinfo{pages}{25--47}.
\newblock


\bibitem[\protect\citeauthoryear{Xue, Dai, Zhang, Huang, and Chen}{Xue
  et~al\mbox{.}}{2017}]%
        {xue2017deep}
\bibfield{author}{\bibinfo{person}{Hong-Jian Xue}, \bibinfo{person}{Xinyu Dai},
  \bibinfo{person}{Jianbing Zhang}, \bibinfo{person}{Shujian Huang}, {and}
  \bibinfo{person}{Jiajun Chen}.} \bibinfo{year}{2017}\natexlab{}.
\newblock \showarticletitle{Deep matrix factorization models for recommender
  systems.}. In \bibinfo{booktitle}{\emph{IJCAI}}, Vol.~\bibinfo{volume}{17}.
  \bibinfo{pages}{3203--3209}.
\newblock


\bibitem[\protect\citeauthoryear{Yang, Gong, and Cai}{Yang
  et~al\mbox{.}}{2017b}]%
        {yang2017fake}
\bibfield{author}{\bibinfo{person}{Guolei Yang},
  \bibinfo{person}{Neil~Zhenqiang Gong}, {and} \bibinfo{person}{Ying Cai}.}
  \bibinfo{year}{2017}\natexlab{b}.
\newblock \showarticletitle{Fake Co-visitation Injection Attacks to Recommender
  Systems.}. In \bibinfo{booktitle}{\emph{NDSS}}.
\newblock


\bibitem[\protect\citeauthoryear{Yang, Cai, and Guan}{Yang
  et~al\mbox{.}}{2016a}]%
        {yang2016estimating}
\bibfield{author}{\bibinfo{person}{Zhihai Yang}, \bibinfo{person}{Zhongmin
  Cai}, {and} \bibinfo{person}{Xiaohong Guan}.}
  \bibinfo{year}{2016}\natexlab{a}.
\newblock \showarticletitle{Estimating user behavior toward detecting anomalous
  ratings in rating systems}.
\newblock \bibinfo{journal}{\emph{KBS}}  \bibinfo{volume}{111}
  (\bibinfo{year}{2016}), \bibinfo{pages}{144--158}.
\newblock


\bibitem[\protect\citeauthoryear{Yang, Cai, and Yang}{Yang
  et~al\mbox{.}}{2017a}]%
        {yang2017spotting}
\bibfield{author}{\bibinfo{person}{Zhihai Yang}, \bibinfo{person}{Zhongmin
  Cai}, {and} \bibinfo{person}{Yuan Yang}.} \bibinfo{year}{2017}\natexlab{a}.
\newblock \showarticletitle{Spotting anomalous ratings for rating systems by
  analyzing target users and items}.
\newblock \bibinfo{journal}{\emph{Neurocomputing}}  \bibinfo{volume}{240}
  (\bibinfo{year}{2017}), \bibinfo{pages}{25--46}.
\newblock


\bibitem[\protect\citeauthoryear{Yang, Sun, Zhang, and Wang}{Yang
  et~al\mbox{.}}{2020}]%
        {yang2020identification}
\bibfield{author}{\bibinfo{person}{Zhihai Yang}, \bibinfo{person}{Qindong Sun},
  \bibinfo{person}{Yaling Zhang}, {and} \bibinfo{person}{Wei Wang}.}
  \bibinfo{year}{2020}\natexlab{}.
\newblock \showarticletitle{Identification of Malicious Injection Attacks in
  Dense Rating and Co-Visitation Behaviors}.
\newblock \bibinfo{journal}{\emph{IEEE Transactions on Information Forensics
  and Security}}  \bibinfo{volume}{16} (\bibinfo{year}{2020}),
  \bibinfo{pages}{537--552}.
\newblock


\bibitem[\protect\citeauthoryear{Yang, Xu, Cai, and Xu}{Yang
  et~al\mbox{.}}{2016b}]%
        {yang2016re}
\bibfield{author}{\bibinfo{person}{Zhihai Yang}, \bibinfo{person}{Lin Xu},
  \bibinfo{person}{Zhongmin Cai}, {and} \bibinfo{person}{Zongben Xu}.}
  \bibinfo{year}{2016}\natexlab{b}.
\newblock \showarticletitle{Re-scale AdaBoost for attack detection in
  collaborative filtering recommender systems}.
\newblock \bibinfo{journal}{\emph{KBS}}  \bibinfo{volume}{100}
  (\bibinfo{year}{2016}), \bibinfo{pages}{74--88}.
\newblock


\bibitem[\protect\citeauthoryear{Ye, Yang, and Zhang}{Ye et~al\mbox{.}}{2021}]%
        {ye2021optimal}
\bibfield{author}{\bibinfo{person}{Dan Ye}, \bibinfo{person}{Bing Yang}, {and}
  \bibinfo{person}{Tian-Yu Zhang}.} \bibinfo{year}{2021}\natexlab{}.
\newblock \showarticletitle{Optimal stealthy linear attack on remote state
  estimation with side information}.
\newblock \bibinfo{journal}{\emph{IEEE Systems Journal}} \bibinfo{volume}{16},
  \bibinfo{number}{1} (\bibinfo{year}{2021}), \bibinfo{pages}{1499--1507}.
\newblock


\bibitem[\protect\citeauthoryear{Yi, Wu, Zhu, Yu, Zhang, Sun, and Xie}{Yi
  et~al\mbox{.}}{2022}]%
        {yi2022ua}
\bibfield{author}{\bibinfo{person}{Jingwei Yi}, \bibinfo{person}{Fangzhao Wu},
  \bibinfo{person}{Bin Zhu}, \bibinfo{person}{Yang Yu}, \bibinfo{person}{Chao
  Zhang}, \bibinfo{person}{Guangzhong Sun}, {and} \bibinfo{person}{Xing Xie}.}
  \bibinfo{year}{2022}\natexlab{}.
\newblock \showarticletitle{UA-FedRec: Untargeted Attack on Federated News
  Recommendation}.
\newblock \bibinfo{journal}{\emph{arXiv preprint arXiv:2202.06701}}
  (\bibinfo{year}{2022}).
\newblock


\bibitem[\protect\citeauthoryear{You, Li, Ding, Zhang, Feng, Pan, and Yang}{You
  et~al\mbox{.}}{2023}]%
        {you2023anti}
\bibfield{author}{\bibinfo{person}{Xiaoyu You}, \bibinfo{person}{Chi Li},
  \bibinfo{person}{Daizong Ding}, \bibinfo{person}{Mi Zhang},
  \bibinfo{person}{Fuli Feng}, \bibinfo{person}{Xudong Pan}, {and}
  \bibinfo{person}{Min Yang}.} \bibinfo{year}{2023}\natexlab{}.
\newblock \showarticletitle{Anti-FakeU: Defending Shilling Attacks on Graph
  Neural Network based Recommender Model}. In \bibinfo{booktitle}{\emph{WWW}}.
  \bibinfo{pages}{938--948}.
\newblock


\bibitem[\protect\citeauthoryear{Yu et~al\mbox{.}}{Yu et~al\mbox{.}}{2021a}]%
        {yu2021reinforcement}
\bibfield{author}{\bibinfo{person}{Chao Yu} {et~al\mbox{.}}}
  \bibinfo{year}{2021}\natexlab{a}.
\newblock \showarticletitle{Reinforcement learning in healthcare: A survey}.
\newblock \bibinfo{journal}{\emph{CSUR}} \bibinfo{volume}{55},
  \bibinfo{number}{1} (\bibinfo{year}{2021}), \bibinfo{pages}{1--36}.
\newblock


\bibitem[\protect\citeauthoryear{Yu, Yin, Gao, Xia, Zhang, and Hung}{Yu
  et~al\mbox{.}}{2021b}]%
        {yu2021socially}
\bibfield{author}{\bibinfo{person}{Junliang Yu}, \bibinfo{person}{Hongzhi Yin},
  \bibinfo{person}{Min Gao}, \bibinfo{person}{Xin Xia},
  \bibinfo{person}{Xiangliang Zhang}, {and} \bibinfo{person}{Nguyen Quoc~Viet
  Hung}.} \bibinfo{year}{2021}\natexlab{b}.
\newblock \showarticletitle{Socially-Aware Self-Supervised Tri-Training for
  Recommendation}.
\newblock \bibinfo{journal}{\emph{arXiv preprint arXiv:2106.03569}}
  (\bibinfo{year}{2021}).
\newblock


\bibitem[\protect\citeauthoryear{Yue, He, Zeng, and McAuley}{Yue
  et~al\mbox{.}}{2021}]%
        {yue2021black}
\bibfield{author}{\bibinfo{person}{Zhenrui Yue}, \bibinfo{person}{Zhankui He},
  \bibinfo{person}{Huimin Zeng}, {and} \bibinfo{person}{Julian McAuley}.}
  \bibinfo{year}{2021}\natexlab{}.
\newblock \showarticletitle{Black-Box Attacks on Sequential Recommenders via
  Data-Free Model Extraction}. In \bibinfo{booktitle}{\emph{RecSys}}.
  \bibinfo{pages}{44--54}.
\newblock


\bibitem[\protect\citeauthoryear{Zhang, Zhang, Zhang, and Wang}{Zhang
  et~al\mbox{.}}{2018}]%
        {zhang2018ud}
\bibfield{author}{\bibinfo{person}{Fuzhi Zhang}, \bibinfo{person}{Zening
  Zhang}, \bibinfo{person}{Peng Zhang}, {and} \bibinfo{person}{Shilei Wang}.}
  \bibinfo{year}{2018}\natexlab{}.
\newblock \showarticletitle{UD-HMM: An unsupervised method for shilling attack
  detection based on hidden Markov model and hierarchical clustering}.
\newblock \bibinfo{journal}{\emph{KBS}}  \bibinfo{volume}{148}
  (\bibinfo{year}{2018}), \bibinfo{pages}{146--166}.
\newblock


\bibitem[\protect\citeauthoryear{Zhang and Zhou}{Zhang and Zhou}{2012}]%
        {zhang2012meta}
\bibfield{author}{\bibinfo{person}{Fuzhi Zhang} {and}
  \bibinfo{person}{Quanqiang Zhou}.} \bibinfo{year}{2012}\natexlab{}.
\newblock \showarticletitle{A Meta-learning-based Approach for Detecting
  Profile Injection Attacks in Collaborative Recommender Systems.}
\newblock \bibinfo{journal}{\emph{J. Comput.}} \bibinfo{volume}{7},
  \bibinfo{number}{1} (\bibinfo{year}{2012}), \bibinfo{pages}{226--234}.
\newblock


\bibitem[\protect\citeauthoryear{Zhang and Zhou}{Zhang and Zhou}{2014}]%
        {zhang2014hht}
\bibfield{author}{\bibinfo{person}{Fuzhi Zhang} {and}
  \bibinfo{person}{Quanqiang Zhou}.} \bibinfo{year}{2014}\natexlab{}.
\newblock \showarticletitle{HHT--SVM: An online method for detecting profile
  injection attacks in collaborative recommender systems}.
\newblock \bibinfo{journal}{\emph{KBS}}  \bibinfo{volume}{65}
  (\bibinfo{year}{2014}), \bibinfo{pages}{96--105}.
\newblock


\bibitem[\protect\citeauthoryear{Zhang, Liu, and Jin}{Zhang
  et~al\mbox{.}}{2020b}]%
        {zhang2020survey}
\bibfield{author}{\bibinfo{person}{Guijuan Zhang}, \bibinfo{person}{Yang Liu},
  {and} \bibinfo{person}{Xiaoning Jin}.} \bibinfo{year}{2020}\natexlab{b}.
\newblock \showarticletitle{A survey of autoencoder-based recommender systems}.
\newblock \bibinfo{journal}{\emph{Frontiers of Computer Science}}
  \bibinfo{volume}{14}, \bibinfo{number}{2} (\bibinfo{year}{2020}),
  \bibinfo{pages}{430--450}.
\newblock


\bibitem[\protect\citeauthoryear{Zhang, Li, Ding, and Gao}{Zhang
  et~al\mbox{.}}{2020a}]%
        {zhang2020practical}
\bibfield{author}{\bibinfo{person}{Hengtong Zhang}, \bibinfo{person}{Yaliang
  Li}, \bibinfo{person}{Bolin Ding}, {and} \bibinfo{person}{Jing Gao}.}
  \bibinfo{year}{2020}\natexlab{a}.
\newblock \showarticletitle{Practical data poisoning attack against next-item
  recommendation}. In \bibinfo{booktitle}{\emph{WWW}}.
  \bibinfo{pages}{2458--2464}.
\newblock


\bibitem[\protect\citeauthoryear{Zhang, Li, Ding, and Gao}{Zhang
  et~al\mbox{.}}{2022a}]%
        {zhang2022loki}
\bibfield{author}{\bibinfo{person}{Hengtong Zhang}, \bibinfo{person}{Yaliang
  Li}, \bibinfo{person}{Bolin Ding}, {and} \bibinfo{person}{Jing Gao}.}
  \bibinfo{year}{2022}\natexlab{a}.
\newblock \showarticletitle{LOKI: A Practical Data Poisoning Attack Framework
  against Next Item Recommendations}.
\newblock \bibinfo{journal}{\emph{IEEE Transactions on Knowledge and Data
  Engineering}} (\bibinfo{year}{2022}).
\newblock


\bibitem[\protect\citeauthoryear{Zhang, Tian, Li, Su, Yang, Zhao, and
  Gao}{Zhang et~al\mbox{.}}{2021a}]%
        {zhang2021data}
\bibfield{author}{\bibinfo{person}{Hengtong Zhang}, \bibinfo{person}{Changxin
  Tian}, \bibinfo{person}{Yaliang Li}, \bibinfo{person}{Lu Su},
  \bibinfo{person}{Nan Yang}, \bibinfo{person}{Wayne~Xin Zhao}, {and}
  \bibinfo{person}{Jing Gao}.} \bibinfo{year}{2021}\natexlab{a}.
\newblock \showarticletitle{Data Poisoning Attack against Recommender System
  Using Incomplete and Perturbed Data}. In \bibinfo{booktitle}{\emph{KDD}}.
  \bibinfo{pages}{2154--2164}.
\newblock


\bibitem[\protect\citeauthoryear{Zhang, Tang, Ma, Tong, Jing, and Li}{Zhang
  et~al\mbox{.}}{2015b}]%
        {zhang2015panther}
\bibfield{author}{\bibinfo{person}{Jing Zhang}, \bibinfo{person}{Jie Tang},
  \bibinfo{person}{Cong Ma}, \bibinfo{person}{Hanghang Tong},
  \bibinfo{person}{Yu Jing}, {and} \bibinfo{person}{Juanzi Li}.}
  \bibinfo{year}{2015}\natexlab{b}.
\newblock \showarticletitle{Panther: Fast top-k similarity search on large
  networks}. In \bibinfo{booktitle}{\emph{KDD}}. \bibinfo{pages}{1445--1454}.
\newblock


\bibitem[\protect\citeauthoryear{Zhang, Chakrabarti, Ford, and Makedon}{Zhang
  et~al\mbox{.}}{2006}]%
        {zhang2006attack}
\bibfield{author}{\bibinfo{person}{Sheng Zhang}, \bibinfo{person}{Amit
  Chakrabarti}, \bibinfo{person}{James Ford}, {and} \bibinfo{person}{Fillia
  Makedon}.} \bibinfo{year}{2006}\natexlab{}.
\newblock \showarticletitle{Attack detection in time series for recommender
  systems}. In \bibinfo{booktitle}{\emph{KDD}}. \bibinfo{pages}{809--814}.
\newblock


\bibitem[\protect\citeauthoryear{Zhang, Yao, Sun, and Tay}{Zhang
  et~al\mbox{.}}{2019}]%
        {zhang2019deep}
\bibfield{author}{\bibinfo{person}{Shuai Zhang}, \bibinfo{person}{Lina Yao},
  \bibinfo{person}{Aixin Sun}, {and} \bibinfo{person}{Yi Tay}.}
  \bibinfo{year}{2019}\natexlab{}.
\newblock \showarticletitle{Deep learning based recommender system: A survey
  and new perspectives}.
\newblock \bibinfo{journal}{\emph{ACM computing surveys (CSUR)}}
  \bibinfo{volume}{52}, \bibinfo{number}{1} (\bibinfo{year}{2019}),
  \bibinfo{pages}{1--38}.
\newblock


\bibitem[\protect\citeauthoryear{Zhang, Yin, Chen, Huang, Nguyen, and
  Cui}{Zhang et~al\mbox{.}}{2022b}]%
        {zhang2021pipattack}
\bibfield{author}{\bibinfo{person}{Shijie Zhang}, \bibinfo{person}{Hongzhi
  Yin}, \bibinfo{person}{Tong Chen}, \bibinfo{person}{Zi Huang},
  \bibinfo{person}{Quoc Viet~Hung Nguyen}, {and} \bibinfo{person}{Lizhen Cui}.}
  \bibinfo{year}{2022}\natexlab{b}.
\newblock \showarticletitle{Pipattack: Poisoning federated recommender systems
  for manipulating item promotion}. In \bibinfo{booktitle}{\emph{WSDM}}.
  \bibinfo{pages}{1415--1423}.
\newblock


\bibitem[\protect\citeauthoryear{Zhang, Yin, Chen, Hung, Huang, and Cui}{Zhang
  et~al\mbox{.}}{2020c}]%
        {zhang2020gcn}
\bibfield{author}{\bibinfo{person}{Shijie Zhang}, \bibinfo{person}{Hongzhi
  Yin}, \bibinfo{person}{Tong Chen}, \bibinfo{person}{Quoc Viet~Nguyen Hung},
  \bibinfo{person}{Zi Huang}, {and} \bibinfo{person}{Lizhen Cui}.}
  \bibinfo{year}{2020}\natexlab{c}.
\newblock \showarticletitle{Gcn-based user representation learning for unifying
  robust recommendation and fraudster detection}. In
  \bibinfo{booktitle}{\emph{SIGIR}}. \bibinfo{pages}{689--698}.
\newblock


\bibitem[\protect\citeauthoryear{Zhang, Tan, Zhang, Liu, Chua, and Ma}{Zhang
  et~al\mbox{.}}{2015a}]%
        {zhang2015catch}
\bibfield{author}{\bibinfo{person}{Yongfeng Zhang}, \bibinfo{person}{Yunzhi
  Tan}, \bibinfo{person}{Min Zhang}, \bibinfo{person}{Yiqun Liu},
  \bibinfo{person}{Tat-Seng Chua}, {and} \bibinfo{person}{Shaoping Ma}.}
  \bibinfo{year}{2015}\natexlab{a}.
\newblock \showarticletitle{Catch the black sheep: unified framework for
  shilling attack detection based on fraudulent action propagation}. In
  \bibinfo{booktitle}{\emph{IJCAI}}.
\newblock


\bibitem[\protect\citeauthoryear{Zhang, Yuan, Li, Lou, Chen, and Tzeng}{Zhang
  et~al\mbox{.}}{2021b}]%
        {zhang2021reverse}
\bibfield{author}{\bibinfo{person}{Yihe Zhang}, \bibinfo{person}{Xu Yuan},
  \bibinfo{person}{Jin Li}, \bibinfo{person}{Jiadong Lou}, \bibinfo{person}{Li
  Chen}, {and} \bibinfo{person}{Nian-Feng Tzeng}.}
  \bibinfo{year}{2021}\natexlab{b}.
\newblock \showarticletitle{Reverse Attack: Black-box Attacks on Collaborative
  Recommendation}. In \bibinfo{booktitle}{\emph{SIGSAC}}.
  \bibinfo{pages}{51--68}.
\newblock


\bibitem[\protect\citeauthoryear{Zhang-Kennedy and Chiasson}{Zhang-Kennedy and
  Chiasson}{2021}]%
        {zhang2021systematic}
\bibfield{author}{\bibinfo{person}{Leah Zhang-Kennedy} {and}
  \bibinfo{person}{Sonia Chiasson}.} \bibinfo{year}{2021}\natexlab{}.
\newblock \showarticletitle{A systematic review of multimedia tools for
  cybersecurity awareness and education}.
\newblock \bibinfo{journal}{\emph{CSUR}} \bibinfo{volume}{54},
  \bibinfo{number}{1} (\bibinfo{year}{2021}), \bibinfo{pages}{1--39}.
\newblock


\bibitem[\protect\citeauthoryear{Zhao, van~der Aa, Nguyen, Nguyen, and
  Weidlich}{Zhao et~al\mbox{.}}{2021}]%
        {zhao2021eires}
\bibfield{author}{\bibinfo{person}{Bo Zhao}, \bibinfo{person}{Han van~der Aa},
  \bibinfo{person}{Thanh~Tam Nguyen}, \bibinfo{person}{Quoc Viet~Hung Nguyen},
  {and} \bibinfo{person}{Matthias Weidlich}.} \bibinfo{year}{2021}\natexlab{}.
\newblock \showarticletitle{Eires: Efficient integration of remote data in
  event stream processing}. In \bibinfo{booktitle}{\emph{Proceedings of the
  2021 International Conference on Management of Data}}.
  \bibinfo{pages}{2128--2141}.
\newblock


\bibitem[\protect\citeauthoryear{Zhao, Zhang, and Cao}{Zhao
  et~al\mbox{.}}{2023}]%
        {zhao2023manipulating}
\bibfield{author}{\bibinfo{person}{Yaru Zhao}, \bibinfo{person}{Jianbiao
  Zhang}, {and} \bibinfo{person}{Yihao Cao}.} \bibinfo{year}{2023}\natexlab{}.
\newblock \showarticletitle{Manipulating vulnerability: Poisoning attacks and
  countermeasures in federated cloud--edge--client learning for image
  classification}.
\newblock \bibinfo{journal}{\emph{Knowledge-Based Systems}}
  \bibinfo{volume}{259} (\bibinfo{year}{2023}), \bibinfo{pages}{110072}.
\newblock


\bibitem[\protect\citeauthoryear{Zhou et~al\mbox{.}}{Zhou
  et~al\mbox{.}}{2020}]%
        {zhou2020recommendation}
\bibfield{author}{\bibinfo{person}{Quanqiang Zhou} {et~al\mbox{.}}}
  \bibinfo{year}{2020}\natexlab{}.
\newblock \showarticletitle{Recommendation attack detection based on deep
  learning}.
\newblock \bibinfo{journal}{\emph{JISA}}  \bibinfo{volume}{52}
  (\bibinfo{year}{2020}), \bibinfo{pages}{102493}.
\newblock


\bibitem[\protect\citeauthoryear{Zhou, Koh, Wen, Alam, and Dobbie}{Zhou
  et~al\mbox{.}}{2014a}]%
        {zhou2014detection}
\bibfield{author}{\bibinfo{person}{Wei Zhou}, \bibinfo{person}{Yun~Sing Koh},
  \bibinfo{person}{Junhao Wen}, \bibinfo{person}{Shafiq Alam}, {and}
  \bibinfo{person}{Gillian Dobbie}.} \bibinfo{year}{2014}\natexlab{a}.
\newblock \showarticletitle{Detection of abnormal profiles on group attacks in
  recommender systems}. In \bibinfo{booktitle}{\emph{SIGIR}}.
  \bibinfo{pages}{955--958}.
\newblock


\bibitem[\protect\citeauthoryear{Zhou, Wen, Koh, Alam, and Dobbie}{Zhou
  et~al\mbox{.}}{2014b}]%
        {zhou2014attack}
\bibfield{author}{\bibinfo{person}{Wei Zhou}, \bibinfo{person}{Junhao Wen},
  \bibinfo{person}{Yun~Sing Koh}, \bibinfo{person}{Shafiq Alam}, {and}
  \bibinfo{person}{Gillian Dobbie}.} \bibinfo{year}{2014}\natexlab{b}.
\newblock \showarticletitle{Attack detection in recommender systems based on
  target item analysis}. In \bibinfo{booktitle}{\emph{IJCNN}}.
  \bibinfo{pages}{332--339}.
\newblock


\bibitem[\protect\citeauthoryear{Zhou, Wen, Koh, Xiong, Gao, Dobbie, and
  Alam}{Zhou et~al\mbox{.}}{2015}]%
        {zhou2015shilling}
\bibfield{author}{\bibinfo{person}{Wei Zhou}, \bibinfo{person}{Junhao Wen},
  \bibinfo{person}{Yun~Sing Koh}, \bibinfo{person}{Qingyu Xiong},
  \bibinfo{person}{Min Gao}, \bibinfo{person}{Gillian Dobbie}, {and}
  \bibinfo{person}{Shafiq Alam}.} \bibinfo{year}{2015}\natexlab{}.
\newblock \showarticletitle{Shilling attacks detection in recommender systems
  based on target item analysis}.
\newblock \bibinfo{journal}{\emph{PloS one}} \bibinfo{volume}{10},
  \bibinfo{number}{7} (\bibinfo{year}{2015}), \bibinfo{pages}{e0130968}.
\newblock


\bibitem[\protect\citeauthoryear{Zhou, Wen, Qu, Zeng, and Cheng}{Zhou
  et~al\mbox{.}}{2018}]%
        {zhou2018shilling}
\bibfield{author}{\bibinfo{person}{Wei Zhou}, \bibinfo{person}{Junhao Wen},
  \bibinfo{person}{Qiang Qu}, \bibinfo{person}{Jun Zeng}, {and}
  \bibinfo{person}{Tian Cheng}.} \bibinfo{year}{2018}\natexlab{}.
\newblock \showarticletitle{Shilling attack detection for recommender systems
  based on credibility of group users and rating time series}.
\newblock \bibinfo{journal}{\emph{PloS one}} \bibinfo{volume}{13},
  \bibinfo{number}{5} (\bibinfo{year}{2018}), \bibinfo{pages}{e0196533}.
\newblock


\bibitem[\protect\citeauthoryear{Zhou, Wen, Xiong, Gao, and Zeng}{Zhou
  et~al\mbox{.}}{2016}]%
        {zhou2016svm}
\bibfield{author}{\bibinfo{person}{Wei Zhou}, \bibinfo{person}{Junhao Wen},
  \bibinfo{person}{Qingyu Xiong}, \bibinfo{person}{Min Gao}, {and}
  \bibinfo{person}{Jun Zeng}.} \bibinfo{year}{2016}\natexlab{}.
\newblock \showarticletitle{SVM-TIA a shilling attack detection method based on
  SVM and target item analysis in recommender systems}.
\newblock \bibinfo{journal}{\emph{Neurocomputing}}  \bibinfo{volume}{210}
  (\bibinfo{year}{2016}), \bibinfo{pages}{197--205}.
\newblock


\end{thebibliography}


%%
%\newpage
\appendix

\section{Model-Intrinsic RecSys}
This section provides a comprehensive exploration of various recommender systems, including ones based on matrix-factorisation, graphs, neighbourhoods, and deep learning, as well as federated recommender systems.

\subsection{Matrix-factorisation-based RecSys} 

\editthree{
In collaborative filtering, the objective is to decompose the user-item interaction matrix into separate matrices representing the latent factors of the users and items ~\cite{koren2009matrix}.
The user-item interaction matrix, denoted as $\mathcal{R}$, represents the interactions between users and items. In this matrix, $r_{ui}$ denotes the specific interaction between user $u$ and item $i$.
The factorisation can be expressed as:}
\editthree{
\begin{equation}
    \mathcal{R} \approx U * V^T
\end{equation}
}
\editthree{
where $U$ denotes the user latent factor matrix with a dimension of $m \times k$, where $m$ represents the number of users and $k$ represents the number of latent factors.
$V$ represents the item latent factor matrix with a dimension of $n \times k$, where $n$ corresponds to the number of items.
The predicted rating for user $u$ and item $i$, denoted as $\Tilde{r}_{ui}$, can be computed as the dot product of the respective latent vectors:
}
\editthree{
\begin{equation}
    \Tilde{r}_{ui} = U[i, :] * V[j, :]
\end{equation}
}
\editthree{
Matrix factorisation-based recommender systems can be enhanced by including regularisation terms, bias terms, and other techniques to improve performance and address additional factors like user/item biases and temporal dynamics~\cite{alhijawi2020recommender}. Nevertheless, the fundamental principle remains centred around decomposing the rating matrix into low-rank user and item factor matrices, which means personalised recommendations can be generated~\cite{koren2009matrix, alhijawi2020recommender}.}

\subsection{Graph-based RecSys} 
\editthree{
In a graph-based recommender system, users, items, and their relationships are modelled using a graph structure~\cite{guo2020survey} denoted as G = (V, E). Such graphs are constructed using the following procedure.}

\begin{compactitem}
    \item \editthree{The user-item relationship can be represented as an edge, denoted as e = (u, i), which indicates that user u has rated item i. Mathematically, this can be defined as:}
    \editthree{
    \begin{equation}
        e = (u, i) \in E, \text{if } r_{ui} \text{ is not null or zero}
    \end{equation}
    }
    \vspace{-0.5cm}
    \item  \editthree{Item-item relationships are also represented as edges, denoted as e = (i, j), which indicates the similarity between items i and j. Mathematically, this can be defined as:}
    \editthree{
    \begin{equation}
        e = (i, j) \in E,\text{if } sim(i, j) > \tau
    \end{equation}
    where $sim(i, j)$ represents a similarity measure between items $i$ and $j$, while the threshold $\tau$ determines the minimum required similarity for an edge to be established.
    }
\end{compactitem}
\editthree{Using the graph structure, recommendations can be generated by leveraging the relationships between users and items, as well as item-item relationships. The specific recommendation algorithms and techniques can vary depending on the goals and characteristics of the recommender system~\cite{huang2002graph}.}

\subsection{Neighbourhood-based RecSys} 
\editthree{
In neighbourhood-based recommender systems, the selection of neighbours holds significant importance. Neighbours are typically chosen based on their similarity to the target user or item~\cite{periyasamy2017analysis}. Several similarity metrics, such as Pearson correlation coefficient or cosine similarity, can be used to quantify the similarity between users or items. After selecting the neighbourhood, the recommender system predicts the rating for a target user-item pair by considering the ratings of the neighbours. One common approach is weighted average rating prediction, where the predicted rating $\Tilde{r}_{ui}$ for a target user $u$ and item $i$ can be calculated using:}
\editthree{
\begin{equation}
    \Tilde{r}_{ui} =  \frac{\sum (sim(u, v) * r_{ui}) }{\sum sim(u, v)}  
\end{equation}
}
\editthree{
where $sim(u, v)$ denotes the similarity between users $u$ and $v$, $r_{ui}$ represents the rating of user $v$ for item $i$, and the summation is performed over the selected neighbourhood of user $u$~\cite{ sachan2013survey}.}

\subsection{Deep-learning-based RecSys} 
\editthree{
Deep-learning-based recommender systems rely on neural network architectures to capture intricate patterns and representations from the data encompassing user-item interactions~\cite{mu2018survey}. Suppose we examine a deep learning model characterised by the parameters $\theta$. This model accepts user ($z_u$) and item ($z_i$) embeddings as its input and generates predictions for ratings or the probabilities of any interactions between the users and items~\cite{zhang2019deep}. The forward propagation in the deep learning model can be expressed as:
}
\editthree{
\begin{equation}
    \Tilde{\mathcal{R}} = f_\theta(z_u, z_i)
\end{equation}
}
\editthree{
where the function $f_\theta$ represents the mapping function, which is parameterised by $\theta$ and is responsible for calculating the predicted rating or the probability of interaction between the user embedding ($z_u$) and the item embedding ($z_i$)~\cite{ali2020deep}.
}

\subsection{Federated Recommender Systems} 
\editthree{The main distinction between a federated recommender system and a traditional recommender system lies in the approach to data sharing~\cite{zhang2021pipattack}. Unlike a traditional recommender system, a federated recommender system does not share the complete rating matrix with any external entity~\cite{wu2022fedattack}. Instead, users, items, and ratings are dispersed across various local data sources. The global model represents a comprehensive collection of knowledge, and recommendations are generated by combining insights from various local models. The collaborative learning process involves aggregating information from these local models:}
% \editthree{
\begin{compactitem}
    \item \editthree{Local Training: This step involves training an ML model using a subset of the rating matrix $\mathcal{R}$ without the need to share the entire rating matrix with any other party.}

    \item \editthree{Global Aggregation: In this step, the global model parameters $\theta$ are obtained by aggregating the local model parameters $\{\theta_1, \theta_2, ..., \theta_k\}$ from individual data sources $\{L_1, L_2, ..., L_k\}$. The aggregation combines these parameters using techniques such as averaging or weighted aggregation. For instance, the global model can be calculated as $M = \frac{1}{N} \sum_{i=1}^{N} M_i$, where $N$ represents the number of participating nodes and $M_i$ corresponds to the local model at node $i$.}
\end{compactitem}

\editthree{
Federated recommender systems leverage a collaborative global model to generate recommendations by using aggregated knowledge from local data sources. This innovative approach to recommender systems shows great promise in tackling many challenges related to privacy and security~\cite{rong2022poisoning}.}

% \newpage
\section{Evaluation Protocols for Countermeasures}

\begin{table}[!h]
\vspace{-1em}
\centering
\caption{Metrics used in countermeasures.}
\label{tbl:countermeasure_metrics}
\vspace{-1em}
\footnotesize
\scalebox{0.75}{
\begin{tabular}{lll@{\hspace{4em}}ll} 
\toprule
\textcolor[rgb]{0.125,0.129,0.141}{\textbf{Abbrv}} & \multicolumn{1}{l}{\textbf{Name}} &  & \textcolor[rgb]{0.125,0.129,0.141}{\textbf{Abbrv}} & \multicolumn{1}{l}{\textbf{Name}}  \\ 
\cmidrule{1-2}\cmidrule{4-5}
\editthree{RDMA}                                                 & 
\editthree{\begin{tabular}[c]{@{}l@{}}Rating Deviation from \\Mean Agreement\end{tabular}}                           &  & 

\editthree{MSEP}                                                 & \editthree{\begin{tabular}[c]{@{}l@{}}Mean Similarity-based \\Expected Profit\end{tabular} } \\ 

%\cmidrule{1-2}\cmidrule{4-5}
\editthree{ASM}                                                 & 
\editthree{Average Similarity Metric}                              &  & 
\editthree{Hit}                                                  & 
\editthree{Hit Rate Ratio}                             \\ 
%\cmidrule{1-2}\cmidrule{4-5}
\editthree{Spe}                                                & \editthree{Specificity}                    &  & 
\editthree{CS}                                                  & 
\editthree{Classification Error}                             \\ 
%\cmidrule{1-2}\cmidrule{4-5}
\editthree{Sen}                                                 & 
\editthree{Sensitivity}                              &  & 
\editthree{DR}                                                 & 
\editthree{Detection Rate}                             \\ 
%\cmidrule{1-2}\cmidrule{4-5}
\editthree{FPR}                                                 & 
\editthree{False Positive Rate}                            &  & 
\editthree{FAR}                                                 & \editthree{False Alarm Rate}                    \\ 
%\cmidrule{1-2}\cmidrule{4-5}
\editthree{Pre}                                                & 
\editthree{Precision}                       &  & 
\editthree{Hv-Score}                                                 & \editthree{Information Gain}                    \\ 
%\cmidrule{1-2}\cmidrule{4-5}
\editthree{Rec}                                                 & \editthree{Recall}                       &  & 
\editthree{MAS}                                                  & 
\editthree{Mean Absolute Shift}                              \\ 
%\cmidrule{1-2}\cmidrule{4-5}
\editthree{Acc}                                                 & 
\editthree{Accuracy}                             &  & 
\editthree{RMSS}                                                  & 
\editthree{Root Mean Square Shift}                                \\ 
%\cmidrule{1-2}\cmidrule{4-5}
\editthree{F1}                                                & 
\editthree{F1-measure}                           &  & 
\editthree{MTP}                                                 & \editthree{Mean of Total Profit}                        \\ 
%\cmidrule{1-2}\cmidrule{4-5}
\editthree{F2}                                                & \editthree{\begin{tabular}[c]{@{}l@{}}F2-measure--recall has \\twice weight of precision\end{tabular}} 
                     &  &

\editthree{PANT}                                                  & 
\editthree{\begin{tabular}[c]{@{}l@{}}Predicted Anomalous is not \\the Target Items\end{tabular}}                           \\ 

%\cmidrule{1-2}\cmidrule{4-5}
\editthree{AUC}                                                & \editthree{Area under the ROC Curve}   
                 &  & 

\editthree{MDD}                                                  & 
\editthree{Mean Detection Delay}                             \\ 

%\cmidrule{1-2}\cmidrule{4-5}
\editthree{MAE}                                                & \editthree{Mean Absolute Error}             &  & \editthree{Time}                                                & 
\editthree{Timeliness}                \\ 
%\cmidrule{1-2}\cmidrule{4-5}

\editthree{PS}                                                 & \editthree{Prediction Shift}                    
              &  &                                                     
&                                    \\

\bottomrule
\end{tabular}
}
\vspace{-1.5em}
\end{table}

The following section presents a comprehensive overview of the evaluation protocols, datasets, and domains used in various methods to counter poison attacks. ~\autoref{tbl:countermeasure_metrics} presents a summary of the evaluation protocols while ~\autoref{tbl:countermeasure_datasets} lists the  most commonly used datasets. And~\autoref{tbl:attack_domains} provides a summary of countermeasures, including their evaluation metrics, domains, and datasets adopted.

\begin{table}[!h]
\vspace{-0.5em}
\centering
\caption{Datasets used in countermeasures.}
\label{tbl:countermeasure_datasets}
\vspace{-1em}
\footnotesize
\scalebox{0.75}{
\begin{tabular}{l l} 
\toprule
\textbf{Abbrv} & \multicolumn{1}{l}{\textbf{Term}}                                                                                        \\ 
\midrule
\editone{SYN}               &  \editone{Synthetic datasets} \\ 
%\hline
\editone{ML}                  & \editone{MovieLens}                                                                                                  \\ 
%\hline
\editone{NF}                   & \editone{Netflix}                                                                                                      \\ 
%\hline
\editone{AMB}                    & \editone{Amazon Book}                                                                                         \\ 
%\hline
\editone{AMM}                    & \editone{Amazon Movie}                                                                                \\ 
%\hline
\editone{AMP}                    & \editone{Amazon Product}                                                                                          \\ 
%\hline
\editone{CI}                  & \editone{Ciao DVD}                                                                                                      \\
\bottomrule
\end{tabular}
}
\vspace{-.5em}
\end{table}

\begin{table*}[!h]
\caption{Domain and Evaluation Comparison.}
\label{tbl:attack_domains}
\vspace{-0.5em}
\centering
\begin{adjustbox}{max width=\textwidth}
\rowcolors{1}{}{lightgray}
\begin{tabular}{cccccc|cccllllllllcccccccccllccccl|ccc|cc} 
\toprule
\textbf{Name}             &                      & \textbf{Authors}                    &                      & \textbf{Year}            & \multicolumn{1}{c}{}  &                      & \multicolumn{25}{c}{\textbf{\editthree{Evaluation Metric}}}                                                                                                                                                                                                                                                                                                                                                                                                                                                                                                                                                                                                                                                                                             & \multicolumn{1}{l}{} &  & \textbf{\editone{Domain}} & \multicolumn{1}{l}{} &  & \textbf{\editone{Dataset}}  \\
\bottomrule
\end{tabular}
\end{adjustbox}
\vspace{-0.5em}
\end{table*}

\sstitle{Evaluation metrics}
\editthree{When evaluating poison attack detection methods in recommender systems, it is crucial to establish acceptable ranges for evaluation metrics. The ranges for metrics like accuracy, precision, recall, and F1-score vary depending on the context of poison attack detection~\cite{you2023anti}. Factors such as the recommender system's nature, the potential consequences of false positives, and the prevalence of poison attacks in the dataset impact these ranges. Simply achieving high accuracy may not suffice due to imbalanced datasets leading to misleading results~\cite{mehta2007lies}. Precision holds particular significance in poison attack detection, especially in scenarios where false positives can promote harmful or malicious content~\cite{hao2021unsupervised}. Although specific benchmarks for poison attack detection in recommender systems may not be explicitly defined, it is reasonable to consider models with performance scores around 0.8 to 0.9 as acceptable\cite{sundar2020understanding}. This range suggests the method correctly identifies poison attacks approximately 80\% to 90\% of the time. An excellent performance would be indicated by a score exceeding 0.9, indicating a robust defence against poison attacks~\cite{rezaimehr2021survey}. In future research, we aim to conduct comparative evaluations and establish concrete acceptable ranges for poison attack detection metrics in recommender systems. These findings will be presented in a technical report.}

\sstitle{Domains of application} \editone{Countermeasures for poison attacks in recommender systems reveals a primary focus on the movie domain and this emphasis highlights the significance of safeguarding movie recommender systems against manipulation~\cite{aktukmak2019quick,zhou2020recommendation, hao2021unsupervised}. However, researchers have also recognised the broader implications of poison attacks and have developed countermeasures that address multiple domains, including products~\cite{hao2021unsupervised}, books~\cite{cai2019trustworthy}, hotels~\cite{yang2020identification}, and synthetic data~\cite{cai2019bs}. Moreover, the inclusion of POI and location~\cite{you2023anti} as domains of interest reflects the concern for protecting location-based recommender systems. By covering these diverse domains, the reviewed countermeasures aim to enhance the security and reliability of recommender systems, preserving user trust and satisfaction in their recommendations across a wide range of applications.}

\end{document}